%% file: lr.tex
\documentclass[11pt,reqno]{amsart}
\usepackage{comment,amssymb,latexsym,upref,enumerate,fouridx}
\usepackage{finalT}

\numberwithin{equation}{section}
\input lr.def

\begin{document}

\input lr_tit.tex
\input intro.tex

\input conf.tex
\input red.tex
\input eul.tex

\input cev.tex

\input stress.tex

\input total.tex

\input idata.tex
\input exist.tex

\input limit.tex
\input main.tex

\bigskip

\noindent \textit{Acknowledgements:} This work was partially supported by the Australian Research Council grant FT1210045 and
is based upon work supported by the National Science Foundation under Grant No. 0932078 000, while the I was in residence at
the Mathematical Science Research Institute in Berkeley, California, during
month of September, 2013. I am grateful to the MSRI for
its support and hospitality during my visit. I would also like to thank Calum Robertson for useful discussions. Finally,
I would like to thank the
referees for their comments and criticisms, which have
served to greatly improve the content and exposition of this article.

\appendix

\input calc.tex

\input elliptic.tex

\bibliographystyle{amsplain}
\bibliography{refs}


\end{document}

%% file: lr_tit.tex
\title[The Newtonian limit on cosmological scales]{The Newtonian limit on cosmological scales}

\author[T.A. Oliynyk]{Todd A. Oliynyk}
\address{School of Mathematical Sciences\\
Monash University, VIC 3800\\
Australia}
\email{todd.oliynyk@sci.monash.edu.au}


\begin{abstract}
We establish the existence of a wide class of inhomogeneous relativistic solutions to the Einstein-Euler equations
that
are well approximated on cosmological scales by solutions of Newtonian gravity. Error estimates
measuring the difference between the Newtonian and relativistic solutions are provided.
\end{abstract}

\maketitle 

%% file: intro.tex
\sect{intro}{Introduction}
The relationship between Newtonian gravity and General Relativity has been the subject of many investigations
over the years going all the way back to the discovery of General Relativity by Einstein. Most interest in
this subject has focused on understanding the relationship in the setting of isolated systems, and the investigations
have almost exclusively involved formal calculations, see
\cite{Blanchet:2014,Blanchet_et_al:2005,Chandrasekhar:1965,Dautcourt:1964,Ehlers:1986,Einstein_et_al:1938,FutamaseItoh:2007,Kunzle:1972,Kunzle:1976,KunzleDuval:1986} and
references therein, with a few exceptions \cite{Oliynyk:CMP_2007,Oliynyk:CMP_2009,Rendall:1994} where rigorous
results were obtained. More recently, interest has shifted to understanding the relationship between Newtonian gravity and General Relativity on cosmological scales \cite{BuchertRasanen:2012,Ellis:2011,Clarkson_etal:2011,Green_Wald:2011,Green_Wald:2012,Hwangetal:2008,HwangNoh:2013,
KopeikinPetrov:2013,KopeikinPetrov:2014,MatarreseTerranova:1996,Rasanen:2010}.
This shift in interest is primarily due to questions surrounding the physical interpretation of large scale cosmological simulations  using Newtonian gravity and
the role of Newtonian gravity in cosmological averaging.

At the level of field equations,
the relationship between Newtonian gravity and General Relativity can be established through the introduction of a small parameter $\ep=v/c$, where
$v$ is a typical speed of the gravitating matter and $c$ is the speed of light, into the Einstein-matter field
equations. The Newtonian-matter
field equations are then recovered in the \emph{singular} limit $\ep \searrow 0$ by assuming a particular
dependence of the metric and matter fields on the parameter $\ep$. At first glance, this type of formal argument
appears to give a clear answer to the relationship between Newtonian gravity and General Relativity. However, it does
not provide any real answers  because the physics is governed by solutions of the Einstein-matter
equations, and consequently, it is the solutions that must be examined in the limit $\ep \searrow 0$.

Understanding how the solutions of Newtonian gravity and General Relativity are related is much more difficult than simply sending
the parameter $\ep$ to zero in the field equations. For isolated dynamical systems, rigorous results concerning the
relationship between solutions have been established in \cite{Oliynyk:CMP_2007,Oliynyk:CMP_2009,Rendall:1994}.
In the articles \cite{Oliynyk:CMP_2010,Oliynyk:JHDE_2010}, we adapted the approach taken in  \cite{Oliynyk:CMP_2007,Oliynyk:CMP_2009} to
the cosmological setting, and we were able to
construct 1-parameter families of $\ep$-dependent solutions to the Einstein-Euler equations
that limit as $\ep \searrow 0$ to solutions of the cosmological Poisson-Euler equations of Newtonian
gravity. However, as pointed out in  \cite{Green_Wald:2012}, the class of solutions that we constructed were
not valid on cosmological scales, and therefore did not address the relationship between Newtonian gravity and
General Relativity on cosmological scales.


To see why the results \cite{Oliynyk:CMP_2010,Oliynyk:JHDE_2010} are not valid on a cosmological scale, we recall the more familiar and well studied setting of isolated systems.
In this setting, we established the existence of 1-parameter families of $\ep$-dependant solutions, $0<\ep<\ep_0$,
to the Einstein-Euler equations that
converge, in a suitable sense, on $[0,T]\times \Rbb^3$  to solutions of the Poisson-Euler equations of Newtonian gravity. In relativistic coordinates, the metrics
from these solutions converge uniformly to the Minkowski metric as $\ep \searrow 0$, while the fluid proper energy densities are of characteristic
size $\sim \ep$ and converge, when suitably rescaled, to Dirac delta functions about their centers of mass. This agrees
with the expected behavior for the Newtonian limit of isolated systems. For readers unfamiliar with this viewpoint
on the Newtonian limit, a more traditional viewpoint is achieved by transforming to Newtonian coordinates by rescaling the spatial coordinates by a power of $\ep$.
Expressed in these coordinates, the metrics converge to degenerate two tensors, while the fluid proper energy densities and spatial components of the four-velocities converge
uniformly to solutions of the Poisson-Euler equations on $[0,T]\times \Rbb^3$.

Returning to the cosmological setting, we established in \cite{Oliynyk:CMP_2010,Oliynyk:JHDE_2010} the existence of 1-parameter families of $\ep$-dependant, $0<\ep <\ep_0$, solutions to the Einstein-Euler
equations on $[0,T]\times \Tbb^n$ that converge to solutions of the cosmological Poisson-Euler equations
as $\ep \searrow 0$. Lifting these solutions to the covering
space, they become periodic solutions of the Einstein-Euler equations on $[0,T]\times \Rbb^n$ with period $\sim \ep$.
This leads, as discussed in \cite{Green_Wald:2012}, to the interpretation of these solutions as being local since the ``size'' of the torus on which the solutions are defined is
of characteristic size $\sim \ep$, and hence, shrinks to zero in the limit $\ep \searrow 0$
in complete analogy with the behavior of isolated systems. One concludes from this that the solutions from \cite{Oliynyk:CMP_2010,Oliynyk:JHDE_2010} do not represent fully relativistic solutions
that converge on cosmological scales to solutions of the cosmological Poisson-Euler equations, but instead, represent solutions that converge on scales comparable to isolated systems.


The main aim of this article is to show that the deficiencies identified by \cite{Green_Wald:2012} in the approach taken in
\cite{Oliynyk:CMP_2007,Oliynyk:CMP_2009} can be fixed.
In order to describe our results, we must first fix our notation. We begin by recalling that
the Einstein-Euler equations for an insentropic perfect fluid with a cosmological constant
are given by
\lalign{EE}{
&\grave{G}^{ij} = 2\bigl(\grave{T}^{ij}-\Lambda \grave{g}^{ij}\bigr), \label{EE.1}\\
&\grave{\nabla}_i \grave{T}^{ij} = 0, \label{EE.2}
}
where $\grave{G}^{ij}$ is the Einstein tensor of the metric
\leqn{gtdef}{
\grave{g} = \grave{g}_{ij} d\xb^i d\xb^j,
}
$\Lambda$ is the cosmological constant, and
\eqn{Ttdef}{
\grave{T}^{ij} = (\rhob + \pb)\vb^i \vb^j + \pb\grave{g}^{ij} \qquad (\vb_i\vb^i = -1)
}
is the perfect fluid stress energy tensors with pressures determined by the equation of
state
\leqn{eosdef}{
\pb = \ep^2 f(\rhob),
}
which we assume is smooth, positive and increasing, that is, $f,f'\in C^\infty(\Rbb_{>0},\Rbb_{>0})$.

Here, and in the following, we use lower case Latin letters, i.e. $i,j,k$ etc.,
to label spacetime coordinate indices that run from 0 to 3, while upper case Latin letters, i.e. $I,J,K$ etc.,
will label spatial coordinate indices that run from 1 to 3. We take the
$(\xb^i)$ $(i=0,1,2,3)$ to denote Cartesian coordinates on a spacetime slab of the form $M=[0,T)\times \Rbb^3$,
and refer to them as \emph{relativistic coordinates}.\footnote{Ultimately, this name is justified by the fact that in the limit $\ep \searrow 0$
the light cones of the metric $\grave{g} = \grave{g}_{ij}d\xb^i d\xb^j$  converge uniformly to the standard light cones
$(\xb^0)^2 - \delta_{IJ}\xb^I\xb^J = 0$ of the Minkowski metric $\eta = \eta_{ij} d\xb^i d\xb^j$ defined by the coordinates $(\xb^i)$.} We assume that
$\xb^0$ is a time coordinate parameterizing the interval $[0,T)$, and we use the
vector field $\delb{0}$ to define a (future) time orientation. The fluid four-velocity
$\vb=\vb^i \delb{i}$
is assumed to be future oriented.

What we establish in this article is the existence of 1-parameter families
of solutions $\{\grave{g}_\ep^{ij},\vb_i^\ep,\rhob_\ep\}$, $0<\ep < \ep_0$, to \eqref{EE.1}-\eqref{EE.2} that are
well approximated on cosmological scales by solutions of the following limit equations:
\lalign{Eulexp0aa}{
\del{0} \mut + \zt^I\del{I}\mut  +  \rhot \del{}^I \zt_I &= 3\betat'\mut, \label{Eulexp0aa.1}\\
\rhot\bigl(\del{0} \zt_J+\zt^I\del{I} \zt_J\bigr) +
f'\bigl(\rhot\bigr)\del{J}\mut
 &= - \Quarter \rhot\Phit_J +\betat'\rhot\zt_J, \label{Eulexp0aa.2} \\
 \del{0} \bigl(e^{-\betat}\Phit_I\bigr) - 4 \Rf_I\Rf_J \Bigl(e^{-3\betat}\rhot\zt^J\Bigr)&=0,
\label{Eulexp0aa.3} \\
\del{0}\Phit - \betat'\Phit + 4 e^{-2\betat} \Rf_J (-\Delta)^{-\frac{1}{2}}(\rhot \zt^J) &= 0,
\label{Eulexp0aa.4} \\
\betat''-\Half\bigl(\betat'\bigr)^2 &= -e^{-2\betat}\Lambda,\label{Eulexp0aa.5}
}
where
\leqn{Eulexp0ab}{
\rhot(t,\mathbf{x}) = e^{3\betat(t)} + \mut(t,\mathbf{x}).
}
Here, $(\cdot)' = d(\cdot)/d{x^0}$ denotes differentiation with respect to time $x^0$, $\mathbf{x}=(x^I)$ are
Cartesian coordinates on $\Rbb^3$, $\del{I}=\del{}/\del{}x^I$, $\Delta=\delta^{IJ}\del{I}\del{J}$ is the Euclidean
Laplacian, $\Rf_I$
is the Riesz transform, see \eqref{Riesztransdef}, $(-\Delta)^{-\frac{1}{2}}$ is the Riesz potential,
see \eqref{Rieszpotdef}, and the spatial indices (i.e. $I,J,K$) are raised and lowered with the Euclidean metric $\delta_{IJ}$.

As we show in Section \ref{limit}, solutions to the above limit equations coincide with
solutions of the \emph{cosmological Poisson-Euler equations} for the type of initial data that we consider.
This provides the interpretation of $\rhot$ as the Newtonian fluid density, and
\eqref{Eulexp0ab} as a decomposition of the density into a sum of homogenous and inhomogeneous components given
  by $e^{3\betat}$ and $\mut$, respectively.
The pressure of the fluid
is determined by the equation of state
\eqn{neos}{\pt=f(\rhot),
}
while $\zt_I$, $\Lambda$ and $\betat$ represent the conformally rescaled fluid three-velocity, the cosmological constant,
and a time dependent conformal factor that accounts for the expansions of space, respectively.
In the following, we refer to the coordinates $(x^i)=(x^0,x^I)$ as \emph{Newtonian
coordinates}. The relationship between the Newtonian and relativistic coordinates is fixed by the simple scaling relation
\eqn{coords}{
\xb^0 = x^0, \quad \xb^J = \ep x^J.
}

In order for solutions of \eqref{Eulexp0aa.1}-\eqref{Eulexp0aa.5} to be cosmologically relevant,
the initial data must be chosen
correctly. The main requirement on the initial data is that the inhomogeneous component of the fluid
density should be composed of localized fluctuations that represent local, near-Newtonian subsystems for which
the light travel time between the localized fluctuations remains bounded away from zero
in the limit $\ep \searrow 0$. In particular, we are looking, in relativistic coordinates, for 1-parameter families
of $\ep$-dependent families of initial data that can be separated into homogenous and inhomogeneous components
where the homogeneous component has a regular limit as $\ep \searrow 0$, while the inhomogeneous component consists of a finite number of
spikes with characteristic width $\sim \ep$ that can be centered at arbitrarily chosen, $\ep$-independent points and
converge, when suitably rescaled, to Dirac delta functions. Initial data of this type represents cosmological
initial data that deviates from homogeneity due to presence of a finite number
of density fluctuations that remain casually separated from each other and behave as isolated systems in the limit $\ep \searrow 0$.
To compare with our previous work, this initial data arose out of our desire to extend the results from \cite{Oliynyk:CMP_2007,Oliynyk:CMP_2009}  to
cosmological scales by essentially “gluing” together multiple,
spatially separated solutions that have local behavior
similar to the solutions from \cite{Oliynyk:CMP_2007,Oliynyk:CMP_2009}.

The starting point for selecting initial data satisfying the above requirements is to separate it into
free and constrained components. For the free initial data, we choose
\lalign{cpeidata}{
\betat(0) &= \betah, \label{cpeidtat.1} \\
\mut|_{x^0=0} &= \mubr_{\ep,\vec{\yv}} := \Quarter e^{2\betah} \tauh_{00}^{\ep,\vec{\yv}},\label{initE.3}\\
\zt_J|_{x^0=0} &= \zbr^{\ep,\vec{\yv}}_J := \frac{1}{4e^{\betah}+\tauh_{00}^{\ep,\vec{\yv}}}\Bigl[\sqrt{\Twothirds e^{-2\betah}\bigl(e^{3\betah}+\Lambda\bigr)}
\Rf_{J}(-\Delta)^{-\frac{1}{2}}\tauh_{00}^{\ep,\vec{\yv}}+\tauh^{0J}_{\ep,\vec{\yv}}\Bigr],
\label{initE.4}
}
where
\leqn{idenD}{
\tauh^{\ep,\vec{\yv}}_{00}(\xv) = \sum_{\lambda =1}^N \tauh^\lambda_{0j}\left(\xv - \frac{\yv_\lambda}{\ep}\right), \quad \tauh_{\ep,\vec{\yv}}^{0J}(\xv) = \sum_{\lambda =1}^N \tauh_\lambda^{0J}\left(\xv - \frac{\yv_\lambda}{\ep}\right)
}
with
\lalign{yvdenBvelB}{
&\betah \in \Rbb, \qquad \vec{\yv} = (\yv_1,\yv_2,\ldots,\yv_N) \in \Rbb^{3N}, \label{yvdef} \\
&\tauh_{00}^\lambda, \tauh^{0J}_\lambda \in L^{\frac{6}{5}}(\Rbb^3)\cap H^{s+1}(\Rbb^3) \qquad 1\leq \lambda \leq N, \label{idenB}
}
and $s\in \Zbb_{> 3/2+1}$.

\begin{rem} \label{sobcoordsrem}
All Sobolev based function spaces employed in this article are defined with respect to the Newtonian coordinates. For example, the $H^s(\Rbb^3)$, $s\in \Zbb_{\geq 0}$, norm is given
by
\eqn{sobcoordsrem1}{
\norm{u}_{H^s(\Rbb^3)}^2 = \sum_{|\alpha|\leq s} \int_{\Rbb^3} |D^\alpha u (\xv)|^2 \, d^3 x
}
where $\alpha = (\alpha_1,\alpha_2,\alpha_3) \in \Zbb_{\geq 0}^3$ is a multi-index,
\eqn{sobcoordsrem2}{
D^\alpha u(\xv) = \del{1}^{\alpha_1} \del{2}^{\alpha_2} \del{3}^{\alpha_3} u(\xv),
}
and as above, $\mathbf{x}=(x^I)$ are Cartesian coordinates on $\Rbb^3$, and $\del{I}=\del{}/\del{}x^I$ are the associated partial derivatives.
\end{rem}

For fixed $\ep >0$, this initial data
represents a fluid that initially consists of a homogenous component $e^{3\betah}$  superimposed with $N$ density\footnote{We can even take $N=\infty$
provided that the $\tauh_{00}^\lambda$ and $\tauh^{0J}_\lambda$ satisfy
$\sum_{\lambda=0}^\infty\bigl( \norm{\tauh^\lambda_{00}}_{L^{6/5}}+\norm{\tauh_{00}^\lambda}_{H^{s+1}}\bigr) < \infty$
and $\sum_{\lambda=0}^\infty\bigl( \norm{\tauh^{0J}_\lambda}_{L^{6/5}}+\norm{\tauh^{0J}_\lambda}_{H^{s+1}}\bigr) < \infty$, respectively.}
fluctuations that are centered at the spatial (Newtonian) points $\mathbf{y}_\lambda/\ep \in \Rbb^3$, $\lambda=1,2,\ldots, N$, and have profiles given by the functions $\tauh^\lambda_{00}$. Switching to relativistic coordinates, it is clear that this
data represents localized density spikes of characteristic width $\sim \ep$ that are centered at the spacetime points
$(0,\yv_\lambda)$, $\lambda=1,\ldots,N$. Since, as we show, the nearby relativistic solutions have light cones that, when expressed in relativistic coordinates, are uniformly close to the standard Minkowskian cones, the light travel time between the different density spikes remains bounded away from zero in the limit $\ep \searrow 0$.


Clearly, we must also select $\betah \in \Rbb$ in a way that is compatible with the
requirement that
\leqn{idenF}{
e^{\betah} + \Quarter \tauh_{00}^{\ep,\vec{\yv}}(\xv) \geq c_0 > 0 \quad \forall\, (\ep,\xv,\vec{\yv}) \in (0,\ep_0)\times \Rbb^3\times \Rbb^{3N}
}
for some positive constant $c_0>0$. There are many choices of $\betah$ and profiles $\tauh^\lambda_{00}$ that
satisfy this condition. For example, we could choose non-negative profiles $\tauh^\lambda_{00} \geq 0$, $1\leq \lambda \leq N$, and pick $\betah$ arbitrarily.

The constrained initial data is determined in terms of the free data by
\lalign{idmatAb}{
\betat'(0)& = \betabr_0 := \sqrt{\Twothirds e^{-2\betah}(e^{3\betah}+\Lambda)},  \label{idmatAb.1} \\
\Phit_I|_{x^0=0} &= \Phibr^{\ep,\vec{\yv}}_I :=  \Rf_I(-\Delta)^{-\frac{1}{2}}(4e^{-2\betah}\mubr_{\ep,\vec{\yv}})
\label{idmatAb.2}
\intertext{and}
\Phit|_{x^0=0} &= \Phibr^{\ep,\vec{\yv}}   := \Delta^{-1}(4e^{-2\betah}\mubr_{\ep,\vec{\yv}}), \label{idmatAc}
}
where $\Delta^{-1}$ is the Newtonian potential, see \eqref{Newtpot}.

With the notation fixed and the setup complete, we are ready to state an informal version of our main result,
which guarantees the existence of 1-parameter families of fully relativistic solutions to the Einstein-Euler
equations that are well approximated, for small $\ep$, by solutions to the IVP consisting of
\eqref{Eulexp0aa.1}-\eqref{Eulexp0aa.5}, \eqref{cpeidtat.1}-\eqref{initE.4} and \eqref{idenF}-\eqref{idmatAc}; see Theorem \ref{mainthm} for the precise version which
includes a characterization of the initial data used to generate the 1-parameter family of relativistic solutions.
This result was announced previously in \cite{Oliynyk:PRD_2014}.
\begin{thm} \label{maininf}
Suppose $s\in \Zbb_{>3/2+1}$, $N\in \Zbb_{\geq 0}$, $\betah\in \Rbb$, $\tauh_{0j}^\lambda \in L^{\frac{6}{5}}(\Rbb^3)\cap H^{s+1}(\Rbb^3)$, $0\leq \lambda \leq N$, and \eqref{idenF} is satisfied.
Then there exist $\ep_0,T>0$ and 1-parameter families of solutions\footnote{See Appendix \ref{Zhidcalc} for a definition of the spaces $K^s(\Rbb^n)$.}
\gath{maininf1}{
\mut_{\ep,\vec{\yv}},\zt_J^{\ep,\vec{\yv}},\in \bigcap_{m=0}^1 C^m([0,T),H^{s+1-m}(\Rbb^3)),\quad
\betat \in C^{\infty}([0,T)),\\
\Phit_J^{\ep,\vec{\yv}} \in \bigcap_{m=0}^1 C^m([0,T),H^{s+1}(\Rbb^3)), \quad  \Phit^{\ep,\vec{\yv}} \in \bigcap_{m=0}^1 C^m([0,T), L^6(\Rbb^3)\cap K^{s+2}(\Rbb^3)),
}
where $(\ep,\vec{\yv})\in (0,\ep_0)\times \Rbb^{3N}$,
to the system \eqref{Eulexp0aa.1}-\eqref{Eulexp0aa.5} on the spacetime region $M=[0,T)\times \Rbb^3$
that satisfy the initial conditions \eqref{cpeidtat.1}-\eqref{initE.4} and \eqref{idenF}-\eqref{idmatAc}.
Moreover, for each $(\ep,\vec{\yv})\in (0,\ep_0)\times \Rbb^{3N}$,
there exists maps
\alin{maininf2}{
&\psi^{\ep,\vec{\yv}},\psi^{\ep,\vec{\yv}}_0, \mu_{\ep,\vec{\yv}}, u^{ij}_{\ep,\vec{\yv}}, z_0^{\ep,\vec{\yv}} \in \bigcap_{\ell=0}^1 C^\ell([0,T),K^{s-\ell}(\Rbb^3)\cap L^6(\Rbb^3)), \\
&\psi^{\ep,\vec{\yv}}_I, z_I^{\ep,\vec{\yv}}, \Phi_I^{\ep,\vec{\yv}}, u_{k,\ep,\vec{\yv}}^{ij} \in \bigcap_{\ell=0}^1 C^\ell([0,T),H^{s-\ell}(\Rbb^3)) ,
\\
&\alpha_\ep, \beta_\ep \in C^\infty([0,T)),
}
that
\begin{enumerate}[(i)]
\item
satisfy the estimates
\alin{maininf3}{
\norm{u^{ij}_{\ep,\vec{\yv}}}_{L^\infty([0,T),K^{s-1}\cap L^6)}+
\norm{u^{ij}_{0,\ep,\vec{\yv}}}_{L^\infty([0,T),K^{s-1}\cap L^6)}
 \hspace{0.8cm} &\notag \\
+ \norm{u^{ij}_{I,\ep,\vec{\yv}}-\delta^i_0\delta^j_0\Phit_I^{\ep,\vec{\yv}}}_{L^\infty([0,T),K^{s-1}\cap L^6)}  &\lesssim \ep, \\
\norm{\psi^{\ep,\vec{\yv}}}_{L^\infty([0,T),K^{s-1}\cap L^6)}
+\norm{\psi^{\ep,\vec{\yv}}_i}_{L^\infty([0,T],K^{s-1}\cap L^6)}
&\lesssim \ep, \\
\norm{\mu_{\ep,\vec{\yv}}-\mut_{\ep,\vec{\yv}}}_{L^\infty([0,T),K^{s-1}\cap L^6)} +\norm{z_0^{\ep,\vec{\yv}}}_{L^\infty([0,T),K^{s-1}\cap L^6)} \hspace{0.8cm}&
\notag \\
 + \norm{z_I^{\ep,\vec{\yv}}-\zt_I^{\ep,\vec{\yv}}}_{L^\infty([0,T),K^{s-1}\cap L^6)} &\lesssim \ep, \\
\norm{\alpha_\ep-e^{3\betat}}_{L^\infty([0,T))}+
 \norm{\beta_\ep-\betat}_{L^\infty([0,T))}
 +\norm{\beta'_\ep-\betat'}_{L^\infty([0,T))}  &\lesssim \ep,
}
and
\item determine a solution
\eqn{maininf4}{
\grave{g} = \grave{g}{}^{ij}_{\ep,\vec{\yv}}(\xb)\delb{i}\delb{j}, \quad
\vb = \vb_i^{\ep,\vec{\yv}}(\xb)d\xb^i, \quad \rhob = \rhob_{\ep,\vec{\yv}}(\xb),
}
in relativistic coordinates
to the Einstein-Euler equations \eqref{EE.1}-\eqref{EE.2} on the spacetime region $M=[0,T)\times \Rbb^3$
according to the formulas:
\lalign{maininf5}{
\grave{g}^{ij}_{\ep,\vec{\yv}}(\xb^0,\mathbf{\xb}) &= \frac{e^{2(\beta_\ep(\xb^0)+\ep\bar{\psi}^{\ep,\vec{\yv}}(\xb^0,\mathbf{\xb}))}}{
\sqrt{-\det\bigl(\gbh_{\ep,\vec{\yv}}^{kl}(\xb^0,\mathbf{\xb})}\bigr)}
\gbh^{ij}_{\ep,\vec{\yv}}(\xb^0,\mathbf{\xb}), \label{maininf5.1} \\
\rhob_{\ep,\vec{\yv}}(\xb^0,\mathbf{\xb}) &= \alpha_\ep(\xb^0) + \mu_{\ep,\vec{\yv}}(\xb^0,\mathbf{\xb}/\ep), \label{maininf5.2} \\
\vb_i^{\ep,\vec{\yv}}(\xb^0,\mathbf{\xb}) &= e^{-\beta_\ep(\xb^0)-\ep\bar{\psi}^{\ep,\vec{\yv}}(\xb^0,\mathbf{\xb})}
\bigl( -\delta_i^0 + \ep z_i^{\ep,\vec{\yv}}(\xb^0,\mathbf{\xb}/\ep)\bigr), \label{maininf5.3}
}
where
\alin{ghat}{
\gbh^{ij}_{\ep,\vec{\yv}}(\xb^0,\mathbf{\xb}) &= \eta^{ij} + \ep u^{ij}_{\ep,\vec{\yv}}(\xb^0,\mathbf{\xb}/\ep),\\
\delb{k} \gbh^{ij}_{\ep,\vec{\yv}}(\xb^0,\mathbf{\xb}) &= \ep u^{ij}_{k,\ep,\vec{\yv}}(\xb^0,\mathbf{\xb}/\ep),\\
\bar{\psi}_{\ep,\vec{\yv}}(\xb^0,\mathbf{\xb}) &= \psi^{\ep,\vec{\yv}}(\xb^0,\mathbf{\xb}/\ep), \\
\delb{k} \bar{\psi}^{\ep,\vec{\yv}}(\xb^0,\mathbf{\xb}) &= \psi^{\ep,\vec{\yv}}_k(\xb^0,\mathbf{\xb}/\ep),
}
and the conformal harmonic gauge condition $\delb{i}\gbh^{ij}_{\ep,\vec{\yv}} = 0$ is satisfied.
\end{enumerate}

\end{thm}

\subsection{Overview} The proof of Theorem \ref{maininf} is based on a conformal version of the Einstein-Euler
equations. More specifically, we introduce a first order formulation
of the harmonically reduced conformal Einstein equations in Section \ref{red}, derive a
symmetric hyperbolic formulation of the conformal Euler equations in Section \ref{ceul}, and
select an evolution equation for the conformal factor in Section \ref{cev}. The particular choices
of variables used to represent the gravitational field and the fluid, and the evolution
equation for the conformal factor are made so that the resulting equations, which are singular in $\ep$, are in a suitable
form to analyze the $\ep \searrow 0$ limit for the class of initial data that is considered
here. In Section \ref{stress}, we also
provide a decomposition of the stress energy tensor $\Tb^{ij}$ that is subsequently used
to show that the components of the stress energy tensor remain in $L^2$ under evolution.
This fact is important for establishing the proof of Theorem \ref{maininf}, and
is not obvious given the evolution equations and initial data that we are considering.

In Section \ref{idata}, suitable families of initial data depending on the parameters
$(\ep,\vec{\yv})$ are constructed using
a fixed point argument that relies on the fact that the constraint equations,
unlike the evolution equations, depend regularly on the parameter $\ep$ for our choice
of gravitational and matter variables. Using this initial data, a local uniform existence and uniqueness result\footnote{That is, local
in time and uniform in the parameters $(\ep,\vec{\yv})$.} for
solutions of the reduced conformal Einstein-Euler equations is established in Section \ref{exist}; see
Proposition \ref{uniprop}. It is also shown that these solutions satisfy the (non-reduced) conformal Einstein-Euler equations.

Proposition \ref{uniprop} is established in 3 steps. In the first step,
we establish, see Proposition \ref{locexist}, the local existence of solutions to the reduced
conformal Einstein-Euler equations, and also the (non-reduced) conformal Einstein-Euler equations in the uniformly local
Sobolev spaces $H^s_{\text{ul}}$. At this point, the time of existence depends on
the parameters $(\ep,\vec{\yv})$. Next, we show in Section \ref{aposteriori} that these solutions
lie in standard Sobolev spaces, and that the time of existence does not depend on the parameters $\vec{\yv}$.
In the final step, we employ a non-local modification of the evolution equations in
conjunction with technqiues from the theory of singular hyperbolic equations to establish, in
Section \ref{unibounds}, that the time of existence is bounded away from zero for $\ep \in (0,\ep_0]$
where $\ep_0>0$ is a suitably small fixed constant.

In Section \ref{limit}, we establish the local existence and uniqueness of solutions
to the cosmological Poisson-Euler equations for suitable initial data that corresponds
to the cosmological Newtonian limit of the relativistic initial data constructed in Section \ref{idata};
see Proposition \ref{limitexist}.
As in the relativistic case, the initial data depends on the parameters $(\ep,\vec{\yv})$, and
the time of existence is shown to be uniform in the parameters $(\ep,\vec{\yv})$.

The final
step in the proof of Theorem \ref{maininf} is contained in Section \ref{main}. There we derive
the equation satisfied by the difference between the fully relativistic solutions and the
Newtonian solutions. This equation is singular in $\ep$, and we use techniques from
the theory of singular hyperbolic equations to show that the difference is of order $\ep$
on a time interval that is independent of the parameters $\vec{\yv}$ and bounded away
from zero for $\ep \in (0,\ep_0]$. This
establishes the error estimate that controls
the difference between the fully relativistic solutions and their Newtonian counterparts, and
shows rigorously that the fully relativistic solutions considered here are approximated on
cosmological scales by solutions of Newtonian gravity. The precise statement of
these results are given in Theorem \ref{mainthm}.

%% file: conf.tex
\sect{conf}{The conformal Einstein-Euler equations}

In order to define the Newtonian limit, some background structure is required. 
The appropriate background structure in the cosmological setting
is provided by a FLRW metric with flat spatial slices. To take advantage
of the fact that such FLRW metrics are conformally flat,  we will work
with a conformally rescaled metric $\gb_{ij}$ given by
\leqn{gbdef}{
\gb_{ij} = e^{2\Psib} \grave{g}_{ij}
}
instead of the physical one. A particular advantage of this transformation is
that it will allow us to work with a flat background and employ 
harmonic coordinates without the need for any gauge source terms.

Under the conformal transformation,
the Einstein equations \eqref{EE.1} transform as
\leqn{cEE.1}{
\Gb^{ij} = 2\Tb^{ij},
}
where $\Gb^{ij}$ is the Einstein tensor of the conformal metric \eqref{gbdef}, and
\leqn{Tbdef}{
\Tb^{ij} = e^{-4\Psib}\grave{T}^{ij} - e^{-2\Psib}\Lambda\gb^{ij}+\Cb^{ij}
}
with
\eqn{Cdefa}{
\Cb^{ij} = -(\nablab^i\nablab^j\Psib + \nablab^i\Psib\nablab^j\Psib)+\bigl(\nablab_k\nablab^k\Psib -
\Half \nablab_k\Psib \nablab^{k}\Psib\bigr)\gb^{ij}.
}
Letting $\grave{\Gamma}{}^k_{ij}$ and $\Gammab^k_{ij}$ denote the Christoffel symbols of the metrics \eqref{gtdef}
and \eqref{gbdef}, respectively, the difference $\grave{\Gamma}{}^k_{ij}-\Gammab^k_{ij}$ is readily calculated to be
\eqn{Gamdiff}{
\grave{\Gamma}{}^k_{ij}-\Gammab^k_{ij} = -\gb^{kl}\bigl(\gb_{il}\nablab_j\Psib
+ \gb_{jl}\nablab_i\Psib - \gb_{ij} \nablab_l\Psib \bigr).
}
Using this, we can express the Euler equations \eqref{EE.2} as
\leqn{cEE.2}{
\nablab_i \grave{T}^{ij} = 6\grave{T}^{ij}\nablab_i\Psib -\gb_{lm}\grave{T}^{lm}
\gb^{ij}\nablab_i\Psib.
}

%% file: red.tex
\subsect{red}{The Reduced conformal Einstein Equations}
Our first step in analyzing the $\ep \searrow 0$ limit is to derive a suitable formulation
for the Einstein equations. The formulation that we use is a slight variation of the one
employed in the articles  \cite{Oliynyk:CMP_2007,Oliynyk:CMP_2010}.

We begin the derivation by
introducing a background metric
\eqn{hdef}{
\hb = \hb_{ij}d\xb^i d\xb^j,
}
and taking the symmetric 2-tensor $\ub^{ij}$, defined by
\leqn{ubdef}{
\ub^{ij} =\frac{1}{\ep} \left( \frac{\sqrt{|\gb|}}{\sqrt{|\hb|}}\gb^{ij}-\hb^{ij}\right),
}
where
\eqn{voldef}{
|\gb| = -\det(\gb_{ij}) \AND |\hb| = -\det(\hb_{ij}),
}
as our primary gravitational variables. The equivalence of the
$\ub^{ij}$ and the conformal metric $\gb_{ij}$ follows
from the formula
\leqn{utog}{
\gb^{ij} = \frac{1}{\sqrt{-|\hb|\det(\gbh^{kl})}}\gbh^{ij},
}
where
\leqn{ghdef}{
\gbh^{ij} = \hb^{ij} + \ep \ub^{ij}.
}
It is clear from this formula, that at each spacetime point $\xb \in M$
the pair
$(\ep,\ub^{ij}(\xb))$, subject to the restriction\footnote{Here, $\Sbb{4}$
denotes the set of symmetric, $4\times 4$ matrices.}
\leqn{VcdefA}{
(\ep\ub)\in \Vc_{\xb} := \{ s=(s^{ij}) \in \Sbb{4}\, | \, -\det(\hb^{ij}(\xb)+ s^{ij}) > 0 \},
}
completely determines the metric at $\xb$, and vice-versa.

Substituting \eqref{utog} into the standard formula for the Christoffel symbols
gives
\lalign{christZ}{
\Gammab^k_{ij} = \gammab^k_{ij}& +\ep\Bigl(-\gbh_{l(i}\Db_{j)}\ub^{kl}+\Half\gbh^{kl}\gbh_{im}\gbh_{jn}\Db_l\ub^{mn} -\Quarter
   \gbh^{kl}\gbh_{ij}\gbh_{mn}\Db_l\ub^{mn} +\Half\gbh_{lm}\delta^k_{(i}\Db_{j)}\ub^{lm}\Bigr), \label{christ}
}
where
\eqn{ghinv}{
(\gbh_{ij})  = (\gbh^{ij})^{-1},
}
$\Db_k$ is Levi-Civita connection of $\hb_{ij}$, and $\gammab^k_{ij}$ are the corresponding
Christoffel symbols.
Using \eqref{christ} to expand the Einstein tensor $\Gb^{ij}$ of the metric $\gb_{ij}$,
we find that
\leqn{Gb1Z.1}{
|\gb|\Gb^{ij} = \frac{\ep}{2}|\hb| \left[ \gbh^{kl} \Db_k\Db_l \ub^{ij} + \ep\bigl(\ab_1^{ij}+ \ab_2^{ij}+ \ab_3^{ij}\bigr) + \bb^{ij} + \cb_1^{ij} +
\ep \cb_2^{ij} + \frac{2}{\ep} \Gcb^{ij}
\right],
}
where
\lalign{Gb2}{
\ab_1^{ij} & = \Half \bigl(\Half \gbh_{k l} \gbh_{mn} - \gbh_{km}
\gbh_{l n} \bigr) \bigl(\gbh^{ip} \gbh^{jq} - \Half \gbh^{ij}
\gbh^{pq} \bigr)\Db_{p} \ub^{k l}\Db_{q} \ub^{mn}
\label{Gb2.1} ,\\
\ab_2^{ij} & = \gbh_{k l}\Bigl( 2\gbh^{n(i }
\Db_{m} \ub^{j) l}\Db_{n} \ub^{k m} - \Half
\gbh^{ij}\Db_{m} \ub^{k n} \Db_{n} \ub^{m l}
- \gbh^{mn}\Db_{m}  \ub^{ik}\Db_{n} \ub^{j l}\Bigr)\label{Gb2.2}, \\
\ab_3^{ij} & = \Db_k\ub^{ij}\Db_{l} \ub^{k l}-
\Db_k\ub^{i l}\Db_l \ub^{jk}\label{Gb2.3}, \\
\bb^{ij} & =  \gbh^{ij}\Db_k\Db_l \ub^{k l} -
2\Db_{l}\Db_k \ub^{k(i} \gbh^{j)l} \label{Gb2.4}, \\
\cb_1^{ij} & = -\bigl(\hb^{ij}\ub^{kl} + \ub^{ij}\hb^{kl}\bigr)\Rcb_{kl}+2\Rcb_{lkm}{}^{(i}\ub^{j)k}\hb^{lm}+2\Rcb_{lkm}{}^{(i}\hb^{j)k}\ub^{lm} \label{Gb2.5}
\intertext{and}
\cb_2^{ij} & = -\ub^{ij}\ub^{kl}\Rcb_{kl}+2\Rcb_{lkm}{}^{(i}\ub^{j)k}\ub^{lm}.\label{Gb2.6}
}
Here, $\Rcb_{ijk}{}^l$, $\Rcb_{kl}$, and $\Gcb^{ij}$ are the Curvature, Ricci, and Einstein tensors of the metric $\hb_{ij}$, respectively.

The gauge is fixed by setting
\eqn{harm1}{
\Db_i\ub^{ij} = 0.
}
For $\ep>0$, it is clear from \eqref{ubdef} that this is equivalent to
\eqn{harm2}{
\Db_i\left(\sqrt{\frac{|\gb|}{|\hb|}}\gb^{ij}\right)=0,
}
which, in turn, is easily seen to be equivalent to the harmonic coordinate condition
\leqn{harm3}{
\gb^{ij}\bigl(\Gammab^k_{ij}-\gammab^k_{ij}\bigr) = 0.
}
With the gauge fixed, we define the \emph{reduced conformal Einstein tensor} $\Gb_R^{ij}$ by
\leqn{redGb}{
\Gb_R^{ij} = \frac{|\gb|}{|\hb|}\Gb^{ij} - \frac{\ep}{2} \bar{b}^{ij} = \frac{\ep}{2}\Bigl(\gbh^{kl}\Db_k\Db_l \ub^{ij}
 + \ep\bigl(\ab_1^{ij} + \ab_2^{ij}+  \ab_3^{ij}\bigr) + \cb_1^{ij} + \ep \cb_2^{ij} +
{\displaystyle \frac{2}{\ep} }\Gcb^{ij} \Bigr),
}
and the \emph{reduced conformal Einstein equations} by
\leqn{redEin}{
\Gb_R^{ij} = 2\frac{|\gb|}{|\hb|}\Tb^{ij}.
}

For the remainder of the article, we assume that $\hb_{ij}$ is flat and that the coordinates $(\xb^i)$ are Minkowskian
with respect to $\hb_{ij}$ so that
\leqn{hbCart}{
\hb_{ij} = \eta_{ij} = \text{diag}(-1,1,1,1), \quad \Db_j = \delb{j}:=\frac{\partial\;}{\partial \xb^j}, \quad \gammab^k_{ij}=0, \quad  \Rcb_{ijk}{}^l = 0,
}
and the set \eqref{VcdefA} is independent of the spacetime point and given by
\eqn{VcdefB}{
\Vc = \{ s=(s^{ij}) \in \Sbb{4}\, | \, -\det(\eta^{ij}+ s^{ij}) > 0 \}.
}
With the above choice for $\hb_{ij}$, the reduced Einstein tensor \eqref{redGb} is well
defined for $(\ep\ub)\in \Vc$ and can be written as
\leqn{redGbA}{
\Gb_R^{ij} = \frac{\ep}{2}\Bigl(\gbh^{kl}\delb{k}\delb{l} \ub^{ij}
 + \ep\bigl(\ab_1^{ij} + \ab_2^{ij}+  \ab_3^{ij}\bigr) \Bigr).
}
From an evolution point of view, it is natural to use a first order formulation. It turns out that the first
 order formulation also happens to be optimal for establishing uniform estimates in the limit $\ep \searrow 0$. This motivates us to introduce the derivatives
\leqn{ubkijdef}{
\ub_k^{ij} = \delb{k} \ub^{ij}
}
as independent variables.

In order to make contact
with the standard interpretation of the Newtonian limit and also to facilitate the analysis of the singular limit $\ep \searrow 0$, we convert the first order variables to Newtonian coordinates by setting
\leqn{udefA}{
u^{ij}(x)=\ub^{ij}(x^0,\ep \xv)\AND u^{ij}_k(x) = (\ub_k^{ij})(x^0,\ep \xv),
}
with the understanding that all variables are implicity $\ep$-dependent.
From these definitions, it follows immediately that
\lgath{udefB}{
u_0^{ij} = \del{0} u^{ij}, \quad u_J^{ij} = \frac{1}{\ep}\del{J} u^{ij},\quad \del{0} u_J^{ij} = \frac{1}{\ep} \del{J} u_0^{ij}, \label{udefB.1}
\intertext{and}
\del{0} u^{ij}_k(x) = (\delb{0}\delb{k} \ub^{ij})(x^0,\ep \xv), \quad \frac{1}{\ep}\del{L}  u_k^{ij}(x) = (\delb{L}\delb{k} \ub^{ij})(x^0,\ep \xv).
\label{udefB.2}
}
Using the definitions \eqref{udefA}, we can write the reduced Einstein tensor \eqref{redGbA} as
\leqn{redGbB}{
\Gb_R^{ij} = \frac{\ep}{2}\Bigl(\gh^{00}\del{0} u_0^{ij}+\frac{2}{\ep}\gh^{0L}\del{L} u_0^{ij}+
 \frac{1}{\ep}\gh^{KL}\del{K} u_L^{ij} + \ep\bigl(a_1^{ij} + a_2^{ij}+  a_3^{ij}\bigr) \Bigr),
}
where
\lalign{redGbC}{
\gh^{ij} & = \eta^{ij} + \ep u^{ij}, \label{redGbC.1}\\
a_1^{ij} & = \Half \bigl(\Half \gh_{k l} \gh_{mn} - \gh_{km}
\gh_{l n} \bigr) \bigl(\gh^{ip} \gh^{jq} - \Half \gh^{ij}
\gh^{pq} \bigr)u_p^{k l} u_q^{mn}
\label{redGbC.2} ,\\
a_2^{ij} & = \gh_{k l}\Bigl( 2\gh^{n(i }
u_m^{j) l}u_n^{k m} - \Half
\gh^{ij}u_m^{k n} u_n^{m l}
- \gh^{mn} u_m^{ik} u_n^{j l}\Bigr)\label{redGbC.3},
\intertext{and}
a_3^{ij} & = u_k^{ij} u_l^{k l}-
u_k^{i l} u_l^{jk}\label{redGbC.4} .
}
The representation \eqref{redGbB} of the reduced Einstein tensor together with \eqref{udefB.1}-\eqref{udefB.2} allow
 us to express the reduced conformal Einstein equations \eqref{redEin}
in the following first order form:
\alin{redEinA}{
-\gh^{00}\del{0} u_0^{ij} & = 2 u^{0I}\del{I} u_0^{ij} +\frac{1}{\ep} \gh^{IJ}\del{I} u_J^{ij}
+\ep\bigl(a^{ij}_1+a^{ij}_2+a^{ij}_3\bigr) - \frac{1}{\ep}\tau^{ij}, 
\\
\gh^{IJ} \del{0} u_J^{ij} & = \frac{1}{\ep} \gh^{IJ}\del{J} u_0^{ij}, 
 \\
\intertext{and}
\del{0} u^{ij} & = u_0^{ij}, %
\label{redEinA.3}
}
where
\leqn{taudef}{
\tau^{ij}(x) := 4\bigl(|\gb|\Tb^{ij}\bigr)(x^0,\ep\xv).
}

%% file: eul.tex
\subsect{ceul}{The conformal Euler equations}
In order to analyze the $\ep\searrow 0$ limit of the fluid, we need to derive a suitable symmetric hyperbolic form
for the Euler equations. To do so, we adapt the approach employed in \cite{Rendall:1992}.
We begin by introducing the conformal fluid four-velocity
\leqn{nbdef}{
\wb^i = e^{-\Psib}\vb^i,
}
and stating a few useful identities. First, we observe that\footnote{In this section, all indices will be raised and lowered with the conformal metric $\gb_{ij}$.}
\leqn{nbn}{
\wb_i\wb^i = \gb_{ij}\wb^i\wb^j =-1
}
holds by \eqref{gbdef} and the normalization $\grave{g}_{ij}\vb^i\vb^j=-1$. Differentiating \eqref{nbn}
gives
\leqn{dnbn}{
\wb_j\nablab_k \wb^j = 0,
}
or equivalently
\leqn{dnbdA}{
\nablab_k \wb^0 = -\frac{\wb_J}{\wb_0} \nablab_k \wb^J.
}
Contracting \eqref{cEE.2} with $\wb_j$ gives 
\alin{pEEa}{
\wb_j\nablab_i \grave{T}^{ij} &= \wb_j\left(6\grave{T}^{ij}\nablab_i\Psib - \gb_{lm}\grave{T}^{lm}
\gb^{ij}\nablab_i\Psib\right).
}
Using the identity \eqref{dnbn}, we can write this equation as
\leqn{pEEb.1}{
\wb^k\delb{k} \rhob + (\rhob+\pb)\nablab_j\wb^j = 3(\rhob+\pb)\wb^k\delb{k} \Psib.
}

Next, we define
\leqn{Lbjk}{
\Lb^k_J = \delta^k_J-\frac{\wb_J}{\wb_0}\delta^k_0,
}
and set
\leqn{LbIj}{
\Lb_{Ij}=\gb_{jk}\Lb^k_I.
}
Applying  $\Lb_{Ij}$ to \eqref{cEE.2} then yields
\leqn{pEEaa}{
\Lb_{Ij}\nablab_k \grave{T}^{kj} = \Lb_{Ij}\left(6\grave{T}^{kj}\nablab_k\Psib -\gb_{lm}\grave{T}^{lm}
\gb^{kj}\nablab_k\Psib\right).
}
Noting that $\Lb_{Ij}$ satisfies
\eqn{LbIja}{
\Lb_{Ij}\wb^j =0,
}
a straightforward calculation shows that \eqref{pEEaa} is equivalent to
\leqn{pEEb.2}{
(\rhob+\pb)\Lb_{Ij}\wb^k\nablab_k\wb^j + \Lb_I^k\delb{k}\pb = (\rhob+\pb)\Lb_I^k\delb{k}\Psib.
}

Using \eqref{dnbdA}, we
note that the divergence of $\wb^j$ can be expressed as
\leqn{divwb}{
\nablab_j\wb^j = \Lb^k_J\nablab_k \wb^J,
}
and we observe that
\lalign{LDww}{
\Lb_{Ij} \wb^k \nablab_k \wb^j &= \left(\Lb_{IJ}-\frac{\wb_J}{\wb_0}\Lb_{I0} \right)\wb^k\nablab_k \wb^J
&&\text{by \eqref{dnbdA}} \notag\\
&= \left(\gb_{Jl}\Lb^l_I-\gb_{0l}\frac{\wb_J}{\wb_0}\Lb^l_I \right)\wb^k\nablab_k \wb^J \notag \\
&= \Mb_{IJ} \wb^k \nablab_k \wb^J \label{LDww.1}
}
where
\leqn{Mbdef}{
\Mb_{IJ} = \gb_{IJ} - \frac{\wb_I}{\wb_0}\gb_{0J} - \frac{\wb_J}{\wb_0}\gb_{0I} + \frac{\wb_I\wb_J}{(\wb_0)^2} \gb_{00}.
}

Next, writing \eqref{nbn} as
\eqn{nbnA}{
\gb^{00}\wb_{0}^2 + 2 \gb^{0J}\wb_0\wb_J + \gb^{IJ}\wb_I\wb_J = -1,
}
we can solve for $\wb_0$ to get
\leqn{nbnB}{
\wb_{0} =\frac{-\gb^{0J}\wb_J + \sqrt{(\gb^{0J}\wb_J)^2-\gb^{00}(\gb^{IJ}\wb_I\wb_J+1)}}{\gb^{00}}.
}
Noting that
\eqn{nbnBa}{
\wb^{0} = \gb^{0i}\wb_i,
}
we see from \eqref{nbnB} that
\eqn{nbnBc}{
\frac{\wb^0}{\sqrt{-\gb^{00}}}\biggl|_{\wb_J=0} = 1,
}
which, in turn, implies that
\leqn{nbnBd}{
\frac{\wb^0}{\sqrt{-\gb^{00}}}-1 = \wc^{0J}\bigl(\gb^{ij},\wb_I\bigr)\wb_J,
}
where $\wc^{0J}\bigl(\gb^{ij},\wb_I\bigr)$ is analytic for\footnote{Here, $\Sbb{3}\!^+$ denotes the set of $3\times 3$ positive definite matrices.}
$(\gb^{00},\gb^{0J},\gb^{IJ},\wb_J) \in \Rbb_{<0}\times \Rbb^3\times \Sbb{3}\!^{+}\times \Rbb^3$.

Introducing a new density variable
\leqn{zetabdef}{
\nub = \nub(\ep,\rhob) := \int_{1}^{\rhob} \frac{d\xi}{\xi + \ep^2 f(\xi)},
}
the identities \eqref{divwb} and \eqref{LDww.1} together with the equation of state \eqref{eosdef} and
the Euler equations given by \eqref{pEEb.1} and \eqref{pEEb.2} show that
pair $(\nub,\wb^J/\ep )$ satisfy the symmetric hyperbolic system
\leqn{pEEc}{
\Cb^k\delb{k}\begin{pmatrix}\nub \\ \begin{displaystyle} \frac{\wb^J}{\ep}\end{displaystyle}\end{pmatrix}
= \Hb,
}
where
\leqn{Cbdef}{
\Cb^k = \begin{pmatrix} f'(\rhob)\wb^k & \ep f'(\rhob)\Lb_J^k  \\
\ep f'(\rhob)\Lb_I^k & \Mb_{IJ}\wb^k \end{pmatrix},
}
\leqn{Hbdef}{
\Hb = \begin{pmatrix}\begin{displaystyle} -\ep f'(\rhob)\Lb_J^k \Gammab^J_{kl} \frac{\wb^l}{\ep} + 3f'(\rhob)\wb^k\delb{k} \Psib
\end{displaystyle}\\
\begin{displaystyle}-\Mb_{IJ}\wb^k\Gammab^J_{kl} \frac{\wb^l}{\ep}+\frac{\Lb_I^k}{\ep}\delb{k}\Psib
\end{displaystyle}\end{pmatrix}
}
and
\leqn{rdefA}{
\rhob = r(\ep,\nub).
}
 Here, $r\in C^\infty\bigl(\bigcup_{\ep\in\Rbb} \{\ep\}\times \nub(\ep,\Rbb_{>0}),\Rbb_{>0}\bigr)$ is
 the inverse of the transformation \eqref{zetabdef}, which we note satisfies
\leqn{rdefB}{
r(0,\nub) = e^{\nub}.
}

Next, we define
\leqn{zbdefA}{
\zb_J = \frac{\wb_J}{\ep} \AND \zetab = \nub-3\Psib,
}
and observe that $\zb_J$ is related to $\wb^J/\ep$ via the transformation
\leqn{zbdefB}{
\frac{\wb^J}{\ep} = \wcb^J := \gb^{JL}\zb_L + \frac{\gb^{J0}}{\ep}(-1+\ep z_0),
}
where
\leqn{zbdefC}{
z_0 := \frac{\wb_0+1}{\ep} = \frac{1}{\ep}\left(\frac{-\gb^{0J}\wb_J + \sqrt{(\gb^{0J}\wb_J)^2-\gb^{00}(\gb^{IJ}\wb_I\wb_J+1)}}{\gb^{00}}+1\right).
}

Multiplying \eqref{pEEc} by the transpose of the matrix
\eqn{Zmatrix}{
\Zcb := \begin{pmatrix} 1 & 0 \\
0 & \begin{displaystyle}\frac{\del{}\wcb^J}{\del{}\zb^I} \end{displaystyle}
\end{pmatrix},
}
a short calculation shows that $(\zetab,\zb_J)$ satisfies
\leqn{pEEd}{
\Zcb^T\Cb^k \Zcb\delb{k}\begin{pmatrix}\zetab \\ \zb_J\end{pmatrix}
=\Zcb^T\left(\Hb -\begin{pmatrix}3f'(\rhob)\wb^k\delb{k}\Psib + \ep f'(\rhob)\Lb^K_J \begin{displaystyle} \frac{\del{}\wcb^J}{\del{}\gb^{lm}} \delb{k}\gb^{lm} \end{displaystyle}
 \\ \ep 3 f'(\rhob)\Lb^k_I \delb{k}\Psib
+ \Mb_{IJ}\wb^k \begin{displaystyle} \frac{\del{}\wcb^J}{\del{}\gb^{lm}} \delb{k}\gb^{lm} \end{displaystyle}\end{pmatrix} \right),
}
which is easily verified as symmetric hyperbolic.

To proceed, we separate the 
conformal factor into homogenous and inhomogenous components as follows
\leqn{betadef}{
\Psib(\xb) = \beta(\xb^0) + \ep \psib(\xb).
}
Using this decomposition, we  split the proper energy density $\rhob$ into homogenous and inhomogenous components according
to
\leqn{alphadef}{
\rhob(\xb) = \alpha(\xb^0) + \mub(\xb),
}
where
\leqn{alphadefB}{
\alpha = r\bigl(\ep,3\beta\bigr)
}
and
\leqn{mudef}{
\mub = q(\ep,\beta,\psib,\zetab):=r\bigl(\ep,3\beta+\ep3\psib+\zetab\bigr)-r\bigl(\ep,3\beta\bigr).
}
Evaluating $q$ at $\ep=0$, we observe, with the help of \eqref{rdefB}, that
\eqn{qep0}{
q(0,\beta,\psib,\zetab) = e^{3\beta}\bigl( e^{\zetab}-1\bigr).
}

Differentiating the transformation \eqref{zetabdef}, we see that
\eqn{dzetab}{
\frac{d\nub}{d\rhob} = \frac{1}{\rhob + \ep^2 f(\rhob)},
}
which in turn, implies that derivative of \eqref{rdefA} satisfies
\eqn{dr}{
\frac{\del{} r}{\del{}\nub}(\ep,\nub) = r(\ep,\nub) + \ep^2 f(r(\ep,\nub)).
}
From this, we see, after differentiating \eqref{alphadefB}, that
\leqn{alphaeqn}{
\alpha^\prime = 3(\alpha + \ep^2 f(\alpha))\beta^{\prime},
}
and using a similar calculation, that
\leqn{Dmub}{
\delb{I}\mub = \bigl(\alpha+\mub+\ep^2 f(\alpha+\mub) \bigr)\bigl(\delb{I}\zetab + \ep 3\delb{I}\psib\bigr).
}

From \eqref{utog}, \eqref{ghdef} and \eqref{hbCart}, we observe that the metric and its inverse can be expanded as
\lalign{gbijexp}{
\gb_{ij} &= \eta_{ij} - \ep\bigl(\eta_{il}\eta_{jm}\ub^{lm}-\Half \eta_{ij}\eta_{lm}\ub^{lm} \bigl) + \gc_{ij}(\ep\ub) \label{gbijexp.1}
\intertext{and}
\gb^{ij} &= \eta^{ij} + \ep\bigl(\ub^{ij}-\Half \eta^{ij}\eta_{lm}\ub^{lm} \bigl) + \gc^{ij}(\ep\ub), \label{gbijexp.2}
}
respectively, where for any $R>0$, there exists an $\ep_0>0$ such that the maps $\gc_{ij}(\ep \ub)$ and $\gc^{ij}(\ep \ub)$ are analytic for  $(\ep,\ub)\in (-\ep_0,\ep_0)\times B_R(\Sbb{4})$ and satisfy
\eqn{gczero}{
|\gc_{ij}(\ep \ub)| + |\gc^{ij}(\ep \ub)| \lesssim \ep^2|\ub|^2.
}
Furthermore, differentiating \eqref{gbijexp.2} shows that
\leqn{Dgijexp}{
\delb{k}\gb^{ij} = \ep\bigl(\ub^{ij}_k-\Half \eta_{lm}\ub^{lm}_k \eta^{ij}\bigr)
+ \ep^2 \begin{displaystyle}\biggl(\frac{1}{\ep}\frac{\del{}\gc^{ij}}{\del{}\xi^{lm}}\biggr)\Bigl|_{\xi=\ep u}\end{displaystyle}\ub^{lm}_k.
}
Using the formula \eqref{christ}, it is clear that the Christoffel symbols can be expanded as
\leqn{christexp}{
\Gammab^{k}_{ij} = \ep\bigl(-\eta_{l(i}\ub_{j)}^{kl}+\Half\eta^{kl}\eta_{im}\eta_{jn}\ub_l^{mn} -\Quarter
   \eta^{kl}\eta_{ij}\eta_{mn}\ub_l^{mn} +\Half\eta_{lm}\delta^k_{(i}\ub_{j)}^{lm} \bigr)    + \ep^2 \Upsilon^{k}_{ij}(\ep,\ub,\ub_l),
}
where the $\Upsilon^{k}_{ij}(\ep,\ub,\ub_l)$ are analytic for $(\ep,\ub,\ub_l) \in (-\ep_0,\ep_0)\times B_R(\Sbb{4})\times (\Sbb{4})^4$
and satisfy
\eqn{chirstexpA}{
|\Upsilon^k_{ij}(\ep,\ub,\ub_l)| \lesssim |\ub_l|.
}

Switching to Newtonian coordinates and defining
\lalign{Nfldef}{
\mu(x^0,\xv) &= \mub(x^0,\ep\xv), \label{Nfldef.1} \\
\zeta(x^0,\xv) &= \zetab(x^0,\ep\xv), \label{Nfldef.2}\\
z_I(x^0,\xv) & = \zb_I(x^0,\ep\xv), \label{Nfldef.3} \\
\psi(x^0,\xv) &= \psib(x^0,\ep\xv), \label{Nfldef.4}
\intertext{and}
\psi_k(x^0,\xv) &= (\delb{k}\psib)(x^0,\ep\xv), \label{Nfldef.5}
}
a straightforward calculation using the decompositions \eqref{betadef} and \eqref{alphadef}, the
equation of motion \eqref{alphaeqn}, the identity \eqref{Dmub}, the expansions \eqref{nbnBd}, \eqref{Lbjk}, \eqref{Mbdef}, \eqref{gbijexp.1},
\eqref{gbijexp.2}, \eqref{Dgijexp} and \eqref{christexp}, and the evolutions equations \eqref{pEEd} shows
that the conformal Euler equations are equivalent to the following symmetric hyperbolic system:
\eqn{regceulmat}{
C^0\del{0}\begin{pmatrix} \zeta \\ z_J \end{pmatrix}+
C^K\del{K}\begin{pmatrix} \zeta \\ z_J \end{pmatrix} = H,
}
where
\alin{Cdef}{
C^0 &= \Ct^0 + \Cc^0(\ep,\ep u,\zeta,z_L,\beta), 
\\
\Ct^0 & = \begin{pmatrix}f'\bigl(e^{3\beta+\zeta}\bigr) & 0\\
0 & \delta_{IJ}\end{pmatrix},
\\
C^K& = \Ct^K + \ep \Cc^K(\ep,u,\zeta,z_L,\beta),
\\
\Ct^K &=  \begin{pmatrix} f'\bigl(e^{3\beta+\zeta}\bigr)(-u^{K0}+\delta^{KL}z_L) &
f'\bigl(e^{3\beta+\zeta}\bigr)\delta^K_J\\
f'\bigl(e^{3\beta+\zeta}\bigr)\delta^K_I & \delta_{IJ}(-u^{K0}+\delta^{KL}z_L)\end{pmatrix}, 
\\
 H &= \Ht + \ep\Hc(\ep,u,u_l,\zeta,z_L,\beta,\beta',\psi_0,\psi_L), \label{Cdef.5}  %
 \\
 \Ht &= \begin{pmatrix}0 \\
 \begin{displaystyle} -\Quarter u_I^{00} - \Quarter \delta_{MN}u^{MN}_I -\delta_{IJ}u^{J0}_0
 +\beta' z_I+ \psi_I
  \end{displaystyle} \end{pmatrix},
}
and
\begin{enumerate}[(i)]
\item for any $R>0$,  there exists an $\ep_0>0$ such that the maps $\Cc^k$ and $\Hc$ are smooth in all variables on
the region defined by
$\alpha+\mu > 0$ and $(\ep,u)\in (-\ep_0,\ep_0)\times B_R(\Sbb{4})$,
\item $\Cc^0$ satisfies
\eqn{Cczero}{
\Cc^0(0,0,\zeta,z_L,\beta)=0
}
\item and $\Hc$ satisfies
\eqn{Hczero}{
\Hc(\ep,u,0,\zeta,0,\beta,\beta',\psi_0,0)=0.
}
\end{enumerate}

%% file: cev.tex
\subsect{cev}{Evolution of the conformal factor}
Thus far, the conformal factor has been arbitrary. We now fix conformal factor, up to
a choice of initial data, by demanding that $\Psib$ satisfy the wave equation
\leqn{cevAa}{
\gb^{ij}\nablab_i\nablab_j \Psib -\Half \gb^{ij}\nablab_i\Psib \nablab_j\Psib + e^{-2\Psib}\bigl(\ep^2f\bigl(r(\ep,3\beta+3\ep\psib+\zetabh)\bigr)-\Lambda)=0,
}
where we have set
\eqn{cevAaa}{
\zetabh = \zetab|_{\xb^0=0}.
}
This choice of evolution equation for $\Psib$ is dictated by the requirement that components of
the stress energy $\Tb^{ij}$,
given by \eqref{Tbdef} (see also \eqref{taudef}), are bounded in $L^2$, which is a non-trivial
requirement given the choice of initial data that is appropriate for our setting.

Assuming the harmonic gauge condition \eqref{harm3}, the wave equation \eqref{cevAa} becomes
\leqn{cevAb}{
\gb^{ij}\delb{i}\delb{j} \Psib -\Half \gb^{ij}\delb{i}\Psib \delb{j}\Psib +
e^{-2\Psib}\bigl(\ep^2f\bigl(r(\ep,3\beta+3\ep\psib+\zetabh)\bigr)-\Lambda).
}
In terms of the decomposition \eqref{betadef},  the wave equation \eqref{cevAb} is easily seen to be equivalent to the system
\lalign{cevA}{
\beta'' &= \Half (\beta')^2 + e^{-2\beta}\bigl(\ep^2 f(\alpha)-\Lambda\bigr), \label{cevA.1}\\
\gb^{ij}\delb{i}\delb{j}\psib &= \beta'\gb^{0j}\delb{j}\psib + \frac{\ep}{2} \gb^{ij}\delb{i}\psib\delb{j}\psib  \notag\\
&\hspace{0.2cm}- e^{-2\beta}\left[\bigl(\ep^2f(\alpha)-\Lambda)
 \begin{displaystyle}\frac{\gb^{00}+e^{-2\ep\psib}}{\ep} \end{displaystyle}+\ep e^{-2\ep\psib}
 \bigl(f\bigl(r(\ep,3\beta+3\ep\psib+\zetabh)\bigr)-f(\alpha)\bigr)\right]. \label{cevA.2}
}

Differentiating \eqref{cevA.2} with respect to the spatial coordinates $\xb^K$ then shows, with the help
of \eqref{Dmub}, that $\delb{K}\psib$ satisfies the wave equation
\lalign{cevB}{
\gb^{ij}&\delb{i}\delb{j}\delb{K}\psib = \beta'\gb^{0j}\delb{j}\delb{K}\psib +\ep \gb^{ij}\delb{i}\psib\delb{j}\delb{K}\psib
-\biggl[\delb{i}\delb{j}\psib - \frac{\ep}{2}\delb{i}\psib\delb{j}\psib -\beta'\delb{j}\psib \delta_{i}^0\biggr]\delb{K}\gb^{ij} \notag \\
&
  - e^{-2\beta}\biggl[\bigl(\ep^2f(\alpha)-\Lambda)\biggl(\begin{displaystyle}\frac{1}{\ep}\end{displaystyle}
  \delb{K}\gb^{00}-2e^{-2\ep\psib}\delb{K}\psib\biggr) +\ep  e^{-2\ep\psib}\biggl\{ -\ep 2
 \Bigl[f\bigl(r(\ep,3\beta+3\ep\psib+\zetabh)\bigr) \notag \\
 &-f(\alpha)\Bigr]\delb{K}\psib + f'(r(\ep,3\beta+3\ep\psib+\zetabh))\Bigl[r(\ep,3\beta+3\ep\psib+\zetabh) \notag \\
 &+ \ep^2 f\bigl(r(\ep,3\beta+3\ep\psib+\zetabh)\bigr)\Bigr]\bigl(\delb{K}\zetabh+\ep 3\delb{K}\psib\bigr)\biggr\}\biggr].\label{cevB.1}
}
Defining
\leqn{ddpsi}{
\psi_{kl}(x^0,\xv) = (\delb{k}\delb{l}\psib)(x^0,\ep\xv),
}
\leqn{zetahdef}{
\zetah = \zeta|_{x^0=0},
}
and
\leqn{ginv}{
g^{ij}(x^0,\xv) = \gb^{ij}(x^0,\ep\xv),
}
we see from equations \eqref{cevA.2} and \eqref{cevB.1}, the expansions \eqref{gbijexp.2} and \eqref{Dgijexp},
and the relation \eqref{rdefB} that $\psi_{lK}$ satisfies
\lalign{cevC}{
-g^{00}\del{0}\psi_{0K} &= \frac{2}{\ep}g^{I0}\del{I}\psi_{0K}+\frac{1}{\ep}g^{IJ}\del{I}\psi_{JK}+\beta'\psi_{0K}+ \Lambda e^{-2\beta}
\Bigl(2\psi_K-\Half \delta_{lm}u^{lm}_K\Bigr) \notag \\
&\quad + f'\bigl(e^{3\beta+\zetah}\bigr)e^{\beta+\zetah}\del{K}\zetah
 + \ep P\bigl(\ep,u,u_L,\zetah,\del{L}\zetah,\beta,\beta',\psi,\psi_0,\psi_L,\psi_{00},\psi_{lM}\bigr), \notag 
  \\
g^{IJ}\del{0}\psi_{JK} &= \frac{1}{\ep}g^{IJ} \del{J}\psi_{0K}, \notag 
\\
\del{0}\psi_0 &=  \psi_{00}, \notag 
\\
\del{0}\psi_K &=  \psi_{0K}, \notag 
 \\
\del{0}\psi   &= \psi_0, \notag 
\intertext{where}
g^{ij} &= \eta^{ij} + \ep\bigl(u^{ij}-\Half \eta^{ij}\eta_{lm} u^{lm} \bigl) + \gc^{ij}(\ep u ),\label{cevC.6}\\
\psi_{00} &= -\frac{1}{g^{00}}\biggl(2 g^{0J}\psi_{0J} + g^{IJ}\psi_{IJ}  -\beta' g^{0j}\psi_j -
\frac{\ep}{2} g^{ij}\psi_i\psi_j \notag \\
&+ e^{-2\beta}\biggl[\bigl(\ep^2f(\alpha)-\Lambda)
 \begin{displaystyle}\frac{g^{00}+e^{-2\ep\psi}}{\ep} \end{displaystyle}+\ep e^{-2\ep\psib}
 \bigl(f\bigl(r(\ep,3\beta+3\ep\psi+\zetah)\bigr)-f(\alpha)\bigr)\biggr]\biggr), \label{cevC.7}
}
and for any $R>0$, there exists an $\ep_0>0$ such that $P$ is smooth in all variables on the region
$(\ep,u)\in (-\ep_0,\ep_0)\times B_R(\Sbb{4})$ and satisfies
\eqn{Pzero}{
\bigl|P\bigl(\ep,u,u_l,\zetah,\del{L}\zetah,\beta,\beta',\psi,\psi_0,\psi_L,\psi_{00},\psi_{lM}\bigr)\bigr|
\lesssim |\del{L}\zetah|+|u_l|+|\psi_L|+|\psi_{lM}|.
}

%% file: stress.tex
\subsect{stress}{Stress energy tensor decomposition}
To fulfill the requirement that the components $\Tb^{ij}$ of the stress energy tensor,
given by \eqref{Tbdef}, or equivalently $\tau^{ij}$, given by \eqref{taudef}, have finite
spatial $L^2$ norm, it is not enough that \eqref{cevAa} is satisfied. In addition to
this, we need that
\eqn{Thetadef}{
\Theta := e^{-2\beta}\bigl(\alpha + \ep^2 f(\alpha)\bigr) - \bigl(\beta''+(\beta')^2\bigr)
\oset{\text{\eqref{cevA.1}}}{=} e^{-2\beta}(\alpha+\Lambda)-\Threehalfs (\beta')^2
}
vanishes for all $x^0\geq 0$. To verify the consistency of this requirement,  we
observe, via
a straightforward computation using the evolution equations \eqref{alphaeqn} and
\eqref{cevA.1} for the homogenous components, that $\Theta$ satisfies
$\Theta' = \beta'\Theta$,
from which we conclude:
\leqn{ThetaevB}{
\Theta(0) = 0 \quad \Longrightarrow \quad \Theta(x^0) = 0 \quad \forall \: x^0 \geq 0.
}
Therefore, by choosing initial data $\{\beta(0),\beta'(0)\}$ so
that the constraint
\leqn{thetazero}{
\Theta(0)=0
}
is satisfied,
we can guarantee that
\leqn{ThetaevC}{
\Theta(x^0) = e^{-2\beta(x^0)}\bigl[\alpha(x^0) + \ep^2 f\bigl(\alpha(x^0)\bigl)\bigr] - \bigl[\beta''(x^0)+\bigl(\beta'(x^0)\bigr)^2\bigr]= 0 \quad \forall\: x^0 \geq 0.
}
In the following, we will always assume that our initial data is chosen to satisfy \eqref{thetazero}.

Next, we observe that the evolution equation \eqref{cevAa} together with the formula \eqref{christ} and the
the definitions  \eqref{ubkijdef}, \eqref{udefA}, \eqref{redGbC.1}, \eqref{nbdef}, \eqref{betadef}, \eqref{alphadef}
and \eqref{Nfldef.1}-\eqref{Nfldef.5} allow
us to write \eqref{taudef} as
\lalign{tauevA}{
\tau_{ij} = -&4\det(\gh)\Bigl\{ e^{-2\beta-2\ep \psi}\bigl(\alpha+\mu + \ep^2 f(\alpha+\mu)\bigr)(-\delta^0_i+\ep z_i)
(-\delta^{0}_j+\ep z_j) \notag \\
& +\ep^2 e^{-2\beta-2\ep\psi}\bigl[f\bigl(r(\ep,3\beta+3\ep\psi+\zeta)\bigr)-f\bigl(r(\ep,3\beta+3\ep\psi+\zetah)\bigr)\bigr]
g_{ij}\notag \\
& - \delta_i^0\delta_j^0\bigl(\beta''+(\beta')^2\bigl)
+\ep \bigl(\Gamma_{ij}^0  - 2\delta^0_{(i}\psi_{j)}\bigr)\beta'
- \ep \bigl(\psi_{(ij)} - \ep \bigl[\Gamma^m_{ij}\psi_m - \psi_i\psi_j\bigr] \bigr)
\Bigr\}, \label{tauevA.1}
}
where
\leqn{GammaevA}{
\Gamma^k_{ij} = -\gh_{l(i}u^{kl}_{j)}+\Half\gh^{kl}\gh_{im}\gh_{jn}u^{mn}_l -\Quarter
   \gh^{kl}\gh_{ij}\gh_{mn}u^{mn}_l +\Half\gh_{lm}\delta^k_{(i} u^{lm}_{j)}.
}
Defining
\eqn{tauevBa}{
\iota_{ij} = e^{-2\beta-2\ep \psi}\bigl(\alpha+\mu + \ep^2 f(\alpha+\mu)\bigr)(-\delta_{i}^{0}+\ep z_i)
(-\delta_{j}^{0}+\ep z_j) - \delta_{i}^{0}\delta_{j}^{0}\bigl(\beta''+(\beta')^2\bigl),
}
we find, using \eqref{zbdefC} and \eqref{ThetaevC}, that $\iota_{ij}$ can be written as
\lalign{tauevBb}{
\iota_{ij} &=  e^{-2\beta}\Bigl[\bigl(e^{-2\ep \psi}-1\bigr))\bigl(\alpha+\mu + \ep^2 f(\alpha+\mu)\bigr)
 +\mu
 + \ep^2 \Bigl( f(\alpha+\mu)- f(\alpha)\Bigr) \Bigr]  \notag \\
 &\times(-\delta_{i}^{0}+\ep z_i)
(-\delta_{j}^{0}+\ep z_j)  + e^{-2\beta}\bigl(\alpha + \ep^2 f(\alpha)\bigr)\Biggl[ \ep(-1+\ep z_0)\bigl(\delta^{0}_{i} \delta_j^K z_{K}
+ \delta^0_j \delta_i^K z_K\bigr)\notag \\
& + \ep^2\delta_i^K\delta_j^L z_K z_L\bigr)
 - \delta^0_i\delta^0_j\frac{\ep}{g^{00}}\Biggl(2g^{0J}(-1+\ep z_0) z_J +\ep g^{IJ}z_I z_J
+\frac{1+g^{00}}{\ep} \Biggr)\Biggr]. \label{tauevBb.1}
}
Additionally, we see from the \eqref{christexp} that \eqref{GammaevA} is given by
\leqn{GammaevB}{
\Gamma^k_{ij} = -\eta_{l(i}u^{kl}_{j)}+\Half\eta^{kl}\eta_{im}\eta_{jn}u^{mn}_l -\Quarter
   \eta^{kl}\eta_{ij}\eta_{mn}u^{mn}_l +\Half\eta_{lm}\delta^k_{(i} u^{lm}_{j)} + \ep \Upsilon^k_{ij}(\ep,u,u_l),
}
while $\det(\gh)$ can be
expanded, using \eqref{udefA} and \eqref{redGbC.1}, as
\leqn{detghexp}{
-\det(\gh)= 1 + \ep \eta_{lm} u^{lm} + \gc(\ep u),
}
where $\gc(\ep u)$ is analytic for $(\ep, u)\in (-\ep_0,\ep_0)\times B_R(\Sbb{4})$, with $\ep_0>0$
is chosen sufficiently small depending on $R>0$, and satisfies
\eqn{detghexpbound}{
|\gc(\ep u)| \lesssim \ep^2 |u|^2.
}

We are now in a position to use the expansions \eqref{cevC.6}, \eqref{GammaevB} and \eqref{detghexp} together with
\eqref{tauevBb.1} to expand the stress energy tensor \eqref{tauevA.1} as follows:
\lalign{tauexpA}{
\tau_{00} &= 4 e^{-2\beta}\mu +
\ep\Tc_{00}\bigl(\ep,u,u_l,\alpha,\mu,z_L,\beta,\beta',\psi,\psi_0,\psi_L,\psi_{00}\bigr) \notag  \\
&\qquad -\ep^2 4\det(\gh) e^{-2\beta-2\ep\psi}\bigl[f\bigl(r(\ep,3\beta+3\ep\psi+\zeta)\bigr)-f\bigl(r(\ep,3\beta+3\ep\psi+\zetah)\bigr)\bigr]g_{00}, \label{tauexpA.1} \\
\tau_{0J} &= \ep\Bigl\{\bigl[-4 e^{-2\beta}(\alpha+\mu)+\ep \Pc(\ep,u,\alpha+\mu,\beta,\psi)\bigr]z_J \notag \\
& \hspace{1.0cm} +   4\bigl[\beta'\bigl( \Quarter \delta_{lm}u^{lm}_J-\psi_J\bigr)-\psi_{0J}\bigr]\Bigr\}
 + \ep^2 \Tc_{0J}(\ep,u,u_l,\beta',\psi_0,\psi_L,\psi_{0L})\notag \\
 &\qquad -\ep^2 4\det(\gh) e^{-2\beta-2\ep\psi}\bigl[f\bigl(r(\ep,3\beta+3\ep\psi+\zeta)\bigr)-f\bigl(r(\ep,3\beta+3\ep\psi+\zetah)\bigr)\bigr]g_{0J},
\label{tauexpA.2}\\
\tau_{IJ} &= \ep4\bigl(-\delta_{L(I}u^{L0}_{J)} - \Half \delta_{IM}\delta_{JN}u^{MN}_{0}
+\Quarter \delta_{IJ}\eta_{mn}u^{mn}_0 - \psi_{(IJ)}\bigr) \notag \\
& \qquad + \ep^2 \Tc_{IJ}\bigl(\ep,u,u_l,\alpha,\mu,z_L,\beta,\beta',\psi,\psi_0,\psi_L,\psi_{LM}\bigr) \notag \\
&\qquad -\ep^2 4\det(\gh) e^{-2\beta-2\ep\psi}\bigl[f\bigl(r(\ep,3\beta+3\ep\psi+\zeta)\bigr)-f\bigl(r(\ep,3\beta+3\ep\psi+\zetah)\bigr)\bigr]g_{IJ},
\label{tauexpA.3}
}
where for any $R>0$, there exists an $\ep > 0$ such that the maps
$\Tc_{ij}$ and $\Pc$ are smooth in all their variables on the region $(\ep, u) \in (-\ep_0,\ep_0)\times B_R(\Sbb{4})$
and satisfy
\lalign{tauexpB}{
\Tc_{00}\bigl(\ep,u,u_l,\alpha,\mu,z_L,\beta,\beta',\psi,\psi_0,\psi_L,\psi_{00}\bigr) &\lesssim
|u|+|u_l|+ |\mu|+|z_L|+ |\psi|+|\psi_0|+|\psi_{00}|, \label{tauexpB.1} \\
|\Tc_{0J}(\ep,u,u_l,\beta',\psi_0,\psi_L,\psi_{0L})| &\lesssim |u_l|+|\psi_L|+|\psi_{0L}| \label{tauexpB.2}
\intertext{and}
\Tc_{IJ}\bigl(\ep,u,u_l,\alpha,\mu,z_L,\beta,\beta',\psi,\psi_0,\psi_L,\psi_{LM}\bigr) &\lesssim
|u_l|+|z_L|+|\psi_L|+|\psi_{LM}|. \label{tauexpB.3}
}

By definition,
\eqn{tauraise}{
\tau^{ij} = g^{ik}g^{jl}\tau_{lm}.
}
For later use, we write this out explicitly as
\lalign{tauexpC}{
\frac{\tau^{00}}{\ep} &= 
\frac{1}{\ep} (g^{00})^2 \tau_{00} + \ep\biggl( 2g^{00}\frac{g^{0K}}{\ep}\frac{\tau_{K0}}{\ep}\biggr) +
\ep^2 \biggl(\frac{g^{0K}}{\ep}\frac{g^{0L}}{\ep}\frac{\tau_{KL}}{\ep}\biggr),
\label{tauexpC.1} \\
\frac{\tau^{0J}}{\ep} &= g^{00}\frac{g^{J0}}{\ep}\tau_{00}+ \biggl(g^{00}g^{JK}+\ep^2\frac{g^{0K}}{\ep}
\frac{g^{J0}}{\ep}\biggr)\frac{\tau_{K0}}{\ep} + \ep\biggl(g^{JK}\frac{g^{0L}}{\ep} \frac{\tau_{KL}}{\ep}\biggr)
\label{tauexpC.2}
\intertext{and}
\frac{\tau^{IJ}}{\ep} &= g^{IK}g^{JL}\frac{\tau_{KL}}{\ep}
+\ep\biggl(\frac{g^{I0}}{\ep}\frac{g^{0J}}{\ep}\tau_{00} + \biggl[g^{JK}\frac{g^{I0}}{\ep}
+ g^{IK}\frac{g^{J0}}{\ep}\biggr]\frac{\tau_{K0}}{\ep}\biggr).
\label{tauexpC.3}
}

%% file: total.tex
\subsect{total}{The complete system}

Collecting the results from the previous sections, the complete set of evolution
equations for the conformal gravitational variables $\{u^{ij},u^{ij}_k\}$, the conformal factor variables
$\{\beta,\beta_0,\psi,\psi_j,\psi_{jK}\}$, and the fluid variables $\{\delta\zeta, z_J\}$ are given by:
\lalign{totalA}{
A^0\del{0} \begin{pmatrix} u^{ij}_0 \\ u^{ij}_J\end{pmatrix}
+ \frac{1}{\ep}E^K\del{K}\begin{pmatrix} u^{ij}_0 \\ u^{ij}_J\end{pmatrix} + A^K\del{K}\begin{pmatrix} u^{ij}_0 \\ u^{ij}_J\end{pmatrix} &= F, \label{totalA.1} \\
\del{0}u^{ij} - u^{ij}_0 &= 0, \label{totalA.2}\\
B^0\del{0} \begin{pmatrix} \psi_{0L} \\ \psi_{JL}\end{pmatrix}
+ \frac{1}{\ep}E^K\del{K}\begin{pmatrix} \psi_{0L} \\ \psi_{JL}\end{pmatrix} + B^K\del{K}\begin{pmatrix} \psi_{0L} \\ \psi_{JL}\end{pmatrix} &= G, \label{totalA.3} \\
\del{0}\psi_0 -  \psi_{00}&=0, \label{totalA.4}\\
\del{0}\psi_K -  \psi_{0K}&=0, \label{totalA.5} \\
\del{0}\psi   - \psi_0&=0, \label{totalA.6}\\
\beta_0'-\Half\beta_0^2-e^{-2\beta}\bigl(\ep^2f(\alpha)-\Lambda)\bigr) & = 0, \label{totalA.7} \\
\beta' -\beta_0 &= 0, \label{totalA.8}\\
C^0\del{0}\begin{pmatrix} \delta\zeta \\ z_J \end{pmatrix}+
C^K\del{K}\begin{pmatrix} \delta\zeta \\ z_J \end{pmatrix} &= H - C^K \del{K}\begin{pmatrix} \zetah \\ 0 \end{pmatrix}, \label{totalA.9}
}
where
\lalign{totalB}{
A^0 &= \begin{pmatrix}1-\ep u^{00} & 0  \\ 0 & \delta^{IJ}+ \ep u^{IJ} \end{pmatrix}, \label{totalB.1}\\
A^K &= \begin{pmatrix} -2u^{0K} & -u^{JK} \\ - u^{IK} & 0 \end{pmatrix}, \label{totalB.2} \\
B^0 &= \begin{pmatrix}1-\ep\bigl(u^{00}+\Half\eta_{lm}u^{lm}\bigr)-\gc^{00}(\ep u^{00}) & 0  \\ 0 & \delta^{IJ}+ \ep\bigl(u^{IJ}-\Half\delta^{IJ}\eta_{lm}u^{lm}\bigr) + \gc^{IJ}(\ep u) \end{pmatrix}, \label{totalB.3}\\
B^K &= \begin{pmatrix}  \begin{displaystyle}-u^{0K}-\frac{2}{\ep}\gc^{0K}(\ep u)
 \end{displaystyle}& \begin{displaystyle}-\bigl(u^{JK}-\Half\delta^{JK}\eta_{lm}u^{lm}\bigr)\end{displaystyle}-
 \begin{displaystyle}\frac{1}{\ep}\gc^{JK}(\ep u)\end{displaystyle} \\
 \begin{displaystyle}-\bigl(u^{IK}-\Half\delta^{IK}\eta_{lm}u^{lm}\bigr)\end{displaystyle} -
 \begin{displaystyle}\frac{1}{\ep}\gc^{IK}(\ep u)\end{displaystyle} & 0 \end{pmatrix}, \label{totalB.4} \\
E^K &= \begin{pmatrix} 0 & -\delta^{JK} \\ -\delta^{IK} & 0 \end{pmatrix}, \label{totalB.5} \\
F &= \begin{pmatrix} -\frac{1}{\ep}\tau^{ij}+\ep(a^{ij}_1+a^{ij}_2+a_3^{ij}) \\ 0\end{pmatrix}, \label{totalB.6} \\
G &= \begin{pmatrix} \beta'\psi_{0L} + \Lambda e^{-2\beta}\bigl(2\psi_L-\Half \delta_{lm}u^{lm}_L\bigr)
 + f'\bigl(e^{3\beta+\zetah}\bigr)e^{\beta+\zetah}\del{L}\zetah + \ep P \\
 0 \end{pmatrix}, \label{totalB.7} \\
\delta \zeta &= \zeta - \zetah, \label{totalB.8}
}
and all other quantities are as previously defined in Sections \ref{red}, \ref{ceul}, \ref{cev} and \ref{stress}.

%% file: idata.tex
\sect{idata}{Initial Data}
As is well known in General Relativity, the initial data cannot be chosen freely on the initial hypersurface
\eqn{Sigmadef}{
\Sigma =\{0\} \times \Rbb^3.
}
The restriction on the choice of initial data is governed by the constraint equations
consisting of
\lalign{ceqns}{
\Gb^{0j} &= 2 \Tb^{0j} && \text{(gravitational constraint equations)}
\label{ceqns.1}
\intertext{and}
\delb{i}\ub^{ij} &= 0, && \text{ (harmonic gauge condition)}
\label{ceqns.2}
}
which must be satisfied by the initial data.
The first step in obtaining solutions to these equations that behave appropriately in the limit $\ep \searrow 0$
is to bring them into a suitable form. To this end, we use
\eqref{Gb2.4}, \eqref{redGb} and \eqref{ceqns.2} to express the
 components
$\Gb^{0j}$ of the conformal Einstein tensor on the hypersurface $\Sigma$ as
\leqn{G0j}{
\frac{2}{\ep}|\gb|\Gb^{0j}= \gbh^{LM}\delb{L}\delb{M}\ub^{0j} - 2  \gbh^{0L}\delb{L}\delb{M} \ub^{Mj}
+ \delta^j_0  \gbh^{00} \delb{L}\delb{M} \ub^{LM} - \delta^j_M \gbh^{00} \delb{L}\delb{0} \ub^{LM}
+ \ep (\ab_1^{0j}+\ab^{0j}_2 + \ab^{0j}_3).
}
Next, we introduce rescaled gravitational variables in Newtonian coordinates via the definitions
\leqn{uhdef}{
\uh^{ij} =  \frac{1}{\ep} u^{ij}|_{x^0=0} \AND \uh_0^{ij} = u_0^{ij}|_{x^0=0},
}
and note that the
harmonic constraint \eqref{ceqns.2} can be expressed as
\leqn{harmuhcA}{
\uh^{0j}_0 = - \del{I}\uh^{Ij}
}
in terms of these variables.
We label the initial data for the homogenous component of the conformal factor by
\leqn{betah}{
\betah = \beta(0) \AND \betah_0 = \beta'(0),
}
and observe that
\eqn{alphah}{
\alphah := \alpha(0) \oset{\text{\eqref{alphadefB}}}{=} r(\ep,3\betah) > 0.
}
Solving the constraint \eqref{thetazero}, we find that the initial data $\betah_0$ is fixed by
\eqn{betah0}{
\betah_0 = \sqrt{\Twothirds e^{-2\betah}(\alphah+\Lambda)}.
}
For the inhomogeneous conformal initial data, which we have complete freedom to specify, we set
\leqn{psidata}{
\psi|_{x^0=0} = \del{0}\psi|_{x^0=0} = 0.
}
This choice of initial data then implies, via the definitions
\eqref{Nfldef.5}, \eqref{ddpsi}, and \eqref{cevC.7}, that
\gath{psidataB}{
\psi_j|_{x^0=0} = \psi_{jL}|_{x^0=0}=0 
\intertext{and}
\psih_{00} := \psi_{00}|_{x^0=0} = - \ep \frac{e^{-2\betah}}{\hat{g}^{00}}\biggl[\bigl(\ep^2 f(\alphah)-\Lambda)\frac{\hat{g}^{00}+1}{\ep^2 }
+ f(\alphah+\muh)-f(\alphah)\biggr],
}
where
\leqn{ghij}{
\hat{g}^{ij} := g^{ij}|_{x^0=0} = \eta^{ij} + \ep^2 \bigl(\uh^{ij}-\Half\eta^{ij}\eta_{lm}\uh^{lm} \bigr)  +
\begin{displaystyle}\ep^4\frac{ \gc^{ij}(\ep^2\uh)}{\ep^4} \end{displaystyle}
}
and
\leqn{muhdef}{
\muh = \mu|_{x^0=0}.
}

Next, we see, using  \eqref{redGbC.1}-\eqref{redGbC.4}, \eqref{taudef}, \eqref{tauexpA.1}-\eqref{tauexpA.3}, \eqref{tauexpC.1}-\eqref{tauexpC.2}, \eqref{ghij} and the rescaled variables  \eqref{uhdef}, that the
constraint equations \eqref{ceqns.1} take the form
\leqn{GconstA}{
\gh^{LM}\del{L}\del{M} \uh^{0j} - 2 \ep^2 \uh^{0L}\del{L}\del{M} \uh^{Mj}
+ \delta^j_0  \gh^{00} \del{L}\del{M} \uh^{LM} - \delta^j_M \gh^{00} \del{L}\uh_0^{LM}
+ \ep^2 \ah^j(\ep,\uh,\del{L}\uh,\uh_0) = \delta^j_0\tauh^{00} + \ep\delta^j_I\tauh^{0I}
}
where
\begin{enumerate}[(i)]
\item
for any $R>0$ there exists an $\ep_0 > 0$ such that
$\ah^j$ is analytic for $(\ep,\uh,\del{L}\uh,\uh_0) \in (-\ep_0,\ep_0)\times B_R(\Sbb{4}) \times
(\Sbb{4})^3\times \Sbb{4}$ and
\eqn{Azero}{
\bigl|\ah^j(\ep,\uh,\del{L}\uh,\uh_0)\bigr| \lesssim |\del{L}\uh|^2 + |\uh_0|^2,
}
\item $\tauh^{00}$ and $\tauh^{0J}$ are given by
\lalign{tauh0jA}{
\tauh^{00} &= (\hat{g}^{00})^2 \tauh_{00}+\ep^3 2\frac{\hat{g}^{0L}}{\ep^2} \hat{g}^{00}\tauh_{0L}
+ \ep^5 \frac{\hat{g}^{0L}}{\ep^2} \frac{\hat{g}^{0L}}{\ep^2}\tauh_{LM}, \label{tauh0jA.1}\\
\tauh^{0J} &=  \ep\hat{g}^{00}\frac{\hat{g}^{J0}}{\ep^2}\tauh_{00} + \Biggl(\ep^4 \frac{\hat{g}^{0L}}{\ep^2}\frac{\hat{g}^{J0}}{\ep^2}+\hat{g}^{00}\hat{g}^{JL}\Biggr)\tauh_{0L}
+ \ep^2 \frac{\hat{g}^{0L}}{\ep^2}\hat{g}^{JM}\tauh_{LM}, \label{tauh0jA.2}
}
with
\lalign{tauh0jB}{
\zh_J &= z_J|_{x^0=0}, \label{tauh0jB.1} \\
\tauh_{00} &= 4 e^{-2\betah}\muh +
\ep\Tc_{00}\bigl(\ep,\ep\uh,\del{L}\uh,\uh_0,\alphah,\muh,\zh_L,\betah,\betah_0,0,0,0,\psih_{00}\bigr),
\label{tauh0jB.2} \\
\tauh_{0J} &= \Bigl\{\bigl[-4 e^{-2\betah}(\alphah+\muh)+\ep \Pc(\ep,\ep\uh,\alphah+\muh,\betah,0)\bigr]\zh_J \notag \\
& \hspace{1.0cm} +  \betah_0\delta_{lm}\del{J}\uh^{lm}\Bigr\}
 + \ep \Tc_{0J}(\ep,\ep \uh,\del{L}\uh,\uh_0,\betah_0,0,0,0), \label{tauh0jB.3}\\
\tauh_{IJ} &= \Tc_{IJ}\bigl(\ep,\ep \uh,
\del{L}\uh, \uh_0,\alphah,\muh,\zh_L,\betah,\betah_0,0,0,0,0\bigr), \label{tauh0jB.4}
}
\item and
\eqn{ghdefa}{
\gh^{ij} = \eta^{ij} + \ep^2 \uh^{ij}.
}
\end{enumerate}

\begin{thm} \label{idatathm}
Suppose $s \in \Zbb_{>3/2}$, $\betah\in \Rbb$,
\gath{idatathm.0}{
\tauh_{00}, \tauh^{0J}, \del{K}\del{L}U^{KL}, \del{L} U^{LM}_0 \in L^{\frac{6}{5}}(\Rbb^3), \\
U^{IJ} \in L^{6}(\Rbb^3)\cap K^{s+2}\cap L^\infty(\Rbb^3), \quad
U_0^{IJ} \in H^{s+1}(\Rbb^3),\\
\tauh_{00}, \tauh^{0J} \in H^s(\Rbb^3), \AND
e^{\betah} + \Quarter\tauh_{00}(\xv) \geq c_0 > 0 \quad \forall\, \xv \in \Rbb^3
}
for some positive constant $c_0$, and let
\alin{idatathm.2}{
R = |\betah| + \norm{\del{K}\del{L}U^{KL}}_{L^{\frac{6}{5}}}&+ \norm{\del{L} U^{LM}_0}_{L^{\frac{6}{5}}}
 \notag \\  +\norm{U^{IJ}}_{L^6\cap K^{s+2}}
 & + \norm{U_0^{IJ}}_{H^{s+1}}
 +\norm{\tauh_{00}}_{H^s\cap L^{\frac{6}{5}}}+
 \norm{\tauh^{0J}}_{H^s\cap L^{\frac{6}{5}}} .
}
Then there exists constants $\ep_0=\ep_0(c_0,R)>0$, $C=C(c_0,R)>0$, and a 1-parameter family of maps
\eqn{idatathm.3}{
\uh^{0j}_\ep \in K^{s+2}(\Rbb^3)\cap L^6(\Rbb^3), \quad \muh_\ep \in K^s(\Rbb^3)\cap L^6(\Rbb^3), \quad \zh_I^\ep \in H^{s}(\Rbb^3) \quad  0 < \ep < \ep_0
}
such that
\begin{enumerate}[(i)]
\item the quadruple $\{\uh^{ij}_\ep,\uh^{ij}_{0,\ep},\muh_\ep,\zh_I^\ep\}$, where
\eqn{idatathm.4}{
\bigl(\uh^{ij}_\ep\bigr) =\begin{pmatrix} \uh^{00}_\ep & \uh^{0J}_\ep \\
\uh^{I0}_\ep & \ep U^{IJ} \end{pmatrix} \AND
\bigl(\uh^{ij}_{0,\ep}\bigr) =\begin{pmatrix} \del{K}\uh^{K0}_\ep & -\ep \del{K} U^{KJ} \\
-\ep \del{K} U^{IK}&  \ep U^{IJ}_0\end{pmatrix},
}
together with the initial values  \eqref{betah} and \eqref{psidata} for the conformal factor determine
a solution of the constraint equations
\eqref{ceqns.1}-\eqref{ceqns.2}
for $0<\ep <\ep_0$,
\item and
\gath{idatathm.6}{
\norm{\muh_\ep-\mubr}_{K^s\cap L^6} + \norm{\zh_I^\ep-\zbr_I}_{H^s}
\leq C \ep, \\
\norm{\uh^{00}_\ep-\ubr^{00}}_{K^{s+2}\cap L^6} + \frac{1}{\ep}\norm{\uh^{0J}_\ep-\ep\ubr^{0J}}_{K^{s+2}\cap L^6}
 \leq C \ep
}
for $0<\ep<\ep_0$,
where
\alin{idatathm.5}{
\ubr^{00} &= \Delta^{-1}\tauh_{00}, \\ 
\ubr^{0J} &= \Delta^{-1}\bigl(\tauh^{0J}-\del{K}U^{KJ}_0\bigr), \\ 
\mubr &= \Quarter e^{2\betah} \tauh_{00}, \\ 
\zbr_J &= \frac{1}{4e^{\betah}+\tauh_{00}}\Bigl[\sqrt{\Twothirds e^{-2\betah}\bigl(e^{3\betah}+\Lambda\bigr)}
\Rf_{J}(-\Delta)^{-\frac{1}{2}}\tauh_{00}+\tauh^{0J}\Bigr],  
}
and
\eqn{idatathm.5a}{
\norm{\ubr^{00}}_{L^6\cap K^{s+2}} + \norm{\ubr^{0J}}_{L^6\cap K^{s+2}}
+ \norm{\mubr}_{L^{\frac{6}{5}}\cap H^{s}} + \norm{\zbr_J}_{H^s} \leq C.
}
\end{enumerate}
\end{thm}
\begin{proof}
Our method for solving the constraint equation \eqref{GconstA} is based on a variation
of the method develop by Lottermoser in \cite{Lottermoser:1992}, and begins with prescribing
the following \textit{free initial data}:
\gath{freeidataA}{
\betah\in \Rbb, \quad
\tauh_{00}, \tauh^{0J}, \del{K}\del{L}U^{LM}, \del{L} U^{LM}_0 \in L^{\frac{6}{5}}(\Rbb^3), 
 \\
\quad U^{IJ} \in L^{6}(\Rbb^3)\cap L^\infty(\Rbb^3),
\quad \del{K}U^{IJ}, U_0^{IJ}\in H^{s+1}(\Rbb^3), \quad \tauh_{00}, \tauh^{0J} \in H^s(\Rbb^3),
}
where $\tauh_{00}$ is chosen to satisfy
\leqn{freeidataB}{
e^{\betah} + \Quarter\tauh_{00}(\xv) \geq c_0 > 0 \quad \forall\, \xv \in \Rbb^3
}
for some positive constant $c_0$. We also set
\lalign{freeidataC}{
R = |\betah| + \norm{\del{K}\del{L}U^{IJ}}_{K^s\cap L^{\frac{6}{5}}}&+ \norm{\del{L} U^{LM}_0}_{K^s\cap L^{\frac{6}{5}}}
 \notag \\  +\norm{U^{IJ}}_{L^6\cap K^s}
 & + \norm{U_0^{IJ}}_{H^{s+1}}
 +\norm{\tauh_{00}}_{K^s\cap L^{\frac{6}{5}}}+
 \norm{\tauh^{0J}}_{K^s\cap L^{\frac{6}{5}}} . \label{iterateC.1}
}

Next, we see from \eqref{tauh0jA.2} that
\leqn{tauh0jD}{
\tauh_{0J} = -\tauh^{0J} + \ep \Tc_{J}\bigl(\ep,\uh,\tauh_{00},\tauh_{LM}\bigr)
}
for a map $\Tc_J$ that (i) for any fixed $\Rh>0$, is smooth  for $(\ep,\uh,\tauh_{00},\tauh_{LM})$$\in $ $(-\ep_0,\ep_0)\times B_{\Rh}(\Sbb{4})
\times \Rbb \times \Sbb{3} $ provided $\ep_0$ is chosen sufficiently small, and (ii) satisfies the estimate
\leqn{tauh0jE}{
\bigl|\Tc_{J}\bigl(\ep,\uh,\tauh_{00},\tauh_{LM}\bigr) \bigr| \lesssim |\uh|\bigl(|\tauh_{00}|+ |\tauh_{LM}|\bigr).
}
From \eqref{tauh0jD} and \eqref{tauh0jE}, it follows, with the help of \eqref{tauexpB.3} and \eqref{tauh0jB.4},
that we can write \eqref{tauh0jA.1} as
\leqn{tauh0jF}{
\tauh^{00} = \tauh_{00} +\ep^2 \Tc_0\bigl(\ep,\tauh_{00},\tauh^{0L},
\uh,
\del{L}\uh, \uh_0,\muh,\zh_L,\betah\bigr)
}
for a map $\Tc_0$ that (i) for any fixed $\Rh>0$, is smooth  for
$(\ep,\tauh_{00},\tauh^{0L},\uh,
\del{L}\uh, \uh_0,\muh,\zh_L,\betah)$ $\in$ $(-\ep_0,\ep_0)\times\Rbb\times \Rbb^3\times B_{\Rh}(\Sbb{4})
\times (\Sbb{4})^3 \times \Sbb{4} \times \Rbb\times \Rbb^3\times \Rbb $ provided $\ep_0$ is chosen sufficiently small, and (ii) satisfies the estimate
\leqn{tauh0jG}{
\bigl|\Tc_{0}\bigl(\ep,\tauh_{00},\tauh^{0L},\uh,
\del{L}\uh, \uh_0,\muh,\zh_L,\betah\bigr) \bigr| \lesssim |\uh|(|\tauh_{00}|+ |\tauh^{0L}|)
+|\uh|^2(|\del{L}\uh|+|\uh_0|+|\zh_L|).
}

From \eqref{tauh0jD}-\eqref{tauh0jG},
it is then clear that we can write \eqref{GconstA}, \eqref{tauh0jB.2} and
 \eqref{tauh0jB.3} as
\lalign{iterateA}{
\Delta \uh^{0j} &= -\ep^2 \uh^{LM}\del{L}\del{M} \uh^{0j} + 2 \ep^2 \uh^{0L}\del{L}\del{M} \uh^{Mj}
 + (1-\ep^2\uh^{00})\bigl( \delta^j_0  \del{L}\del{M} \uh^{LM}  \notag \\
 &\hspace{1.5cm}- \delta^j_M \del{L}\uh_0^{LM}\bigr)
+ \delta^j_0\tauh_{00} + \ep\delta^j_I\tauh^{0I} +\ep^2 \Ah^j\bigl(\ep,\uh,\del{L}\uh,\uh_0,\muh,\zh_L,\betah,
\tauh_{00},\tauh^{0L}\bigr), \label{iterateA.1}\\
\muh &= \Quarter e^{2\betah}\tauh_{00} +
\ep\hat{\tc}_{0}\bigl(\ep,\uh,\del{L}\uh,\uh_0,\muh,\zh_L,\betah\bigr), \label{iterateA.2}
\intertext{and}
\zh_J &=  \frac{1}{4e^{-2\betah}(e^{3\betah}+\Quarter e^{2\betah}\tauh_{00})}\biggl[\sqrt{\Twothirds e^{-2\betah}\bigl(e^{3\betah}+\Lambda\bigr)}\delta_{lm}\del{J}\uh^{lm}+\tauh^{0J}
  \notag \\
&\hspace{6.5cm} + \ep \hat{\tc}_{J}(\ep,\uh,\del{L}\uh,\uh_0,\muh,\zh_L,\betah,\tauh_{00},\tauh^{0L})\biggr], \label{iterateA.3}
}
respectively,
where, for any fixed $\Rh>0$, $\Ah^j$, $\hat{\tc}_{J}$ are smooth for
$(\ep,\uh,\del{L}\uh,\uh_0,$ $\muh,\zh_L,\betah,
\tauh_{00},\tauh^{0L})$ $\in$ $(-\ep_0,\ep_0)\times B_{\Rh}(\Sbb{4})\times (\Sbb{4})^3\times \Sbb{4}
\times \Rbb  \times  \Rbb^3 \times \Rbb \times \Rbb \times \Rbb^3 $, and  $\tauh_{00}$
is smooth for
$(\ep,\uh,\del{L}\uh,\uh_0,\muh,\zh_L,\betah)$ $\in$
$(-\ep_0,\ep_0)\times B_{\Rh}(\Sbb{4})\times (\Sbb{4})^3\times \Sbb{4} \times \Rbb  \times  \Rbb^3 \times \Rbb$
provided $\ep_0$ is chosen sufficiently small. Moreover, these maps satisfy the estimates
\lalign{iterateB}{
|\Ah^j\bigl(\ep,\uh,\del{L}\uh,\uh_0,\muh,\zh_L,\betah,
\tauh_{00},\tauh^{0L}\bigr) &\lesssim |\uh|(|\tauh_{00}|+ |\tauh^{0L}|)
 \notag \\
 &\hspace{0.5cm} +|\uh|^2|(|\del{L}\uh|+|\uh_0|+|\zh_L|) + |\del{L}\uh|^2 + |\uh_0|^2, \label{iterateB.1}\\
|\hat{\tc}_{0}\bigl(\ep,\uh,\del{L}\uh,\uh_0,\muh,\zh_L,\betah\bigr)|
&\lesssim |\uh| + |\del{L}\uh|+|\uh_0|+|\muh|+|\zh_L|, \label{iterateB.2}
\intertext{and}
|\hat{\tc}_{J}(\ep,\uh,\del{L}\uh,\uh_0,\muh,\zh_L,\betah,\tauh_{00},\tauh^{0J})| &\lesssim
|\del{L}\uh|+|\uh_0|+|\zh_L| + |\tauh_{00}|+|\tauh^{0J}|. \label{iterateB.3}
}

We solve the system \eqref{iterateA.1}-\eqref{iterateA.3} using a contraction map starting
with the seed solution
\eqn{seedsolC}{
\xibr_\ep = \bigl(\ubr^{00}_\ep,\ep^{-1}\ubr^{0J}_\ep,\mubr,\zbr_J\bigr),
}
where
\alin{seedsolA}{
\ubr_\ep &= \bigl(\ubr^{ij}\bigr) =\begin{pmatrix} \Delta^{-1}\bigl(\tauh_{00}+\ep\del{K}\del{L}U^{KL}\bigr)\ & \ep\Delta^{-1}\bigl(\tauh^{0J}-\del{K}U^{KJ}_0\bigr) \\
\ep\Delta^{-1}\bigl(\tauh^{0I}-\del{K}U^{IK}_0\bigr) & \ep U^{IJ} \end{pmatrix}, 
\\
\ubr_{0,\ep} &= \bigl(\ubr^{ij}_{0}\bigr) =\begin{pmatrix} \ep\Rf_{K}(-\Delta)^{-\frac{1}{2}}\bigl(\tauh^{0K}-\del{L}U^{KL}_0\bigr) & -\ep \del{K} U^{KJ} \\
-\ep \del{K} U^{IK}&  \ep U^{IJ}_0\end{pmatrix}, 
\\
\mubr &= \Quarter e^{2\betah} \tauh_{00}, 
\intertext{and}
\zbr_J &= \frac{1}{4e^{\betah}+\tauh_{00}}\Bigl[\sqrt{\Twothirds e^{-2\betah}\bigl(e^{3\betah}+\Lambda\bigr)}
\Rf_{J}(-\Delta)^{-\frac{1}{2}}\tauh_{00}+\tauh^{0J}\Bigr], 
}
which is well defined due to Theorem \ref{ellipDthm}, and the mapping properties of Riesz potentials
and transforms given in Theorems \ref{Rieszpotthm} and \ref{Riesztransthm}.

On quadruples of the form
\eqn{xitdef}{
\xit = \bigl(\ut^{00},\ep^{-1}\ut^{0J},\mut,\zt_J\bigr),
}
we define a norm
\eqn{Xnormdef}{
\norm{\xit}_X = \norm{\ut^{00}}_{L^6\cap K^{s+2}} + \norm{D^2 \ut^{00}}_{L^\frac{6}{5}}
+ \norm{\ep^{-1}\ut^{0J}}_{L^6\cap K^{s+2}} + \norm{ D^2 \ep^{-1}\ut^{0J}}_{L^\frac{6}{5}}
+ \norm{\mut}_{H^s} + \norm{\zt_J}_{H^s},}
and let
\eqn{Xspacedef}{
X = \{\, \xit \in H^{s+2}_{\text{loc}}(\Rbb^3)\times H^{s+2}_{\text{loc}}(\Rbb^3,\Rbb^3)
\times H^s_{\text{loc}}(\Rbb^3) \times H^s_{\text{loc}}(\Rbb^3,\Rbb^3) \,|\, \norm{\xit}_X < \infty \}.
}
In terms of this norm, the seed solution $\xibr_{\ep}$ satisfies the estimate
\leqn{idataboundsA}{
\norm{\xibr_{\ep}} \leq \breve{C}(c_0,R)R \qquad 0\leq \ep < \ep_0
}
for some positive constant $\breve{C}(c_0,R)>0$.

For fixed $\Rh>0$, $\ep_0$ chosen sufficiently small depending on $\Rh$, $\ep \in (0,\ep_0)$, and
\eqn{JdefA}{
\xit = \bigl(\ut^{00},\ep^{-1}\ut^{0J},\mut,\zt_J\bigr) \in B_{\Rh}(X),
}
we define
\eqn{JdefB}{
J(\xit) := \xih = \bigl(\uh^{00},\ep^{-1}\uh^{0J},\muh,\zh_J\bigr)
}
by requiring $\{\uh_\ep^{0j},\muh_\ep,\zh^\ep_J\}$ to be the unique solution of
\alin{JdefC}{
\Delta \uh^{0j} &= -\ep^2 \ut^{LM}\del{L}\del{M} \ut^{0j} + 2 \ep^2 \ut^{0L}\del{L}\del{M} \ut^{Mj}
 + (1-\ep^2\ut^{00})\bigl( \delta^j_0  \del{L}\del{M} \ut^{LM}  \notag \\
 &- \delta^j_M \del{L}\ut_0^{LM}\bigr)
+ \delta^j_0\tauh_{00} + \ep\delta^j_I\tauh^{0I} +\ep^2 \Ah^j\bigl(\ep,\ut,\del{L}\ut,\ut_0,\mut,\zt_L,\betah,
\tauh_{00},\tauh^{0L}\bigr), 
\\
\muh &= \Quarter e^{2\betah}\tauh_{00} +
\ep\hat{\tc}_{0}\bigl(\ep,\ut,\del{L}\ut,\ut_0,\mut,\zt_L,\betah\bigr), 
 \\
\zh_J &=  \frac{1}{4e^{\betah}+\tauh_{00}}\biggl[\sqrt{\Twothirds e^{-2\betah}\bigl(e^{3\betah}+\Lambda\bigr)}\biggl(\Rf_{J}(-\Delta)^{-\frac{1}{2}}\tauh_{00} \notag \\ &\qquad +\ep \frac{\delta_{lm}\del{J}\uh^{lm}-\Rf_{J}(-\Delta)^{-\frac{1}{2}}\tauh_{00}}{\ep}\biggr)
+\tauh^{0J}\biggr]
 + \ep \hat{\tc}_{J}(\ep,\ut,\del{L}\ut,\ut_0,\mut,\zt_L,\betah,\tauh_{00}), 
}
where
\eqn{JdefDa}{
\ut = \bigl(\ut^{ij}\bigr) =\begin{pmatrix} \ut^{00} & \ut^{0J} \\
\ut^{I0} & \ep U^{IJ} \end{pmatrix} \AND
\ut_{0} = \bigl(\ut^{ij}_0\bigr) =\begin{pmatrix} \del{K}\ut^{K0} & -\ep \del{K} U^{KJ} \\
-\ep \del{K} U^{IK}&  \ep U^{IJ}_0\end{pmatrix}.
}
That $J(\xit)$ is well-defined for $\xit \in B_{\Rh}(X)$  follows
from Theorem \ref{ellipDthm}, the mapping properties of Riesz potentials
and transforms given in Theorems \ref{Rieszpotthm} and \ref{Riesztransthm},
the bounds on the initial data given by \eqref{freeidataB} and \eqref{iterateC.1},
and the calculus inequalities from Appendix \ref{calc}.
It also follows directly from
these estimates that $J$ satisfies a Lipschitz estimate of the form
\leqn{JlipA}{
\norm{J(\xit_1)-J(\xit_2)}_{X} \leq \ep C(c_0,R,\Rh)
\norm{\xit_1-\xit_2}_X
}
for all $\xit_1,\xit_2 \in B_{\Rh}(X)$ and $\ep \in (0,\ep_0)$.

Choosing
\eqn{Rhfix}{
\Rh = 2\breve{C}(c_0,R)R,
}
where $\breve{C}(c_0,R)$ is the constant from \eqref{idataboundsA}, we see that $J(\xibr_\ep)$ is well defined.
Moreover, from the definition of $\xibr_\ep$ and \eqref{idataboundsA}, it is clear that
\eqn{seesol}{
J(\xibr_\ep) = \xibr_\ep + C(c_0,R)\ep \qquad 0<\ep < \ep_0,
}
for some constant $C(c_0,R)$, and so, it follows from \eqref{JlipA} that there exists a
positive constant $C(c_0,R)>0$ such that
\eqn{JlipB}{
\norm{J(\xit)-\xibr_\ep}_{X} \leq \ep C(c_0,R)
(\norm{\xit-\xibr_\ep}_X+1)
}
for all $\xi \in B_{\Rh}(0;X)$ and $\ep \in (0,\ep_0)$. Choosing $\ep_0>0$ small enough so that
\eqn{ep0fix}{
\ep_0 C(c_0,R,\Rh) < 1 \AND \ep_0 C(c_0,R)\biggl(\frac{\Rh}{4} + 1\biggr) < \frac{\Rh}{4}
}
shows that $J$ defines a contraction map on $B_{\Rh/4}(\xibr_\ep;X)$, and consequently, for
each $\ep \in (0,\ep_0)$, establishes the existence of
a unique fixed point $\xih_\ep \in B_{\Rh/4}(\xibr_\ep;X)$, that is, $\xih_\ep$ satisfies
$J(\xih_\ep) = \xih_\ep$
and
\leqn{xihfixB}{
\norm{\xih_\ep-\xibr_\ep}_{X} = \norm{J(\xih_\ep)-\xibr_\ep}_{X} \lesssim \ep.
}
Letting
\eqn{xihfixC}{
\xih_\ep = \bigl(\uh^{00}_\ep,\ep^{-1}\uh^{0J}_\ep,\muh_\ep,\zh^\ep_J\bigr)
}
denote the fixed point, where
\eqn{JdefD}{
\bigl(\uh^{ij}_\ep\bigr) =\begin{pmatrix} \uh^{00}_\ep & \uh^{0J}_\ep \\
\uh^{I0}_\ep & \ep U^{IJ} \end{pmatrix} \AND
\bigl(\uh^{ij}_{0,\ep}\bigr) =\begin{pmatrix} \del{K}\uh^{K0}_\ep & -\ep \del{K} U^{KJ} \\
-\ep \del{K} U^{IK}&  \ep U^{IJ}_0\end{pmatrix},
}
it follows from the definition of the map $J$ that $\{\uh^{ij}_\ep,\uh^{ij}_{0,\ep},\muh_\ep,\zh^\ep_J\}$
solves \eqref{iterateA.1}-\eqref{iterateA.3}, and it is easy to verify that these
solutions satisfy the harmonic
constraint \eqref{harmuhcA}.  This result together with the estimate \eqref{xihfixB} completes the proof.
\end{proof}

%% file: exist.tex
\sect{exist}{Uniform local existence and uniqueness}
The next step in analyzing the $\ep \searrow 0$ limit is to evolve the 1-parameter families
of initial data generated by Theorem \ref{idatathm} for a particular choice of
the free data, and show that this initial data leads to
the existence of 1-parameter families of solutions to the Einstein-Euler equations
that exist on a common spacetime region of the form $M=[0,T)\times \Rbb^3$. We begin this
process by fixing our choice of free initial data in the next section.

\subsect{init}{Initial data}

Our choice of initial data begins by fixing the free components $\betah$, $\tauh^{\ep,\vec{\yv}}_{00}$ and
$\tauh^{0J}_{\ep,\vec{\yv}}$ as in \eqref{idenD}, \eqref{yvdef} and \eqref{idenB}. We also assume that
this initial data is specified so that \eqref{idenF} holds. From the
triangle inequality and the translational invariance of the $L^p$ norms, we observe the
bounds
\leqn{idenE}{
\norm{\tauh^{\ep,\vec{\yv}}_{00}}_{L^{\frac{6}{5}}\cap H^{s+1}} \leq \sum_{\lambda=1}^N \norm{\tauh^{\lambda}_{00}}_{L^{\frac{6}{5}}\cap H^{s+1}} \lesssim 1   \qquad  \forall \, (\ep,\vec{\yv})\in (0,\infty)\times \Rbb^{3N}
}
and
\leqn{ivelC}{
\norm{\tauh_{\ep,\vec{\yv}}^{0J}}_{L^{\frac{6}{5}}\cap H^{s+1}} \leq \sum_{\lambda=1}^N \norm{\tauh_{\lambda}^{0J}}_{L^{\frac{6}{5}}\cap H^{s+1}} \lesssim 1   \qquad  \forall \, (\ep,\vec{\yv})\in (0,\infty)\times \Rbb^{3N}.
}

For the remainder of the free initial data, we can take essentially arbitrary bounded $\ep$-dependent sequences. To be definite, we fix 1-parameter families
\eqn{iremA}{
U^{IJ}_\ep,\;  U^{IJ}_{0,\ep} \qquad \ep \in (0,\ep_0)
}
satisfying
\leqn{iremB}{
\norm{U^{IJ}_\ep}_{L^6\cap K^{s+3}} + \norm{\del{I}\del{J}U^{IJ}_\ep}_{L^\frac{6}{5}} + \norm{U^{IJ}_{0,\ep}}_{H^{s+2}} +
\norm{\del{I} U^{IJ}_{0,\ep}}_{L^{\frac{6}{5}}} \lesssim 1 \quad 0<\ep < \ep_0,
}
with the simplest possibility being $U^{IJ}_\ep = U^{IJ}_{0,\ep}$ $=$ $0$.

Shrinking  $\ep_0 > 0$ if necessary, the  bounds \eqref{idenF}, \eqref{idenE}, \eqref{ivelC} and \eqref{iremB}
on the free initial data
are enough to imply via
Theorem \ref{idatathm} the existence of corresponding constrained initial data
\leqn{initA}{
\uh^{0j}_{\ep,\vec{\yv}} \in K^{s+3}(\Rbb^3)\cap L^6(\Rbb^3),\; \muh_{\ep,\vec{\yv}} \in K^{s+1}(\Rbb^3)\cap L^6(\Rbb^3), \;
\zh^{\ep,\vec{\yv}}_I\in H^{s+1}(\Rbb^3) \quad \forall \; (\ep,\vec{\yv})\in (0,\ep_0)\times \Rbb^{3N}
}
such that the free and constrained components
\begin{enumerate}[(i)]
\item determine a solution of
the constraint equations \eqref{ceqns.1}-\eqref{ceqns.2} for all  $(\ep,\vec{\yv})\in (0,\ep_0)\times \Rbb^{3N}$,
\item yield the following initial data for the system \eqref{totalA.1}-\eqref{totalA.9}:
\lalign{initB}{
\bigl(u^{ij}_{0}\bigr)|_{x^0=0} &=\begin{pmatrix} \del{K}\uh^{K0}_{\ep,\vec{\yv}} & -\ep \del{K} U^{KJ}_\ep \\
-\ep \del{K} U^{IK}_{\ep}&  \ep U^{IJ}_{0,\ep}\end{pmatrix}, \label{initB.1} \\
\bigl(u^{ij}_K\bigr)|_{x^0=0} &=\begin{pmatrix} \del{K}\uh^{00}_{\ep,\vec{\yv}} & \del{K}\uh^{0J}_{\ep,\vec{\yv}} \\
\del{K}\uh^{I0}_{\ep,\vec{\yv}} & \ep \del{K}U^{IJ}_\ep \end{pmatrix}, \label{initB.2} \\
\bigl(u^{ij}\bigr)|_{x^0=0} &=\begin{pmatrix} \ep \uh^{00}_{\ep,\vec{\yv}} & \ep\uh^{0J}_{\ep,\vec{\yv}} \\
\ep\uh^{I0}_{\ep,\vec{\yv}} & \ep^2 U^{IJ}_\ep \end{pmatrix}, \label{initB.3} \\
\psi_{jL}|_{x^0=0} &= 0, \label{initB.4} \\
\psi_k|_{x^0=0} &= 0, \label{initB.5} \\
\psi|_{x^0=0} &= 0, \label{initB.6}\\
\beta_0(0) &= \sqrt{\Twothirds e^{-2\betah}(\alphah_\ep+\Lambda)} , \label{initB.7} \\
\beta(0) &= \betah, \label{initB.8} \\
\delta\zeta|_{x^0=0} &= 0 \label{initB.9}
\intertext{and}
z_J|_{x^0=0} &= \zh^{\ep,\vec{\yv}}_J \label{initB.10},
}
where
\lgath{initC}{
\alphah_\ep = r(\ep,3\betah) \label{initC.1}
\intertext{and}
\zetah =  \int_{1}^{\alphah_\ep+\muh_{\ep,\vec{\yv}}}\frac{d\xi}{\xi + \ep^2 f(\xi)} -3\betah =  \int_{\alphah_\ep}^{\alphah_\ep+\muh_{\ep,\vec{\yv}}}\frac{d\xi}{\xi + \ep^2 f(\xi)}
\label{initC.2},
}
\item and satisfy the bounds
\lgath{initD}{
\norm{\muh_{\ep,\vec{\yv}}-\mubr_{\ep,\vec{\yv}}}_{K^{s+1}\cap L^6} +
\norm{\zh^{\ep,\vec{\yv}}_I-\zbr^{\ep,\vec{\yv}}_I}_{H^{s+1}}
\lesssim \ep,  \label{initD.1} \\
\norm{\uh^{00}_{\ep,\vec{\yv}}-\ubr^{00}_{\ep,\vec{\yv}}}_{K^{s+3}\cap L^6} + \frac{1}{\ep}\norm{\uh^{0J}_{\ep,\vec{\yv}}-\ep\ubr^{0J}_{\ep,\vec{\yv}}}_{K^{s+3}\cap L^6}
 \lesssim \ep \label{initD.2}
}
for all $(\ep,\vec{\yv})\in (0,\ep_0)\times \Rbb^{3N}$,
where
\alin{initE}{
\ubr^{00}_{\ep,\vec{\yv}} &:= \Delta^{-1}\tauh_{00}^{\ep,\vec{\yv}}, 
\\
\ubr^{0J}_{\ep,\vec{\yv}} &:= \Delta^{-1}\bigl(\tauh^{0J}_{\ep,\vec{\yv}}-\del{K}U^{KJ}_{0,\ep}\bigr), 
}
$\mubr_{\ep,\vec{\yv}}$ and $\zbr^{\ep,\vec{\yv}}_J$ are given by \eqref{initE.3} and \eqref{initE.4},
respectively,
and
\leqn{initF}{
\norm{\ubr^{00}_{\ep,\vec{\yv}}}_{L^6\cap K^{s+3}} + \norm{\ubr^{0J}_{\ep,\vec{\yv}}}_{L^6\cap K^{s+3}}
 + \norm{\zbr^{\ep,\vec{\yv}}_J}_{H^{s+1}} \lesssim 1
}
for all $(\ep,\vec{\yv})\in (0,\ep_0)\times \Rbb^{3N}$.
\end{enumerate}

As a final remark for this section, we note that \eqref{idenF} and \eqref{idenE} together with Moser's estimates from Corollary \ref{calccorA} and
the definition \eqref{initE.3} imply
that $\zetah$, as defined by \eqref{initC.2}, satisfies the estimate
\leqn{initG}{
\norm{\zetah-\zetabr_{\ep,\vec{\yv}}}_{L^6\cap K^{s+1}} \lesssim \ep \quad \forall \; (\ep,\vec{\yv})\in (0,\ep_0)\times \Rbb^{3N},
}
where
\leqn{initH}{
 \zetabr_{\ep,\vec{\yv}} = \ln\bigl( e^{3\betah}+\mubr_{\ep,\vec{\yv}}\bigr)
 -3\betah \AND \norm{\zetabr_{\ep,\vec{\yv}}}_{L^{\frac{6}{5}}\cap K^{s+1}} \lesssim 1.
}

\subsect{loc}{Local existence and continuation}
With the initial data fixed, we now turn to proving the existence of solutions to the conformal
Einstein-Euler equations generated by the initial data of the previous section. Our basic local existence result is
contained in the following proposition.

\begin{prop} \label{locexist}
For each $\ep \in (0,\ep_0)$ and $\vec{\yv}\in \Rbb^{3N}$, there exists a $T_{\ep,\vec{\yv}}>0$ and
a unique solution\footnote{$H_{\textrm{ul}}^s(\Rbb^3)$ denotes the uniformly local Sobolev spaces; see
 \cite[\S 2.3]{Majda:1984} for a definition.}
\gath{locexist1}{
u^{ij}_{k,\ep,\vec{\yv}}, u^{ij}_{\ep,\vec{\yv}}, \psi^{\ep,\vec{\yv}}_{jL}, \psi^{\ep,\vec{\yv}}_j,
\psi^{\ep,\vec{\yv}}, \delta\zeta_{\ep,\vec{\yv}}, z_J^{\ep,\vec{\yv}} \in
L^\infty([0,T_{\ep,\vec{\yv}}),H_{\emph{ul}}^{s}(\Rbb^3))\cap  \bigcap_{m=0}^1 C^m\bigl([0,T_{{\ep,\vec{\yv}}}),H_{\emph{loc}}^{s-m}(\Rbb^3)\bigl),\\
\beta_0^\ep,\beta^\ep \in C^1([0,T_{\ep,\vec{\yv}}))
}
to the IVP consisting of the evolution equation \eqref{totalA.1}-\eqref{totalA.9} and
the initial data \eqref{initB.1}-\eqref{initB.10} on the spacetime region $[0,T_{\ep,\vec{\yv}})\times \Rbb^3$. Moreover,
\begin{enumerate}[(i)]
\item if the solution remains bounded in the $W^{1,\infty}(\Rbb^3)$ norm on the interval $[0,T_{\ep,\vec{\yv}})$, the matrices $A^0$ and
$B^0$, see \eqref{totalB.1} and  \eqref{totalB.3},  satisfy $A^0,B^0 \geq \kappa \id$ for some $\kappa > 0$ on the interval $[0,T_{\ep,\vec{\yv}})$, and the solution remains inside
a compact subregion of the open region on which the coefficient functions of the system, e.g. $A^k$, $B^k$, $F$, $G$, etc., are smooth, then there exists
a unique continuation of the solution to a time $T^*> T_{\ep,\vec{\yv}}$, and
\item
the collection $\{u^{ij}_{k,\ep,\vec{\yv}},\delta\zeta_{\ep,\vec{\yv}}, z_J^{\ep,\vec{\yv}},\beta^\ep, \psi^{\ep,\vec{\yv}}\}$
determine, via \eqref{utog}, \eqref{ghdef}, \eqref{udefA}, \eqref{nbdef}, \eqref{zetabdef}, \eqref{zbdefA}, \eqref{betadef},\eqref{Nfldef.2},
\eqref{Nfldef.3}, \eqref{Nfldef.4}, \eqref{zetahdef} and \eqref{totalB.8}, a classical solution of the conformal Einstein-Euler equations, given by \eqref{cEE.1}
and \eqref{cEE.2}, on $[0,T_{\ep,\vec{\yv}})\times \Rbb^3$.
\end{enumerate}
\end{prop}
\begin{proof}
For fixed $\ep \in (0,\ep_0)$ and $\vec{\yv}\in \Rbb^{3N}$, it follows directly from \eqref{idenF}, \eqref{ivelC}, \eqref{iremB}, \eqref{initA},
\eqref{initG} and \eqref{initH} that the initial data for the fields $u^{ij}_{k,\ep,\vec{\yv}}$, $u^{ij}_{\ep,\vec{\yv}}$, $\psi^{\ep,\vec{\yv}}_{jL}$, $\psi^{\ep,\vec{\yv}}_j$,
$\psi^{\ep,\vec{\yv}}$, $\delta\zeta_{\ep,\vec{\yv}}$ and $z_J^{\ep,\vec{\yv}}$ lie in the uniformly local Sobolev space $H^s_{\textrm{ul}}(\Rbb^3)$, and that $\zetah$, defined by \eqref{initC.2},
and $\del{K}\zetah$, which enter as time-independent functions in the system \eqref{totalA.1}-\eqref{totalA.9}, also lie in $H^s_{\textrm{ul}}(\Rbb^3)$. Since the system
\eqref{totalA.1}-\eqref{totalA.9} is symmetric hyperbolic and $s\in \Zbb_{>3/2+1}$, standard existence theorems, see Theorems 2.1 and 2.2  in \cite[\S 2.3]{Majda:1984}, guarantee the existence and uniqueness of
a solution
\lgath{locexist2}{
u^{ij}_{k,\ep,\vec{\yv}}, u^{ij}_{\ep,\vec{\yv}}, \psi^{\ep,\vec{\yv}}_{jL}, \psi^{\ep,\vec{\yv}}_j,
\psi^{\ep,\vec{\yv}}, \delta\zeta_{\ep,\vec{\yv}}, z_J^{\ep,\vec{\yv}} \in L^\infty([0,T_{\ep\vec{\yv}}),H_{\textrm{ul}}^{s}(\Rbb^3))\cap  \bigcap_{m=0}^1 C^m\bigl([0,T_{{\ep,\vec{\yv}}}),H_{\textrm{loc}}^{s-m}(\Rbb^3)\bigl), \label{locexist2.1}\\
\beta_0^\ep,\beta^\ep \in C^1([0,T_{\ep,\vec{\yv}})) \label{locexist2.2}
}
to the IVP consisting of the evolution equation \eqref{totalA.1}-\eqref{totalA.9} and
the initial data \eqref{initB.1}-\eqref{initB.9} for some positive time $T_{\ep,\vec{\yv}} > 0$. The stated continuation principle is also a direct consequence of these theorems.
Moreover, from the derivation of the system \eqref{totalA.1}-\eqref{totalA.9}, it is clear that the solution
\eqref{locexist2.1}-\eqref{locexist2.2} will also solve  reduced conformal Einstein-Euler equations, given by \eqref{cEE.2} and \eqref{redEin}, on $[0,T_{\ep,\vec{\yv}})\times \Rbb^3$ provided that
the constraints
\leqn{cprop1}{
\cf = \begin{pmatrix}\cf^1 \\ \cf^2_K \\ \cf^3 \\ \cf^4 \\ \cf^5 \\ \cf^6_I  \\ \cf^7_I  \\ \cf^8_{IJ} \end{pmatrix} =\begin{pmatrix}\del{0}u^{ij}_{\ep,\vec{\yv}}-u^{ij}_{0,\ep,\vec{\yv}} \\ u^{ij}_{K,\ep,\vec{\yv}} - \frac{1}{\ep}\del{K}u^{ij}_{\ep,\vec{\yv}}\\ \Theta_\ep \\ {\beta_{\ep}}'-\beta_{0,\ep} \\ \del{0}\psi^{\ep,\vec{\yv}}-
\psi^{\ep,\vec{\yv}}_0 \\ \psi_{I}^{\ep,\vec{\yv}}-\frac{1}{\ep}\del{I}\psi^{\ep,\vec{\yv}} \\
\del{0}\psi_{I}^{\ep,\vec{\yv}}-\psi_{0I}^{\ep,\vec{\yv}} \\  \psi_{IJ}^{\ep,\vec{\yv}} -
\frac{1}{\ep} \del{I}\psi^{\ep,\vec{\yv}}_J \end{pmatrix}
}
vanish for $0\leq x^0 < T_{\ep,\vec{\yv}}$, where
\eqn{locexist3}{
\Theta_\ep = e^{-2\beta_\ep}(\alpha_\ep+\Lambda)-\Threehalfs ({\beta_{\ep}}')^2 \AND \alpha_\ep = r(\ep,3\beta_\ep).
}

To establish the vanishing of the constraints \eqref{cprop1}, we first
observe that the field equations \eqref{totalA.2}, \eqref{totalA.5}, \eqref{totalA.6}
and \eqref{totalA.8} imply that
\leqn{cprop2}{
\cf^1(x^0,\xv) = \cf^4(x^0) = \cf^5(x^0,\xv) = \cf^7_I(x^0,\xv) = 0\quad \forall \; (x^0,\xv)\in  [0,T_{\ep,\vec{\yv}})\times \Rbb^3.
}
Furthermore, from \eqref{ThetaevB} and the fact that the initial data is chosen so that $\cf^3(0)=0$, we have that
\eqn{cprop3}{
\cf^3(x^0)= 0 \quad \forall \; x^0\in  [0,T_{\ep,\vec{\yv}}).
}
Next, a short computation using the evolution equations \eqref{totalA.1} and \eqref{totalA.3} shows that
\eqn{cprop4}{
\del{0}\cf^2_K = 0 \AND \del{0}\cf^{8}_{IJ} = 0.
}
Since $\cf^2_K|_{x^0=0}=\cf^8_{IJ}|_{x^0=0} = 0$ is satisfied by our choice of initial data, it follows immediately
that
\leqn{cprop5}{
\cf^2_K(x^0,\xv)= \cf^8_{IJ}(x^0,\xv) = 0 \quad \forall \; (x^0,\xv) \in [0,T_{\ep,\vec{\yv}})\times \Rbb^3,
}
and in particular, that
\leqn{cprop6}{
\del{[I}\psi_{J]K}(x^0,\xv) = 0  \quad \forall \; (x^0,\xv) \in [0,T_{\ep,\vec{\yv}})\times \Rbb^3.
}
A short computation then shows that
\leqn{cprop7}{
\del{0}(\psi_{IJ}-\psi_{JI}) = \frac{1}{\ep}(\del{I}\psi_{0J} - \del{J}\psi_{0I}) = \frac{1}{\ep}
\del{0}(\del{I}\psi_J - \del{J}\psi_I)
}
follows from the field equations \eqref{totalA.3}.  Setting,
\eqn{cprop8}{
\cf^{9}_{IJ} = \psi_{IJ}-\psi_{JI},
}
equation \eqref{cprop7} and the fact that $\cf^6_I|_{x^0=0}=\cf^{9}_{IJ}|_{x^0=0}=0$,
by our choice of initial data, imply that
\lgath{cprop9}{
\cf^{9}_{IJ}(x^0,\xv)
= \frac{1}{\ep}\bigl(\del{I}\psi_J(x^0,\xv) - \del{J}\psi_I(x^0,\xv)\bigr) \quad \forall \; (x^0,\xv) \in [0,T_{\ep,\vec{\yv}})\times \Rbb^3.\label{cprop9.2}
}

Continuing on, we define
\eqn{cprop10}{
\cf^{10}_K = \psi_{0K}-\frac{1}{\ep}\del{K}\psi_0,
}
and observe that the evolution equations \eqref{totalA.3} imply, with the help
of \eqref{cprop2}, \eqref{cprop5}, \eqref{cprop6} and  \eqref{cprop9.2}, that the triplet $\{\cf^6_K,\cf^{9}_{JK},\cf^{10}_K\}$
satisfies the system
\lalign{cprop11}{
-g^{KL}g^{00}\del{0}\cf^{10}_K &= \frac{2}{\ep}g^{KL}g^{0J}\del{0}\cf^{9}_{JK} +\frac{1}{2\ep}g^{KL}g^{IJ}\del{I}\bigl(\cf^{9}_{JK}-\cf^9_{KJ}\bigr)
+ \pf^{KL} \cf^{10}_K + \qf^{KLM} \cf^{9}_{KM} + \rf^{KL}\cf^6_K, \label{cprop11.1} \\
\frac{1}{2}g^{IJ}g^{KL}\del{0}\cf^9_{JK} &=\frac{1}{2\ep} g^{KL}g^{IJ}\bigl(\del{J}\cf^{10}_K - \del{K}\cf^{10}_J\bigr),\label{cprop11.2} \\
\del{0}\cf^6_K &= \cf^{10}_K, \label{cprop11.3}
}
where the coefficients $\pf^{KL}$, $\qf^{KLM}$ and $ \rf^{KL}$ depend smoothly on the solution. Introducing the change of variables
\eqn{cprop11a}{
\cf^{11}_K = \cf^{10}_K + \frac{2g^{0J}}{\ep g^{00}}\cf^9_{JK}
}
then allows us, using \eqref{cprop6} and \eqref{cprop9.2}, to write \eqref{cprop11.1}-\eqref{cprop11.3}  in the following  symmetric hyperbolic form
\alin{cprop12}{
-g^{KL}g^{00}\del{0}\cf^{11}_K &=\frac{1}{2\ep}g^{KL}g^{IJ}\del{I}\bigl(\cf^{9}_{JK}-\cf^9_{KJ}\bigr)
+ \pf^{KL}\biggl( \cf^{11}_K - \frac{2g^{0M}}{\ep g^{00}}\cf^9_{MK}\biggr) \notag \\
& \hspace{2.5cm}  +\biggl( \qf^{KLM} -g^{KM}g^{00}\del{0}\biggl(\frac{2g^{0K}}{\ep g^{00}}\biggr)\biggr) \cf^{9}_{KM}+ \rf^{KL}\cf^6_K, 
\\
\frac{1}{2}g^{IJ}g^{KL}\del{0}\cf^9_{JK} &= - \frac{2 g^{KL}g^{IJ}g^{0M}}{\ep^2 g^{00}} \del{M}\cf^0_{JK} + \frac{1}{2\ep} g^{KL}g^{IJ}\bigl(\del{J}\cf^{11}_K - \del{K}\cf^{11}_J\bigr)
\notag \\
& \hspace{2.5cm} + \frac{g^{KL}g^{IJ}}{2\ep}\biggl(\del{K}\biggl(\frac{2g^{0M}}{\ep g^{00}}\biggr)\cf^9_{MJ} - \del{J}\biggl(\frac{2g^{0M}}{\ep g^{00}}\biggr)\cf^9_{MK}\biggr),
 \\
\del{0}\cf^6_K &= \cf^{10}_K. 
}
But, by our choice of initial data,
$\cf^6_J|_{x^0=0}$ $=$ $\cf^9_{IJ}|_{x^0=0}$ $=$ $\cf^{11}_J|_{x^0=0}$ $=$ $0$, and so,
it follows from the uniqueness of solutions to hyperbolic systems that
\eqn{cprop14}{
\cf^6_J(x^0,\xv)= \cf^9_{IJ}(x^0,\xv) = \cf^{11}_J(x^0,\xv) = 0 \quad \forall \; (x^0,\xv) \in [0,T_{\ep,\vec{\yv}})\times \Rbb^3,
}
thereby verifying the vanishing of the constraints \eqref{cprop1}. Thus, we have established
that the solution
\eqref{locexist2.1}-\eqref{locexist2.2} solves the reduced conformal Einstein-Euler equations.

To see that the solution \eqref{locexist2.1}-\eqref{locexist2.2} also solves the full conformal Einstein-Euler equations,
we employ a variation of a well known argument due originally to Y. Choquet-Bruhat, see \cite[Ch. VI, \S 8]{ChoquetBruhat:2009}
for details. Writing the reduced conformal Einstein equations as
\leqn{CB1}{
\Gb^{ij} - \frac{\ep}{2} \frac{1}{|\gb|}\bb^{ij} = 2\Tb^{ij},
}
which  we know from above are satisfied in $[0,T_{\ep,\vec{\yv}})\times \Rbb^3$,
we see, after taking the divergence and using the contracted Bianchi identity $\nablab_i \Gb^{ij}=0$, that $\bb^{ij}$ satisfies
\leqn{CB2}{
-\frac{\ep}{2}\nablab_i \biggl(\frac{1}{|\gb|}\bb^{ij}\biggr) = 2 \nablab_i \Tb^{ij} \hspace{0.4cm}\text{in $[0,T_{\ep,\vec{\yv}})\times \Rbb^3$}.
}
Using the Euler equations \eqref{cEE.2}, which we know are satisfied in $[0,T_{\ep,\vec{\yv}})\times \Rbb^3$, a short calculation
shows, with the help of the identity $\nablab_i\nablab_j\nablab_k\Psib - \nablab_j \nablab_i\nablab_k \Psib = \Rb_{ijk}{}^l\nablab_l\Psib$,
that we can express \eqref{CB2} as
\lalign{CB3}{
-\frac{\ep}{4}\nablab_i \biggl(\frac{1}{|\gb|}\bb^{ij}\biggr) = e^{-4\Psib}\bigl(2\grave{T}^{ij}-\gb_{kl}\grave{T}^{kl}\gb^{ij}\bigr)&\nablab_i\Psib
+2\Lambda e^{-2\Psib}\nablab^j\Psib   \notag \\  - \Rb^{ij}\nablab_i\Psib
 - &\nablab^{i}\nablab_i \nablab^j \Psib - 2 \nablab^i\nablab^j\Psib \nablab_i\Psib \hspace{0.4cm} \text{in $[0,T_{\ep,\vec{\yv}})\times \Rbb^3$}. \label{CB3.1}
}
Next, we express the Ricci curvature tensor as follows:
\lalign{CB5}{
\Rb^{ij} &= \Gb^{ij} - \frac{1}{2} \gb_{kl}\Gb^{kl} \gb^{ij}\notag \\
&= \frac{1}{|\gb|}\biggl[\Gb_R^{ij}+\frac{\ep}{2}\bb^{ij}- \frac{1}{2}\biggr(\gb_{kl}\Gb_R^{kl}+\frac{\ep}{2}\gb_{kl}\bb^{kl}\biggr)\gb^{ij}\biggr] && \text{( by \eqref{redGb} and  \eqref{hbCart}) }\notag \\
&= 2\Tb^{ij} - \gb_{kl}\Tb^{kl}\gb^{ij}+\frac{\ep}{2}\frac{1}{|\gb|}\biggl(\bb^{ij}-\frac{1}{2} \gb_{kl}\bb^{kl}\gb^{ij}\biggr) && \text{ (by \eqref{CB1})}. \notag
}
Substituting this into \eqref{CB3.1}, we find, after a straightforward computation, that
\leqn{CB6}{
\nablab_i \biggl(\frac{1}{|\gb|}\bb^{ij}\biggr) = \frac{2}{|\gb|}\biggl(\bb^{ij}-\frac{1}{2}\gb_{kl}\bb^{kl}\gb^{ij}\biggr) \hspace{0.4cm}\text{in $[0,T_{\ep,\vec{\yv}})\times \Rbb^3$}.
}
Setting
\eqn{CB7}{
\beta^j = \delb{i}\ub^{ij},
}
it is not difficult to verify that \eqref{CB6} implies that $\beta^j$ satisfies a wave equation of the form
\eqn{CB8}{
\gb^{ij}\delb{i}\delb{j}\beta^k + \Upsilon^k_j \beta^j = 0  \hspace{0.4cm}\text{in $[0,T_{\ep,\vec{\yv}})\times \Rbb^3$},
}
where the coefficients $\Upsilon^k_j$ depend analytically on the fields $(\gb_{ij},\delb{k}\gb_{ij},\delb{i}\Psib)$ whenever
$\det(\gb_{ij}) < 0$. By our choice of initial data, we know that\footnote{That $\beta^j|_{\xb^0=0} = 0$ is satisfied is
obvious from our choice of initial data. That $\delb{0}\beta^j|_{\xb^0=0} = 0$ is also satisfied is
not so obvious. It is a consequence of the fact that our initial data satisfies the constraint equations; see
\cite[Ch. VI, \S 8]{ChoquetBruhat:2009} for details.}
\eqn{CB9}{
\beta^j|_{\xb^0=0} = 0 \AND \delb{0}\beta^j|_{\xb^0=0} = 0.
}
From the uniqueness of solutions to wave equations, it follows that $\beta^j = 0$ in $[0,T_{\ep,\vec{\yv}})\times \Rbb^3$,
and hence, that  $\Gb_R^{ij} = \Gb^{ij}$. We therefore conclude that the full conformal Einstein-Euler equations are
satisfied in the spacetime region $[0,T_{\ep,\vec{\yv}})\times \Rbb^3$.

\end{proof}

\begin{rem} \label{locexistrem}
Due to singular $\frac{1}{\ep}$ terms that appear in the system \eqref{totalA.1}-\eqref{totalA.9}, see in particular \eqref{totalA.1}, \eqref{totalA.3} and
\eqref{totalB.6}, the existence result given by Proposition \ref{locexist} leaves open the possibility that the time of existence $T_{\ep,\vec{\yv}}$
shrinks to zero in the limit $\ep \searrow 0$. This behaviour needs to be explicitly ruled out if we are to extract a Newtonian limit on a non-trivial spacetime region. We also need to establish a lower bound on the time existence that is independent of the
parameters $\vec{\yv}\in \Rbb^{3N}$.
\end{rem}

\subsect{aposteriori}{A posteriori bounds} Next, we show that the solutions from Proposition \ref{locexist} satisfy better bounds than just being in the
uniformly local spaces $H^s_{\textrm{ul}}(\Rbb^3)$. To this end, we take the divergence of
\eqn{tbardef}{
\taub_{ij} = 4\frac{|\gb|}{|\hb|}\Tb_{ij}
}
to get
\eqn{divTden}{
\nablab^i\taub_{ij} = \nablab_i\biggl(4\frac{|\gb|}{|\hb|}\biggr)\Tb^i{}_j+ 4\frac{|\gb|}{|\hb|}\nablab^i\Tb_{ij} = 2(\Gammab^k_{ki}-\gammab^k_{ki})\taub^{i}{}_j + 4\frac{|\gb|}{|\hb|}\nablab^i\Tb_{ij}.
}
Since $\nablab_i T^{ij} = 0$ for solutions of the conformal Einstein equation \eqref{cEE.1} by virtue of contracted Bianchi identity, we see that
\eqn{divTdenA}{
\nablab^i\taub_{ij} - 2(\Gammab^k_{ki}-\gammab^k_{ki})\taub^{i}{}_j =0,
}
which, in turn, implies via \eqref{harm3} and \eqref{hbCart} that
\eqn{divTdenB}{
\gb^{ik}\delb{k}\taub_{ij}-\Gammab^l_{kj}\taub^k{}_l - 2\Gammab^k_{ki}\taub^i{}_j = 0.
}
Switching to Newtonian coordinates, this implies that the solutions
from Proposition \ref{locexist} satisfy
\leqn{divTdenC}{
-g^{00}_{\ep,\vec{\yv}}\del{0} \tau^{\ep,\vec{\yv}}_{0j} - \frac{g^{0I}_{\ep,\vec{\yv}}}{\ep}\del{I}\tau^{\ep,\vec{\yv}}_{0j}-g_{\ep,\vec{\yv}}^{IJ}\del{I}\frac{\tau^{\ep,\vec{\yv}}_{Jj}}{\ep} = \ep \biggl(\ep \frac{g^{0I}_{\ep,\vec{\yv}}}{\ep}\frac{\del{0}\tau^{\ep,\vec{\yv}}_{Ij}}{\ep} - \Gamma^l_{kj,{\ep,\vec{\yv}}}\tau^k{}_{l,{\ep,\vec{\yv}}} - 2\Gamma^k_{ki,{\ep,\vec{\yv}}}\tau^i{}_{j,{\ep,\vec{\yv}}}\biggr)  .
}

\begin{rem} \label{omitrem}
In the following, we will treat the solution from Proposition \ref{locexist} as being bounded in certain stronger norms. All of the following calculations are easily justified by using
the finite speed of propagation to first prove them on truncated spacetime cones followed by letting the width of the cone go to infinity to obtain
estimates on the spacetime slab
$[0,T_{\ep,\vec{\yv}})\times \Rbb^3$. For reasons of economy, we omit these easily reproducible details.
\end{rem}

Setting
\eqn{Xidef}{
\Xi_{\ep,\vec{\yv}} =  \bigl(u^{ij}_{k,\ep,\vec{\yv}}, u^{ij}_{\ep,\vec{\yv}}, \psi^{\ep,\vec{\yv}}_{jL}, \psi^{\ep,\vec{\yv}}_j,
\psi^{\ep,\vec{\yv}}, \delta\zeta_{\ep,\vec{\yv}}, z_J^{\ep,\vec{\yv}},\beta_{0,\ep},\beta_\ep\bigr)
}
and
\alin{nTheta}{
\nnorm{\Xi_{\ep,\vec{\yv}}}_{s}= &\norm{u^{ij}_{k,\ep,\vec{\yv}}}_{H^s} + \norm{u^{ij}_{\ep,\vec{\yv}}}_{L^6\cap K^s} +
\norm{\psi^{\ep,\vec{\yv}}_{jL}}_{H^s}
+  \norm{\psi^{\ep,\vec{\yv}}_0}_{L^6\cap K^{s}}   \notag \\ & + \norm{\psi^{\ep,\vec{\yv}}_L}_{H^s}+
\norm{\psi^{\ep,\vec{\yv}}}_{L^6\cap K^s}
+ \norm{\delta\zeta_{\ep,\vec{\yv}}}_{H^s} + \norm{z_J^{\ep,\vec{\yv}}}_{H^s}+ |\beta_{0,\ep}| + |\beta_\ep|, 
}
it follows directly from the bounds \eqref{idenE}, \eqref{ivelC}, \eqref{iremB}, \eqref{initD.1}, \eqref{initD.2},
\eqref{initF}, \eqref{initG} and \eqref{initH} on the initial data that
\eqn{Thetab}{
\nnorm{\Xi_{\ep,\vec{\yv}}(0)}_s \leq R \qquad \forall \: (\ep,\vec{\yv})\in (0,\ep_0)\times \Rbb^{3N}
}
for some positive constant $R>0$.

From \eqref{cevC.6}-\eqref{cevC.7}, \eqref{GammaevB}, \eqref{tauexpA.1}-\eqref{tauexpA.3}, the evolution
equations \eqref{totalA.1}-\eqref{totalA.9}, the estimates \eqref{initG}-\eqref{initH}, Sobolev's inequality, Moser's estimates as given in Theorems \ref{calcpropC} and \ref{calcpropE}, and the assumption $s-1\in \Zbb_{>3/2}$, we observe that
\eqn{apostA}{
\Tf_{\ep,\vec{\yv}} = -\frac{1}{g^{00}_{\ep,\vec{\yv}}}\biggl[\frac{g^{0I}_{\ep,\vec{\yv}}}{\ep}\del{I}\tau^{\ep,\vec{\yv}}_{00} + g_{\ep,\vec{\yv}}^{IJ}\del{I}\frac{\tau^{\ep,\vec{\yv}}_{J0}}{\ep} + \ep \biggl(\ep \frac{g^{0I}_{\ep,\vec{\yv}}}{\ep}\frac{\del{0}\tau^{\ep,\vec{\yv}}_{I0}}{\ep} - \Gamma^l_{k0,{\ep,\vec{\yv}}}\tau^k{}_{l,{\ep,\vec{\yv}}} - 2\Gamma^k_{ki,{\ep,\vec{\yv}}}\tau^i{}_{0,{\ep,\vec{\yv}}}\biggr)
\biggr]
}
satisfies
\leqn{apostB}{
\norm{\Tf_{\ep,\vec{\yv}}(x^0)}_{L^2} \leq C(\nnorm{\Xi_{\ep,\vec{\yv}}(x^0)}_s)\nnorm{\Xi_{\ep,\vec{\yv}}(x^0)}_s
\qquad 0\leq x^0 < T_{\ep,\vec{\yv}}
}
for some positive constant $C(\nnorm{\Xi_{\ep,\vec{\yv}}}_s)>0$ independent of $(\ep,\vec{\yv})\in (0,\ep_0)\times \Rbb^{3N}$. Writing \eqref{divTdenC} with $j=0$ as
\eqn{apostC}{
\del{0} \tau^{\ep,\vec{\yv}}_{00} =  \Tf_{\ep,\vec{\yv}},
}
we see, after integrating in time and employing the estimate \eqref{apostB}, 
that
\eqn{apostD}{
\norm{\tau^{\ep,\vec{\yv}}_{00}(x^0)}_{L^2} \leq \norm{\tauh^{\ep,\vec{\yv}}_{00}}_{L^2} + \int_{0}^{x^0} C(\nnorm{\Xi_{\ep,\vec{\yv}}(\tau)}_s)\nnorm{\Xi_{\ep,\vec{\yv}}(\tau)}_s\, d\tau.
}
Combining this with the initial data bound \eqref{idenE} then yields the estimate
\leqn{apostE}{
\norm{\tau^{\ep,\vec{\yv}}_{00}(x^0)}_{L^2} \lesssim 1 + \int_{0}^{x^0} C(\nnorm{\Xi_{\ep,\vec{\yv}}(\tau)}_s)\nnorm{\Xi_{\ep,\vec{\yv}}(\tau)}_s\, d\tau .
}
We also observe it follows directly from Moser's estimates, Theorem \ref{calcpropE}, together
with \eqref{cevC.6}-\eqref{cevC.7}, \eqref{tauexpA.1}, \eqref{tauexpB.1} and the estimates \eqref{initG}-\eqref{initH} that
\leqn{apostF}{
\norm{D\tau^{\ep,\vec{\yv}}_{00}(x^0)}_{H^{s-1}} \leq C(\nnorm{\Xi_{\ep,\vec{\yv}}(x^0)}_s)\nnorm{\Xi_{\ep,\vec{\yv}}(x^0)}_s.
}

Next, we see that the expansions \eqref{tauexpC.1}-\eqref{tauexpC.3} for $\tau^{ij}$ together with the formulas \eqref{cevC.6}-\eqref{cevC.7}, \eqref{detghexp}, \eqref{tauexpA.2}-\eqref{tauexpA.3},
 \eqref{tauexpB.2}-\eqref{tauexpB.3},
and the estimates \eqref{initG}-\eqref{initH} and \eqref{apostE}-\eqref{apostF} imply, via  Moser's estimates from Theorems
\ref{calcpropC} and \ref{calcpropE} and Corollary \ref{calccorA},
the following estimate for $F_{\ep,\vec{\yv}}$ (see \eqref{totalB.6}):
\leqn{apostG}{
\norm{F_{\ep,\vec{\yv}}(x^0)}_{H^s} \leq \frac{1}{\ep}\biggl[ C(\nnorm{\Xi_{\ep,\vec{\yv}}(x^0)}_s)\nnorm{\Xi_{\ep,\vec{\yv}}(x^0)}_s + \int_{0}^{x^0} C(\nnorm{\Xi_{\ep,\vec{\yv}}(\tau)}_s)\nnorm{\Xi_{\ep,\vec{\yv}}(\tau)}_s\, d\tau\biggr].
}
Similar considerations also lead to the following estimates for the quantities defined by
\eqref{totalB.1}-\eqref{totalB.4}, \eqref{totalB.7} and the right hand side of \eqref{totalA.9}:
\lalign{apostGa}{
&\frac{1}{\ep}\norm{A^0_{\ep,\vec{\yv}}(x^0)-\id}_{K^s}+\frac{1}{\ep}\norm{B^0_{\ep,\vec{\yv}}(x^0)-\id}_{ K^s} \notag \\
&\hspace{3.5cm} +\norm{A^K_{\ep,\vec{\yv}}(x^0)}_{K^s}+\norm{B^K_{\ep,\vec{\yv}}(x^0)}_{K^s}\leq
C(\norm{u^{ij}_{\ep,\vec{\yv}}(x^0)}_{K^s})\norm{u^{ij}_{\ep,\vec{\yv}}(x^0)}_{K^s}
\label{apostGa.1}\\
& \hspace{3.0cm} \norm{C^k_{\ep,\vec{\yv}}(x^0)}_{K^s}  \lesssim 1+C(\nnorm{\Xi_{\ep,\vec{\yv}}(x^0)}_s)\nnorm{\Xi_{\ep,\vec{\yv}}(x^0)}_s \label{apostGa.2}
\intertext{and}
& \norm{G_{\ep,\vec{\yv}}(x^0)}_{H^s}+\norm{H_{\ep,\vec{\yv}}(x^0)}_{H^s} + \biggl\| C^K_{\ep,\vec{\yv}}(x^0)\del{K}\begin{pmatrix} \zetah \\ 0 \end{pmatrix}\biggr\|_{H^s} \lesssim 1+C(\nnorm{\Xi_{\ep,\vec{\yv}}(x^0)}_s)\nnorm{\Xi_{\ep,\vec{\yv}}(x^0)}_s.
 \label{apostGa.3}
}

Due to the $\frac{1}{\ep}E^K$ terms and also the source term $F$, the system \eqref{totalA.1}-\eqref{totalA.9}
is clearly singular. However, ignoring the singular term $F$ for the moment, \eqref{totalA.1}-\eqref{totalA.9}
is of a type that has been well-studied beginning with the pioneering work of Browning,
Klainerman, Kreiss and Majda \cite{BrowningKreiss:1982,KlainermanMajda:1981,KlainermanMajda:1982,Kreiss:1980}.
The energy estimates, which are non-singular in $\ep$, developed in these works are based on the fact that for symmetric
hyperbolic systems of the form 
\eqn{KKM}{
a^0(\ep \phi)\del{0}\phi + \frac{1}{\ep}c^K\del{K}\phi + a^K(\ep,\phi)\del{K}\phi = f(\ep,\phi) \qquad (c^K=\text{const.})
}  
only the divergence $\frac{1}{\ep}\del{K}c^K$  of the
singular term $\frac{1}{\ep}c^K$  appears in the estimates, and this vanishes because the matrices
$c^K$ are constant by assumption. Because the matrices $E^K$ in \eqref{totalA.1} and\eqref{totalA.3} are constant , and the $A^0$ and $B^0$ are functions
of $\ep u^{ij}$ only, we can apply the estimates from \cite{BrowningKreiss:1982,KlainermanMajda:1981,KlainermanMajda:1982,Kreiss:1980}\footnote{A slight generalization of these results is
required to take account of the fact that the $u^{ij}$, which appear in the coefficient matrices $A^k$, $B^k$
and $C^k$, lie in the spaces $K^s(\Rbb^3)$ instead of $H^s(\Rbb^3)$. Since the
 modification required is straightforward and amounts
 to a substitution of the analogous calculus inequality from Appendix \ref{calc} that are valid on the $K^s(\Rbb^3)$ spaces,
 we omit the details.}, for example, see Theorem 1' in \cite{KlainermanMajda:1982}, to 
\eqref{totalA.1}, \eqref{totalA.3} and \eqref{totalA.9}, while observing \eqref{apostGa.1}-\eqref{apostGa.3}, to obtain the estimates
\alin{apostH}{
\del{0}(\norm{u^{ij}_{0,{\ep,\vec{\yv}}}(x^0)}^2_{H^s} + \norm{u^{ij}_{J,{\ep,\vec{\yv}}}(x^0)}^2_{H^s}) & \lesssim \bigl(1+C(\norm{u^{ij}_{\ep,\vec{\yv}}(x^0)}_{K^s})
\norm{u^{ij}_{\ep,\vec{\yv}}(x^0)}_{K^s} \\
&\hspace{1.0cm} + \norm{F_{\ep,\vec{\yv}}(x^0)}_{H^s}\bigr)
(\norm{u^{ij}_{0,{\ep,\vec{\yv}}}(x^0)}^2_{H^s}  + \norm{u^{ij}_{J,{\ep,\vec{\yv}}}(x^0)}^2_{H^s})^{\frac{1}{2}}, 
\\
\del{0}(\norm{\psi^{\ep,\vec{\yv}}_{0L}(x^0)}^2_{H^s} +
\norm{\psi^{\ep,\vec{\yv}}_{JL}(x^0)}^2_{H^s}) & \lesssim \bigl(1+C(\norm{u^{ij}_{\ep,\vec{\yv}}(x^0)}_{K^s})
\norm{u^{ij}_{\ep,\vec{\yv}}(x^0)}_{K^s}  \\
&\hspace{1.0cm}+
\norm{G_{\ep,\vec{\yv}}(x^0)}_{H^s}\bigr)
(\norm{\psi^{\ep,\vec{\yv}}_{0L}(x^0)}^2_{H^s} + \norm{\psi^{\ep,\vec{\yv}}_{LJ}(x^0)}^2_{H^s})^{\frac{1}{2}} 
\intertext{and}
\del{0}(\norm{\delta\zeta_{\ep,\vec{\yv}}(x^0)}^2_{H^s} +
\norm{z^{\ep,\vec{\yv}}_{J}(x^0)}^2_{H^s}) & \lesssim \bigl(1 +
C(\nnorm{\Xi_{\ep,\vec{\yv}}(x^0)}_s\nnorm{\Xi_{\ep,\vec{\yv}}(x^0)}_s\bigr)
(\norm{\psi^{\ep,\vec{\yv}}_{0L}(x^0)}^2_{H^s} + \norm{\psi^{\ep,\vec{\yv}}_{LJ}(x^0)}^2_{H^s})^{\frac{1}{2}}.
}
Integrating these estimates in time,
we find, with the help of the bounds \eqref{iremB}, \eqref{initD.1}-\eqref{initD.2} and \eqref{initF}-\eqref{initH}  on the initial data, that
\lalign{apostHa}{
\norm{u^{ij}_{0,{\ep,\vec{\yv}}}(x^0)}_{H^s} + \norm{u^{ij}_{J,{\ep,\vec{\yv}}}(x^0)}_{H^s} & \lesssim 1 + \int_0^{x^0} \bigl(1+C(\norm{u^{ij}_{\ep,\vec{\yv}}(\tau)}_{K^s})\bigr)\norm{u^{ij}_{\ep,\vec{\yv}}(\tau)}_{K^s} + \norm{F_{\ep,\vec{\yv}}(\tau)}_{H^s}
\, d\tau, \label{apostHa.1} \\
\norm{\psi^{\ep,\vec{\yv}}_{0L}(x^0)}_{H^s} +
\norm{\psi^{\ep,\vec{\yv}}_{JL}(x^0)}_{H^s} & \lesssim  1 + \int_0^{x^0}\bigl( 1+C(\norm{u^{ij}_{\ep,\vec{\yv}}(\tau)}_{K^s})
\bigr)\norm{u^{ij}_{\ep,\vec{\yv}}(\tau)}_{K^s} +
\norm{G_{\ep,\vec{\yv}}(\tau)}_{H^s}
\, d \tau  \label{apostHa.2}
\intertext{and}
\norm{\delta\zeta_{\ep,\vec{\yv}}(x^0)}_{H^s} +
\norm{z^{\ep,\vec{\yv}}_{J}(x^0)}_{H^s} & \lesssim  1 + \int_0^{x^0} \bigl( 1+C(\nnorm{\Xi_{\ep,\vec{\yv}}(x^0)}_s)\bigr)\nnorm{\Xi_{\ep,\vec{\yv}}(x^0)}_s\, d\tau.
 \label{apostHa.3}
}

Integrating the remaining evolution equations \eqref{totalA.2}, \eqref{totalA.4}-\eqref{totalA.8} in time, we
obtain, using the bounds on the initial data  and the calculus inequalities from Appendix \ref{calc},
the estimates
\lalign{apostIa}{
\norm{u^{ij}_{\ep,\vec{\yv}}(x^0)}_{L^6\cap K^s} &\lesssim 1 + \int_{0}^{x^0}  \norm{u^{ij}_{0,\ep,\vec{\yv}}(\tau)}_{H^s}\,
d\tau, \label{apostI.1} \\
\norm{\psi^{\ep,\vec{\yv}}_0(x^0)}_{L^6\cap K^s} &\lesssim 1 + \int_{0}^{x^0} 1+
C(\nnorm{\Xi_{\ep,\vec{\yv}}(\tau)}_s)\nnorm{\Xi_{\ep,\vec{\yv}}(\tau)}_s\,
d\tau, \label{apostI.2} \\
\norm{\psi^{\ep,\vec{\yv}}_K(x^0)}_{H^s} &\lesssim 1 + \int_{0}^{x^0}  \norm{\psi^{\ep,\vec{\yv}}_{0K}(\tau)}_{H^s}\,
d\tau, \label{apostI.3} \\
\norm{\psi^{\ep,\vec{\yv}}(x^0)}_{L^6\cap K^s} &\lesssim 1 + \int_{0}^{x^0}
\norm{\psi^{\ep,\vec{\yv}}_{0}(\tau)}_{L^6\cap K^s}\,
d\tau, \label{apostI.4} \\
\intertext{and}
|\beta_{0,\ep}(x^0)| + |\beta_\ep(x^0)| &\lesssim 1 + \int_{0}^{x^0} C(|\beta_{0,\ep}(\tau)|,|\beta_\ep(\tau)|)
\bigl(\beta_{0,\ep}(\tau)| + |\beta_\ep(\tau)| \bigr) \, d\tau \label{apostI.5}.
}

Combining the estimates \eqref{apostHa.1}-\eqref{apostHa.3} and \eqref{apostI.1}-\eqref{apostI.5}, we
 arrive at the estimate
\leqn{apostJ}{
\nnorm{\Xi_{\ep,\vec{\yv}}(x^0)}_s \leq \Rc(1+x^0) + \int_0^{x^0} C(\nnorm{\Xi_{\ep,\vec{\yv}}(\tau)}_s)\nnorm{\Xi_{\ep,\vec{\yv}}(\tau)}_s
+\frac{1}{\ep}\int_0^\tau C(\nnorm{\Xi_{\ep,\vec{\yv}}(\sigma)}_s)\nnorm{\Xi_{\ep,\vec{\yv}}(\sigma)}_s \, d\sigma \, d\tau
}
for the solution $\Xi_{\ep,\vec{\yv}}$, which holds
for some positive constant $\Rc \geq R > 0$ independent
of $(\ep,\vec{\yv})\in (0,\ep_0)\times \Rbb^{3N}$. Defining
\leqn{kappadef}{
\kappa(x^0) = \sup_{0\leq \tau < x^0}\nnorm{\Xi_{\ep,\vec{\yv}}(\tau)}_s,
}
we then have that
\eqn{apostHaa}{
\nnorm{\Xi_{\ep,\vec{\yv}}(x^0)}_s \leq \Rc(1+x^0)+\frac{1}{\ep}C(\kappa(x^0))(x^0)^2 + C(\kappa(x^0))\int_0^{x^0} \nnorm{\Xi_{\ep,\vec{\yv}}(\tau)}_s\, d\tau,
}
which, after applying Grownwall's inequality, yields
\leqn{apostI}{
\kappa(x^0) \leq \biggl[\Rc(1+x^0)+\frac{1}{\ep}C(\kappa(x^0))(x^0)^2\biggr]e^{C(\kappa(x^0))x^0}
\quad 0\leq x^0 < T_{\ep,\vec{\yv}}.
}
Since $\kappa(0)\leq \Rc$, it follows, from the continuation principle from Proposition \ref{locexist}
 and the estimate \eqref{apostI} that there exists a $T_\ep >0$ such
that the solution  $\Xi_{\ep,\vec{\yv}}$ exists on the time interval $[0, T_\ep)$ and satisfies
\leqn{apostK}{
\nnorm{\Xi_{\ep,\vec{\yv}}(x^0)}_s \leq 2 \Rc \qquad 0\leq x^0 < T_\ep.
}

\subsect{unibounds}{Uniform bounds} With the time of existence bounded below independently of
$\vec{\yv}\in \Rbb^{3N}$, we are left with bounding the time of existence below by some positive
time $T>0$ that is independent of  $0<\ep <\ep_0$.
We accomplish this by using a non-local modification of equation
\eqref{totalA.1} of the type used in
\cite{Oliynyk:CMP_2007,Oliynyk:CMP_2009}. The modification removes the singular $\frac{1}{\ep}$ terms from the source $F$ in equation
\eqref{totalA.1} while retaining an overall structure for the evolution equations. This
allows us derive $\ep$-independent energy estimates using the techniques from \cite{BrowningKreiss:1982,KlainermanMajda:1981,KlainermanMajda:1982,Kreiss:1980}.

The modification begins with the introduction of the quantity
\eqn{PhiIdef}{
\Phi_J^{\ep,\vec{\yv}} = (-\Delta)^{-\frac{1}{2}}\Rf_J\tau^{00}_{\ep,\vec{\yv}},
}
which, as we shall see below, is closely related to the Newtonian gravitational force.
First, we note that $\Phi_J^{\ep,\vec{\yv}}$ is well-defined by virtue of the estimates \eqref{apostE} and
\eqref{apostF}, and Theorems \ref{Rieszpotthm} and \ref{Riesztransthm}. We also note that the estimates \eqref{apostE} and
\eqref{apostF} together with \eqref{tauexpA.2}-\eqref{tauexpA.3}, Sobolev's inequality and Moser's estimates
 from Theorem \ref{calcpropE} imply that
\lalign{unibA}{
\norm{\tau^{00}_{\ep,\vec{\yv}}(x^0)}_{H^s} &\lesssim 1 + C(\nnorm{\Xi_{\ep,\vec{\yv}}(x^0)}_s)\biggl(\nnorm{\Xi_{\ep,\vec{\yv}}(x^0)}_s +
 \int_0^{x^0} C(\nnorm{\Xi_{\ep,\vec{\yv}}(\tau)}_s)
\nnorm{\Xi_{\ep,\vec{\yv}}(\tau)}_s\, d\tau \biggr)  \label{unibA.1}
\intertext{and}
\frac{1}{\ep}\norm{\tau^{iJ}_{\ep,\vec{\yv}}(x^0)}_{H^s} & \lesssim 1 + C(\nnorm{\Xi_{\ep,\vec{\yv}}(x^0)}_s)\biggl(\nnorm{\Xi_{\ep,\vec{\yv}}(x^0)}_s +
 \int_0^{x^0} C(\nnorm{\Xi_{\ep,\vec{\yv}}(\tau)}_s)
\nnorm{\Xi_{\ep,\vec{\yv}}(\tau)}_s\, d\tau \biggr),
\label{unibA.2}}
which, in turn, imply, with the help of \eqref{GammaevB},  H\"{o}lder's inequality,  Sobolev's inequality and Moser's estimates, that
\leqn{unibB}{
\norm{\Gamma_{ij,\ep,\vec{\yv}}^k(x^0) \tau^{lm}_{\ep,\vec{\yv}}(x^0)}_{L^1\cap H^s}
\leq  C(\nnorm{\Xi_{\ep,\vec{\yv}}(x^0)}_s)\nnorm{\Xi_{\ep,\vec{\yv}}(x^0)}_s \biggl(1 + \int_0^{x^0} C(\nnorm{\Xi_{\ep,\vec{\yv}}(\tau)}_s)\nnorm{\Xi_{\ep,\vec{\yv}}(\tau)}_s \, d\tau \biggr).
}

Similar calculations used to derive \eqref{divTdenC} show that
\eqn{divTdenD}{
\del{0} \tau^{0j}_{\ep,\vec{\yv}} =  -\frac{1}{\ep}\del{I}\tau^{Ij}_{\ep,\vec{\yv}} -\ep\bigl( \Gamma^{j}_{ik,{\ep,\vec{\yv}}}\tau^{ik}_{\ep,\vec{\yv}} - \Gamma^k_{ik,{\ep,\vec{\yv}}}\tau^{ij}_{\ep,\vec{\yv}} \bigr),
}
from which we see that $\Phi_{I}^{\ep,\vec{\yv}}$ satisfies
\leqn{divTdenE}{
\del{0}\Phi_{I}^{\ep,\vec{\yv}} = -\Rf_I\Rf_J \frac{1}{\ep}\tau^{J0}_{\ep,\vec{\yv}}  - \ep\del{I}\Delta^{-1}\bigl( \Gamma^{0}_{ik}\tau^{ik,{\ep,\vec{\yv}}}_{\ep,\vec{\yv}} - \Gamma^k_{ik,{\ep,\vec{\yv}}}\tau^{i0}_{\ep,\vec{\yv}}
 \bigr).
}
From this equation, the estimates \eqref{unibA.1}-\eqref{unibB}, and Theorems \ref{Rieszpotthm} and \ref{Riesztransthm}, it follows that
\leqn{unibC}{
\norm{\del{0}\Phi_{I}^{\ep,\vec{\yv}}(x^0)}_{H^s} \lesssim 1 + C(\kappa(x^0))(\nnorm{\Xi_{\ep,\vec{\yv}}(x^0)}_s+x^0),
}
with $\kappa(x^0)$ as defined above by \eqref{kappadef}.
Moveover, integrating \eqref{divTdenE} in time shows, with the help of initial data bound \eqref{idenE}
and the estimates \eqref{unibA.1}-\eqref{unibB}, that
\leqn{unibD}{
\norm{\Phi_{I}^{\ep,\vec{\yv}}(x^0)}_{H^{s+1}} \lesssim 1 + C(\kappa(x^0))(\nnorm{\Xi_{\ep,\vec{\yv}}(x^0)}_s+x^0+(x^0)^2).
}

Next, we define
\eqn{ucdef}{
\uc^{ij}_{J,{\ep,\vec{\yv}}} := u^{ij}_{J,{\ep,\vec{\yv}}} - \delta^i_0\delta^j_0 \Phi_J^{\ep,\vec{\yv}}.
}
Substituting this into \eqref{totalA.1} then gives
\leqn{nonlocA}{
A^0_{\ep,\vec{\yv}}\del{0} \begin{pmatrix} u^{ij}_{0,{\ep,\vec{\yv}}} \\ \uc^{ij}_{J,{\ep,\vec{\yv}}}\end{pmatrix}
+ \frac{1}{\ep}E^K\del{K}\begin{pmatrix} u^{ij}_{0,{\ep,\vec{\yv}}} \\ \uc^{ij}_{J,{\ep,\vec{\yv}}}\end{pmatrix} + A^K_{\ep,\vec{\yv}}\del{K}\begin{pmatrix} u^{ij}_{0,{\ep,\vec{\yv}}}
\\ \uc^{ij}_{J,{\ep,\vec{\yv}}}\end{pmatrix} = \Fc_{\ep,\vec{\yv}} ,
}
where
\leqn{nonlocB}{
\Fc_{\ep,\vec{\yv}} = \begin{pmatrix}
u^{JK}_{\ep,\vec{\yv}}\delta^i_0\delta^j_0\del{K}\Phi_J^{\ep,\vec{\yv}}-\frac{1}{\ep}\bigl(\tau^{ij}_{\ep,\vec{\yv}}-\delta^{i}_0\delta^j_0\tau^{00}_{\ep,\vec{\yv}}\bigr)
+\ep\bigl(a^{ij}_{1,{\ep,\vec{\yv}}}+a^{ij}_{2,{\ep,\vec{\yv}}}+a^{ij}_{3,{\ep,\vec{\yv}}}\bigr) \\
-\bigl(\delta^{IJ}+\ep u^{IJ}_{\ep,\vec{\yv}}\bigr)\bigl(\delta^i_0\delta^j_0 \del{0}\Phi_J^{\ep,\vec{\yv}}\bigr).
\end{pmatrix}.
}
From the estimates \eqref{unibA.1}, \eqref{unibA.2}, \eqref{unibC} and \eqref{unibD}, and
the calculus inequalities from Appendix \ref{calc}, we obtain the estimate
\leqn{unibE}{
\norm{\Fc_{\ep,\vec{\yv}}(x^0)}_{H^s} \lesssim 1 + C(\kappa(x^0))(\nnorm{\Xi_{\ep,\vec{\yv}}(x^0)}_s+x^0+(x^0)^2),
}
with the point being that, unlike the estimate \eqref{apostG} for
$F_{\ep,\vec{\yv}}$, this estimate is independent of $\ep \in (0,\ep_0)$.

Next, we set
\eqn{unibF}{
\xi_{{\ep,\vec{\yv}}} = \bigl(\delta_k^0u^{ij}_{0,\ep,\vec{\yv}}+\delta_k^J \uc^{ij}_{J,\ep,\vec{\yv}} , u^{ij}_{\ep,\vec{\yv}}, \psi^{\ep,\vec{\yv}}_{jL}, \psi^{\ep,\vec{\yv}}_j,
\psi^{\ep,\vec{\yv}}, \delta\zeta_{\ep,\vec{\yv}}, z_J^{\ep,\vec{\yv}},\beta_{0,\ep},\beta_\ep\bigr),
}
and note that
\leqn{unibG}{
\norm{\Phi_{I}^{\ep,\vec{\yv}}}_{H^{s}} +
\nnorm{\xi_{\ep,\vec{\yv}}}_s \approx \nnorm{\Xi_{\ep,\vec{\yv}}}_s.
}
From the arguments of the previous section, it is clear that the $\ep$ dependence of the time $T_\ep$
for which the bound \eqref{apostK} holds is due to the $\ep$ dependence in the estimate
\eqref{apostG}. Because of the estimate \eqref{unibE}, we can remove this $\ep$ dependence by
 using the evolution equation \eqref{nonlocA} in favour of \eqref{totalA.1}. Doing so, it
then follows from similar arguments used in the previous section, the bound
\eqref{unibD} and the equivalence of norms \eqref{unibG} that there exists a
$T>0$ such that the solution $\Xi_{\ep,\vec{\yv}}$ exists on the time interval $[0,T)$
and satisfies
\leqn{unibH}{
\norm{\Phi_{I}^{\ep,\vec{\yv}}(x^0)}_{H^{s+1}} +
\nnorm{\xi_{\ep,\vec{\yv}}(x^0)}_{s} \lesssim 1 \quad \forall \; (\ep,\vec{\yv},x^0)\in (0,\ep_0)\times \Rbb^{3N}\times [0,T).
}

Finally, we note by virtute of our choice of initial data that the time derivative $\del{0}\Xi_{\ep,\vec{\yv}}$
at $x^0=0$ satisfies
\eqn{unibI}{
\nnorm{\del{0}\Xi_{\ep,\vec{\yv}}(0)}_{s-1} \lesssim 1 \quad \forall \;(\ep,\vec{\yv})\in (0,\ep_0)\times \Rbb^{3N}.
}
Then differentiating the evolution equation \eqref{nonlocA} and \eqref{totalA.2}-\eqref{totalA.9} with respect
to $x^0$, we obtain an equation for $\del{0}\xi_{\ep,\vec{\yv}}$ with a similar structure to that satisfied
by $\xi_{\ep,\vec{\yv}}$. The same arguments used to derive the estimate \eqref{unibH} go through to show
that  $\del{0}\xi_{\ep,\vec{\yv}}$ satisfies a similar estimate, that is
\eqn{unibHa}{
\norm{\del{0}\Phi_{I}^{\ep,\vec{\yv}}(x^0)}_{H^{s}} +
\nnorm{\del{0}\xi_{\ep,\vec{\yv}}(x^0)}_{s-1} \lesssim 1 \quad \forall \; (\ep,\vec{\yv},x^0)\in (0,\ep_0)\times \Rbb^{3N}\times [0,T).
}

We formalize the above results in the following proposition:
\begin{prop} \label{uniprop}
Let
\eqn{uniprop1}{
\Xi_{\ep,\vec{\yv}} =  \bigl(u^{ij}_{k,\ep,\vec{\yv}}, u^{ij}_{\ep,\vec{\yv}}, \psi^{\ep,\vec{\yv}}_{jL}, \psi^{\ep,\vec{\yv}}_j,
\psi^{\ep,\vec{\yv}}, \delta\zeta_{\ep,\vec{\yv}}, z_J^{\ep,\vec{\yv}},\beta_{0,\ep},\beta_\ep\bigr)
}
denote the solution from Proposition \ref{locexist} to \eqref{totalA.1}-\eqref{totalA.9}.
Then there exists
$T>0$ such that the solution $\Xi_{\ep,\vec{\yv}}$ exists on the spacetime
region $[0,T)\times \Rbb^3$ and satisfies the estimate
\eqn{uniprop2}{
\nnorm{\Xi_{\ep,\vec{\yv}}(x^0)}_{s} +
\nnorm{\del{0}\Xi_{\ep,\vec{\yv}}(x^0)}_{s-1} \lesssim 1
}
for all $(\ep,\vec{\yv},x^0)\in (0,\ep_0)\times \Rbb^{3N}\times [0,T)$.

Moreover, $\Phi_J^{\ep,\vec{\yv}}$, given by
\gath{uniprop3}{
\Phi_J^{\ep,\vec{\yv}} = (-\Delta)^{-\frac{1}{2}}\Rf_J\tau^{00}_{\ep,\vec{\yv}},
}
is well defined, and the following estimates hold:
\gath{uniprop4}{
\norm{\Phi_{I}^{\ep,\vec{\yv}}(x^0)}_{H^{s+1}} + \norm{\del{0}\Phi_{I}^{\ep,\vec{\yv}}(x^0)}_{H^{s}} \lesssim 1
\intertext{and}
\norm{\tau^{00}_{\ep,\vec{\yv}}(x^0)}_{H^s}+
\norm{\Gamma^k_{ij,\ep,\vec{\yv}}(x^0)\tau^{lm}_{\ep,\vec{\yv}}(x^0)}_{L^1\cap H^s} \lesssim 1
}
for all $(\ep,\vec{\yv},x^0)\in (0,\ep_0)\times \Rbb^{3N}\times [0,T)$.
\end{prop}

%% file: limit.tex
\sect{limit}{Cosmological Poisson-Euler equations}
In this section, we establish the local existence of solutions to the limit
equations \eqref{Eulexp0aa.1}-\eqref{Eulexp0aa.2}, and show that these solutions
satisfy the cosmological Poisson-Euler
equations of Newtonian gravity. These solutions will be shown in the next
section to provide an accurate approximation for small values of
$\ep$ to the 1-parameter family of solutions to the Einstein-Euler
equations from Proposition \ref{locexist}.

Before proceeding with the existence proof, we first make some observations. Writing \eqref{Eulexp0aa.1} as
\eqn{Eulexp1}{
\del{0}\bigl( e^{-3\betat}\mut\bigr) +\del{J}\bigl(e^{-3\betat}\rhot\zt^J\bigr) = 0
}
and applying $4\Rf_I(-\Delta)^{-\frac{1}{2}}$, it follows that
\eqn{Eulexp2}{
\del{0}\bigl( e^{-\betat}\Rf_I(-\Delta)^{-\frac{1}{2}} (4e^{-2\betat}\mut)\bigr) -4\Rf_I\Rf_J\bigl(e^{-3\betat}\rhot\zt^J\bigr) = 0.
}
Subtracting \eqref{Eulexp0aa.3} from this equation, we observe that
\eqn{Eulexp3}{
\del{0}\bigl( e^{-\betat} \Rf_I(-\Delta)^{-\frac{1}{2}}(4e^{-2\betat}\mut) - e^{-\betat}\Phit_I \bigr) = 0.
}
In particular, this shows that the constraint $\Phit_I =  \Rf_I(-\Delta)^{-\frac{1}{2}}(4e^{-2\betat}\mut)$ propagates, that is, if initially
\leqn{Eulexp4b}{
 \Phit_I|_{x^0=0} =  \Rf_I(-\Delta)^{-\frac{1}{2}}(4e^{-2\betat}\mut)|_{x^0=0},
}
then
\leqn{Eulexp4a}{
\Phit_I =  \Rf_I(-\Delta)^{-\frac{1}{2}}(4e^{-2\betat}\mut)  \quad \forall \: x^0\geq 0.
}
Taking the divergence of this expression, we see that
\leqn{Eulexp5}{
\del{}^I\Phit_I = 4e^{-2\betat}\mut \quad \forall \: x^0\geq 0
}
holds.
A similar calculation shows that the constraint $\Phit =\Delta^{-1}(4e^{-2\beta}\mut)$ also propagates:
\leqn{Eulexp5a}{
 \Phit|_{x^0=0} =  \Delta^{-1}(4e^{-2\betat}\mut)|_{x^0=0} \quad \Longrightarrow \quad  \Phit =  \Delta^{-1}(4e^{-2\betat}\mut) \quad \forall \: x^0\geq 0.
}
Together, the two constraints \eqref{Eulexp4b} and \eqref{Eulexp5a} imply that
\eqn{Eulexp5b}{
\Phit_I = \del{I} \Phit,
}
from which, we obtain
\leqn{Eulexp6b}{
\Delta\Phit = 4e^{-2\beta}\mut
}
by \eqref{Eulexp5}. Defining
\leqn{Eulexp6}{
\taut_{00} = 4e^{-2\betat}\mut \AND
\taut_{J0} =  -4e^{-2\betat}\rhot\zt_J +  \betat'\Phit_J,
}
a short calculation using, \eqref{Eulexp0aa.1} and \eqref{Eulexp5}, verifies the conservation law
\leqn{Eulexp7}{
\eta^{ij}\del{i}\taut_{j0} = 0 \quad \forall \; x^0\geq 0.
}

The equations \eqref{Eulexp0aa.1}, \eqref{Eulexp0aa.2}, \eqref{Eulexp0aa.4}, \eqref{Eulexp0ab}
and \eqref{Eulexp6b} collectively define the \emph{cosmological Poisson-Euler equations} of Newtonian
gravity. The above calculations show that any solution of the limit equations \eqref{Eulexp0aa.1}-\eqref{Eulexp0aa.5}
that satisfy initially the constraints \eqref{Eulexp4b} and \eqref{Eulexp5a} determine a solution
of the cosmological Poisson-Euler equations. Since the solutions to the limit equations
that we are interested satisfy
these constraints, our approximate solutions are always solutions to the cosmological Poisson-Euler equations.
With the preliminaries out of the way, we turn to establishing the existence and uniqueness of solutions
to the cosmological Poisson-Euler equations.
\begin{prop} \label{limitexist}
Let $\betah$, $\mubr_{\ep,\vec{\yv}}$, $\zbr_J^{\ep,\vec{\yv}}$, $\betabr_0$,
$\Phibr^{\ep,\vec{\yv}}_I$ and $\Phibr^{\ep,\vec{\yv}}$  be as defined
\eqref{cpeidtat.1}, \eqref{initE.3}, \eqref{initE.4}, \eqref{idmatAb.1}, \eqref{idmatAb.2}
and \eqref{idmatAc}, respectively and assume that \eqref{idenF} is satisfied.
Then there exists a $T>0$ and a unique solution
\gath{limitexist2}{
\mut_{\ep,\vec{\yv}},\zt_J^{\ep,\vec{\yv}},\in \bigcap_{m=0}^1 C^m([0,T),H^{s+1-m}(\Rbb^3)),\quad
\betat \in C^{2}([0,T)),\\
\Phit_J^{\ep,\vec{\yv}} \in \bigcap_{m=0}^1 C^m([0,T),H^{s+1}(\Rbb^3)), \quad  \Phit^{\ep,\vec{\yv}} \in \bigcap_{m=0}^1 C^m([0,T), L^6(\Rbb^3)\cap K^{s+2}(\Rbb^3))
}
to the system \eqref{Eulexp0aa.1}-\eqref{Eulexp0aa.5} on the spacetime region $[0,T)\times \Rbb^3$
that satisfies the initial conditions
\eqn{limitexist3}{
\bigl(\mut_{\ep,\vec{\yv}},\zt_J^{\ep,\vec{\yv}},\Phit^{\ep,\vec{\yv}}_{I},\Phit^{\ep,\vec{\yv}},
\betat,\betat' \bigr)|_{x^0=0} = \bigl(\mubr_{\ep,\vec{\yv}}, \zbr_J^{\ep,\vec{\yv}},
\Phibr^{\ep,\vec{\yv}}_I,\Phibr^{\ep,\vec{\yv}},\betah,\betabr_0\bigr),
}
and the bounds
\gath{limitexist4}{
\sup_{0\leq x^0 < T}\bigl(\norm{\mut_{\ep,\vec{\yv}}(x^0)}_{H^{s+1}}+
\norm{\zt^{\ep,\vec{\yv}}_J(x^0)}_{H^{s+1}} + \norm{\Phit^{\ep,\vec{\yv}}_J(x^0)}_{H^{s+1}}
+ \norm{\Phit^{\ep,\vec{\yv}}(x^0)}_{L^6\cap K^{s+2}} + |\betat(x^0)|
\bigr) \lesssim 1,\\
\sup_{0\leq x^0 < T}\bigl(\norm{\del{0}\mut_{\ep,\vec{\yv}}(x^0)}_{H^{s}}+
\norm{\del{0}\zt^{\ep,\vec{\yv}}_J(x^0)}_{H^{s}} + \norm{\del{0}\Phit^{\ep,\vec{\yv}}_J(x^0)}_{H^{s+1}}
+ \norm{\del{0}\Phit^{\ep,\vec{\yv}}(x^0)}_{L^6\cap K^{s+2}} + |\betat'(x^0)|
\bigr) \lesssim 1
}
for all $(\ep,\vec{\yv})\in (0,\ep_0)\times \Rbb^{3N}$.
Moreover, the relations
\eqn{limitexist5}{
\Phit^{\ep,\vec{\yv}}_I = \del{I} \Phit^{\ep,\vec{\yv}} \AND \Delta \Phit^{\ep,\vec{\yv}} =
4e^{-2\betat}\mut_{\ep,\vec{\yv}}
}
hold,
and
\eqn{limitexist6}{
\zetat_{\ep,\vec {\yv}} = \ln\bigl(e^{-3\betat}\mut_{\ep,\vec{\yv}}+1\bigr) \in \bigcap_{m=0}^1 C^m([0,T),H^{s+1-m}(\Rbb^3))
}
is also bounded, that is,
\eqn{limitexist7}{
\sup_{0\leq x^0 < T}\bigl(\norm{\zeta_{\ep,\vec{\yv}}(x^0)}_{H^{s+1}} + \norm{\del{0}\zeta_{\ep,\vec{\yv}}(x^0)}_{H^{s+1}}\bigr)
\lesssim 1
}
for all $(\ep,\vec{\yv})\in (0,\ep_0)\times \Rbb^{3N}$.
\end{prop}
\begin{proof}
By construction, see Section \ref{init}, the initial data satisfies the bound
\leqn{limitexist8}{
\norm{\mut_{\ep,\vec{\yv}}(0)}_{H^{s+1}}+
\norm{\zt^{\ep,\vec{\yv}}_J(0)}_{H^{s+1}} + \norm{\Phit^{\ep,\vec{\yv}}_J(0)}_{H^{s+1}}
+ \norm{\Phit^{\ep,\vec{\yv}}(0)}_{L^6\cap K^{s+2}} + |\betat(0)| + |\betat'(0)|
 \lesssim 1
}
for all $(\ep,\vec{\yv})\in (0,\ep_0)\times \Rbb^{3N}$. Due to the estimates for
the Riesz transform from Theorem \ref{Riesztransthm}, it is clear that we can treat
the evolution equation \eqref{Eulexp0aa.3} as an ODE on $H^{s+1}(\Rbb^3)$. This allows us to
view \eqref{Eulexp0aa.1}-\eqref{Eulexp0aa.3} and \eqref{Eulexp0aa.5}
as a non-local, symmetrizable hyperbolic system.
Standard local existence and uniqueness theorems for such systems,
see Theorems 2.1 and 2.2 in \cite[\S 2.3]{Majda:1984}, together with the initial data bound \eqref{limitexist8}
then imply
the existence of a $T>0$ and a unique solution
\gath{limitexist9}{
\mut_{\ep,\vec{\yv}},\zt_J^{\ep,\vec{\yv}},\in \bigcap_{m=0}^1 C^m([0,T),H^{s+1-m}(\Rbb^3)),\quad
\Phit_J^{\ep,\vec{\yv}} \in \bigcap_{m=0}^1 C^m([0,T),H^{s+1}(\Rbb^3)), \\
\betat \in C^{2}([0,T))
}
to the system \eqref{Eulexp0aa.1}-\eqref{Eulexp0aa.3} and \eqref{Eulexp0aa.5} on the spacetime region $[0,T)\times \Rbb^3$
that coincides with the chosen initial data at $x^0=0$ and satisfies the estimate
\eqn{limitexist10}{
\sup_{0\leq x^0 < T}\bigl(\norm{\mut_{\ep,\vec{\yv}}(x^0)}_{H^{s+1}}+
\norm{\zt^{\ep,\vec{\yv}}_J(x^0)}_{H^{s+1}} + \norm{\Phit^{\ep,\vec{\yv}}_J(x^0)}_{H^{s+1}}
+ |\betat(x^0)|
\bigr) \lesssim 1
}
for all $(\ep,\vec{\yv})\in (0,\ep_0)\times \Rbb^{3N}$. The remainder statements of the proposition
follow from a straightforward application of the calculus inequalities from
Appendix \ref{calc} and the elliptic estimates from Appendix \ref{elliptic}.
\end{proof}

For use in the next section, we note that it can be verified via a straightforward
calculation that solutions $\bigl(\mut_{\ep,\vec{\yv}},\zt_J^{\ep,\vec{\yv}},\Phit^{\ep,\vec{\yv}}_{I},\Phit^{\ep,\vec{\yv}},
\betat,\betat_{0}=\betat' \bigr)$ from Proposition \ref{limitexist} satisfies the following
system of equations:
\lalign{limitA}{
E^K\del{K}\omega_{\ep,\vec{\yv}} &= \Ft_{\ep,\vec{\yv}}, \label{limitA.1} \\
E^K\del{K}\sigma_{\ep,\vec{\yv}} &= \Gt_{\ep,\vec{\yv}}, \label{limitA.2} \\
\betat_{0}'-\Half\betat_{0}^2 +e^{-2\betat}\Lambda & = 0, \label{limitA.3} \\
\betat' -\betat_{0} &= 0, \label{limitA.4}\\
\Ct_{\ep^0,\vec{\yv}}\del{0}\begin{pmatrix} \delta \zetat_{\ep,\vec{\yv}} \\ \zt^{\ep,\vec{\yv}}_J \end{pmatrix}+
\Ct_{\ep,\vec{\yv}}^K\del{K}\begin{pmatrix} \delta\zetat_{\ep,\vec{\yv}} \\
\zt^{\ep,\vec{\yv}}_J \end{pmatrix} &= \Ht_{\ep,\vec{\yv}} - \Ct_{\ep,\vec{\yv}}^K\del{K}\begin{pmatrix}
\zetabr_{\ep,\vec{\yv}} \\ 0\end{pmatrix} , \label{limitA.9}
}
where
\alin{limitB}{
\Ct^0_{\ep,\vec{\yv}} &= \begin{pmatrix}f'\bigl(\rhot_{\ep,\vec{\yv}}\bigr) & 0\\
0 & \delta_{IJ}\end{pmatrix}, 
\\
\Ct^K_{\ep,\vec{\yv}} & =  \begin{pmatrix} f'\bigl(\rhot_{\ep,\vec{\yv}}\bigr)\delta^{KL}
\zt^{\ep,\vec{\yv}}_L &
f'\bigl(\rhot_{\ep,\vec{\yv}}\bigr)\delta^K_J\\
f'\bigl(\rhot_{\ep,\vec{\yv}}\bigr)\delta^K_I &
\delta_{IJ}\delta^{KL}\zt^\ep_L\end{pmatrix}, 
\\
\Ft_{\ep,\vec{\yv}} & =
\begin{pmatrix}2\delta^{(i}_0\delta^{j)L}\taut^{\ep,\vec{\yv}}_{0L} \\
-\delta^i_0\delta^j_0\delta^{IJ}\del{0}\Phit^{\ep,\vec{\yv}}_J \end{pmatrix} ,
\\
\Gt_\ep & = \begin{pmatrix} e^{-2\betat}\Bigl(-\frac{\Lambda}{2} \Phit^{\ep,\vec{\yv}}_L + f'(\rhot_{\ep,\vec{\yv}})
\rhot_{\ep,\vec{\yv}}\del{L}\zetabr_\ep\Bigr) \\ 0 \end{pmatrix},
\\
\Ht_{\ep,\vec{\yv}} & = \begin{pmatrix} 0 \\
 \begin{displaystyle} -\Quarter \Phit^{\ep,\vec{\yv}}_{I}
 +\betat_{0} \zt^\ep_I
  \end{displaystyle} \end{pmatrix},
  \\
\delta \zetat_{\ep,\vec{\yv}} &= \zetat_{\ep,\vec{\yv}} - \zetabr_{\ep,\vec{\yv}}, 
\\
\omega_{\ep,\vec{\yv}} &=  \begin{pmatrix} \omega_{0,\ep,\vec{\yv}}^{ij} \\ \omega_{J,\ep,\vec{\yv}}^{ij} \end{pmatrix}
:=\begin{pmatrix}
\delta^i_0\delta^j_0\del{0}\Phit^{\ep,\vec{\yv}} \\
-2\delta_0^{(i}\delta^{j)L}\Rf_J (-\Delta)^{-\frac{1}{2}}\taut^{\ep,\vec{\yv}}_{L0}
\end{pmatrix}
\intertext{and}
\sigma^{\ep,\vec{\yv}} &= \begin{pmatrix} \sigma^{\ep,\vec{\yv}}_{0L} \\
\sigma^{\ep,\vec{\yv}}_{JL} \end{pmatrix} := \begin{pmatrix}
0 \\ e^{-2\betat} \Rf_J(-\Delta)^{-\frac{1}{2}}\bigl[\frac{\Lambda}{2}\Phit^{\ep,\vec{\yv}}_{L} - f'(\rhot_{\ep,\vec{\yv}})\rhot_{\ep,\vec{\yv}}\del{L}\zetabr_\ep\bigr]
\end{pmatrix}.
}

We conclude this section with a technical lemma that will be used in the following section.
\begin{lem} \label{Nlem}
Suppose \eqn{Nlem0}{
\bigl(u^{ij}_{k,\ep,\vec{\yv}}, u^{ij}_{\ep,\vec{\yv}}, \psi^{\ep,\vec{\yv}}_{jL}, \psi^{\ep,\vec{\yv}}_j,
\psi^{\ep,\vec{\yv}}, \delta\zeta_{\ep,\vec{\yv}}, z_J^{\ep,\vec{\yv}},\beta_{0,\ep},\beta_\ep\bigr)
}
and
\eqn{Nlem1}{
\bigl(\mut_{\ep,\vec{\yv}},\zt_J^{\ep,\vec{\yv}},\Phit^{\ep,\vec{\yv}}_{I},\Phit^{\ep,\vec{\yv}},
\betat\bigr)
}
are the solutions from Proposition \ref{uniprop} and \ref{limitexist}, respectively. Then
\alin{Nlem2}{
&\norm{D\ep^{-1}\tau^{J0}_{\ep,\vec{\yv}}-D\taut_{J0}^{\ep,\vec{\yv}}}_{H^{s-2}}+
\norm{\del{0}D \Phi_I^{\ep,\vec{\yv}}(x^0)-\del{0} D\Phit_I^{\ep,\vec{\yv}}(x^0)}_{H^{s-2}}
\lesssim |\beta_\ep(x^0)-\betat(x^0)| \\
&\hspace{0.2cm}+ |\beta_{0,\ep}(x^0)-\betat'(x^0)|
 +\norm{D\delta \zeta_{\ep,\vec{\yv}}(x^0)-D\delta\zetat_{\ep,\vec{\yv}}(x^0)}_{H^{s-2}}
+ \norm{D z^{\ep,\vec{\yv}}_I(x^0)-D\zt^{\ep,\vec{\yv}}_I(x^0)}_{H^{s-s}} \\
&\hspace{0.8cm} + \norm{D u^{ij}_{K,\ep,\vec{\yv}}(x^0)-\delta^i_0\delta^j_0D \Phit_K^{\ep,\vec{\yv}}(x^0)}_{H^{s-2}}
+ \norm{D\psi_J^{\ep,\vec{\yv}}(x^0)}_{H^{s-2}} + \norm{D\psi_{0J}^{\ep,\vec{\yv}}(x^0)}_{H^{s-2}}
+\ep
}
for all $(\ep,\vec{\yv},x^0)\in (0,\ep_0)\times \Rbb^{3N}\times [0,T)$ and $k=0,1$.
\end{lem}
\begin{proof}
We begin by observing the bounds on the fully relativistic solution
from Proposition \ref{uniprop}, the expansions \eqref{cevC.6}, \eqref{tauexpA.1}-\eqref{tauexpA.3} and
\eqref{tauexpC.1}-\eqref{tauexpC.2}, and the calculus inequalities from Appendix \ref{calc} imply
that the difference $D\ep^{-1}\tau^{J0}_{\ep,\vec{\yv}}-D\taut_{J0}^{\ep,\vec{\yv}}$ can be
estimated as
\lalign{Nlem3}{
&\norm{D\ep^{-1}\tau^{J0}_{\ep,\vec{\yv}}-D\taut_{J0}^{\ep,\vec{\yv}}}_{H^{s-2}}
\lesssim |\beta_\ep(x^0)-\betat(x^0)| + |\beta_{0,\ep}(x^0)-\betat'(x^0)|
\notag \\
&\qquad +\norm{D\delta \zeta_{\ep,\vec{\yv}}(x^0)-D\delta\zetat_{\ep,\vec{\yv}}(x^0)}_{H^{s-2}}
+ \norm{D z^{\ep,\vec{\yv}}_I(x^0)-D\zt^{\ep,\vec{\yv}}_I(x^0)}_{H^{s-2}}+
\notag \\
&\quad \norm{D u^{ij}_{K,\ep,\vec{\yv}}(x^0)-\delta^i_0\delta^j_0 D \Phit_K^{\ep,\vec{\yv}}(x^0)}_{H^{s-2}}+
 \norm{D\psi_J^{\ep,\vec{\yv}}(x^0)}_{H^{s-2}} +\norm{D\psi_{0J}^{\ep,\vec{\yv}}(x^0)}_{H^{s-2}}
+\ep \label{Nlem3.1}
}
for all $(\ep,\vec{\yv},x^0)\in (0,\ep_0)\times \Rbb^{3N}\times [0,T)$.
Since $\taut_{j0}^{\ep,\vec{\yv}}$ satisfies
\eqn{Nlem4}{
\del{0}\taut_{00}^{\ep,\vec{\yv}} = \del{}^J \taut_{J0}^{\ep,\vec{\yv}}
}
by \eqref{Eulexp7},
we can apply the operator $\Delta^{-1}\del{I}$ to this expression to obtain, see \eqref{Eulexp4a} and \eqref{Eulexp6},
\eqn{Nlem5}{
\del{0} \Phit_I^{\ep,\vec{\yv}} =
\Rf_I\Rf^J \taut_{J0}^{\ep,\vec{\yv}}.
}
Comparing this to \eqref{divTdenE}, it follows directly from the bounds from
Proposition \ref{uniprop}, Theorems \ref{Rieszpotthm} and \ref{Riesztransthm},
and the estimate \eqref{Nlem3.1}
that the difference
$\del{0} D\Phi_I^{\ep,\vec{\yv}}-\del{0} D\Phit_I^{\ep,\vec{\yv}}$ satisfies the estimate
\alin{Nlem6}{
&\norm{\del{0} D \Phi_I^{\ep,\vec{\yv}}(x^0)-\del{0}D \Phit_I^{\ep,\vec{\yv}}(x^0)}_{H^{s-2}}
\lesssim |\beta_\ep(x^0)-\betat(x^0)| + |\beta_{0,\ep}(x^0)-\betat'(x^0)|
\notag \\
&\qquad +\norm{D\delta \zeta_{\ep,\vec{\yv}}(x^0)-D\delta\zetat_{\ep,\vec{\yv}}(x^0)}_{H^{s-2}}
+ \norm{D z^{\ep,\vec{\yv}}_I(x^0)-D\zt^{\ep,\vec{\yv}}_I(x^0)}_{H^{s-2}}+
\notag \\
&\quad \norm{D u^{ij}_{K,\ep,\vec{\yv}}(x^0)-\delta^i_0\delta^j_0 D \Phit_K^{\ep,\vec{\yv}}(x^0)}_{H^{s-2}}+
 \norm{D\psi_J^{\ep,\vec{\yv}}(x^0)}_{H^{s-2}} +\norm{D\psi_{0J}^{\ep,\vec{\yv}}(x^0)}_{H^{s-2}}
+\ep
}
for all $(\ep,\vec{\yv},x^0)\in (0,\ep_0)\times \Rbb^{3N}\times [0,T)$.
\end{proof}

%% file: main.tex
\sect{main}{The cosmological Newtonian limit}

We are now ready to prove the main results of this article.
\begin{thm} \label{mainthm}
Let
\eqn{main0}{
\bigl(u^{ij}_{k,\ep,\vec{\yv}},u^{ij}_{\ep,\vec{\yv}},\psi^{\ep,\vec{\yv}}_{jL},
\psi^{\ep,\vec{\yv}}_j,\psi^{\ep,\vec{\yv}},\delta\zeta_{\ep,\vec{\yv}},z_J^{\ep,\vec{\yv}},\beta_{0,\ep},
\beta_\ep\bigr) \AND \bigl(\mut_{\ep,\vec{\yv}},\zt_J^{\ep,\vec{\yv}},\Phit^{\ep,\vec{\yv}}_I,
\Phit^{\ep,\vec{\yv}},\betat\bigr)
}
denote the solutions defined on the spacetime region $[0,T)\times \Rbb^3$ from Proposition \ref{locexist}
(see also Proposition \ref{uniprop}) and Proposition \ref{limitexist}, respectively. Then
\gath{main1}{
\norm{u^{ij}_{\ep,\vec{\yv}}}_{L^\infty([0,T),K^{s-1}\cap L^6)}+\norm{u^{ij}_{0,\ep,\vec{\yv}}}_{L^\infty([0,T),K^{s-1}\cap L^6)}+ \norm{u^{ij}_{K,\ep,\vec{\yv}}-\delta^i_0\delta^j_0\Phit^{\ep,\vec{\yv}}_K}_{L^\infty([0,T),K^{s-1}\cap L^6)}
\lesssim \ep,\\
\norm{\psi^{\ep,\vec{\yv}}}_{L^\infty([0,T),K^{s-1}\cap L^6)}+\norm{\psi_j^{0,\ep,\vec{\yv}}}_{L^\infty([0,T),K^{s-1}\cap L^6)}+ \norm{\psi_{jK}^{\ep,\vec{\yv}}}_{L^\infty([0,T),K^{s-1}\cap L^6)}
\lesssim \ep,\\
\norm{\delta\zeta_{\ep,\vec{\yv}}-\delta\zetat_{\ep,\vec{\yv}}}_{L^\infty([0,T),K^{s-1}\cap L^6)}+\norm{z_I^{0,\ep,\vec{\yv}}-\zt_I^{0,\ep,\vec{\yv}}}_{L^\infty([0,T),K^{s-1}\cap L^6)}\lesssim \ep
\intertext{and}
|\beta_\ep -\betat|_{L^\infty([0,T))} + |\beta_{0,\ep} -\betat'|_{L^\infty([0,T))}\lesssim \ep
}
for all $(\ep,\vec{\yv})\in (0,\ep_0)\times \Rbb^{3N}$.
\end{thm}
\begin{proof}
We define
\eqn{Zdef}{
Z_{\ep,\vec{\yv}} =\begin{pmatrix} u^{ij}_{0,\ep,\vec{\yv}}\\\uc^{ij}_{J,\ep,\vec{\yv}}\\\psi^{\ep,\vec{\yv}}_{0L}
\\ \psi^{\ep,\vec{\yv}}_{JL}\\\delta\zeta_{\ep,\vec{\yv}} \\
z^{\ep,\vec{\yv}}_J \\
u^{ij}_{\ep,\vec{\yv}} \\ \psi^{\ep,\vec{\yv}}_0 \\ \psi^{\ep,\vec{\yv}}_L\\
\psi^{\ep,\vec{\yv}} \end{pmatrix}
-\ep\begin{pmatrix}\omega^{ij}_{0,\ep,\vec{\yv}} \\ \omega^{ij}_{J,\ep,\vec{\yv}} \\
0 \\ \sigma^\ep_{JL,\vec{\yv}} \\ 0 \\ 0 \\ 0 \\0  \\0 \\0\end{pmatrix} -
\begin{pmatrix} 0 \\ 0 \\ 0 \\0  \\ \delta\zetat_{\ep,\vec{\yv}} \\ \zt^{\ep,\vec{\yv}}_J \\ 0 \\ 0 \\0 \\0  \end{pmatrix},
}
\eqn{Ydef}{
Y_\ep = \begin{pmatrix} \beta'_\ep \\ \beta_\ep \end{pmatrix} - \begin{pmatrix} \betat' \\ \betat\end{pmatrix},
}
and
\eqn{Xdef}{
X_{\ep,\vec{\yv}} = (\Phi_I^{\ep,\vec{\yv}})-(\Phit_I^{\ep,\vec{\yv}} ),
}
and observe that the estimate
\leqn{XYZidata}{
\norm{Z_{\ep,\vec{\yv}}(0)}_{L^6} + \norm{DZ_{\ep,\vec{\yv}}(0)}_{H^{s}}
+ |Y_\ep(0)|+ \norm{X_{\ep,\vec{\yv}}(0)}_{H^s} \lesssim \ep
}
is a direct consequence of our choice of initial data from Section \ref{init}.

Next, a straightforward calculation using the evolution equations
\eqref{totalA.2}-\eqref{totalA.9} and \eqref{nonlocA}-\eqref{nonlocB} in conjunction
with \eqref{limitA.1}-\eqref{limitA.9} shows
that $Z_{\ep,\vec{\yv}}$, $Y_\ep$ and $X_{\ep,\vec{\yv}}$ satisfy
equations of the form
\lalign{Zev}{
\Af^0_{\ep,\vec{\yv}}\del{0}Z_{\ep,\vec{\yv}} + \frac{1}{\ep}\Cf^K\del{K}Z_{\ep,\vec{\yv}} +
\Af^K_{\ep,\vec{\yv}}\del{K}Z_{\ep,\vec{\yv}} &= \Lc_\ep Z_{\ep,\vec{\yv}} +  \Rc_{\ep,\vec{\yv}}
+\Lf_\ep X_{\ep,\vec{\yv}} +  \Bf(V_{\ep,\vec{\yv}},Z_{\ep,\vec{\yv}})+
\ep\Rf_{\ep,\vec{\yv}}, \label{Zev.1}
\\
Y_\ep' + \mathfrak{m}_\ep Y_\ep  &=  \bfr(W_\ep,Y_\ep) + \ep\rfr_\ep \label{Zev.2} ,
\intertext{and}
\del{0} X_{\ep,\vec{\yv}} &= \lf \Rc_{\ep,\vec{\yv}}, \label{Zev.3}
}
respectively, where:
\begin{enumerate}[(i)]
\item
\alin{ZevdefA}{
\Af^0_{\ep,\vec{\yv}} &= \begin{pmatrix} A^0\bigr(\ep u_{\ep,\vec{y}}\bigl) & 0 & 0 & 0 \\
0& B^0\bigr(\ep u_{\ep,\vec{y}}\bigl) & 0 & 0 \\
0 & 0 & C^0\bigl(\ep,\ep u_{\ep,\vec{y}},\zeta_{\ep,\vec{y}},z^{\ep,\vec{y}},\beta_\ep\bigr) & 0\\
0 & 0 & 0 & \id \end{pmatrix} ,
\\
\Af^K_{\ep,\vec{\yv}} &= \begin{pmatrix} A^K\bigr(\ep, u_{\ep,\vec{y}}\bigl) & 0 & 0 & 0 \\
0& B^K\bigr(\ep, u_{\ep,\vec{y}}\bigl) & 0 & 0 \\
0 & 0 & C^K\bigl(\ep, u_{\ep,\vec{y}},\zeta_{\ep,\vec{y}},z^{\ep,\vec{y}},\beta_\ep\bigr) & 0\\
0 & 0 & 0 & 0 \end{pmatrix}
\intertext{and}
\Cf^K &= \begin{pmatrix} E^K & 0 & 0 & 0 \\
0 & E^K & 0 & 0 \\
0 & 0 & 0 & 0 \\
0 & 0 & 0 & 0 \end{pmatrix}.
}
\item $\Lf_\ep$,  $\Lc_\ep$, and $\mathfrak{m}_\ep$  are $x^0$-dependent matrices (independent of the spatial coordinates $(x^I)$) satisfying
\leqn{ZevdefC}{
|\Lf_\ep(x^0)|+|\Lc_\ep(x^0)|+|\mathfrak{m}_\ep(x^0)|
\lesssim 1
}
for all $(x^0,\ep)\in [0,T)\times (0,\ep_0)$ and
\eqn{ZevdefCa}{
 \Pbb \Lf_\ep = 0,
}
where
\eqn{Pbbdef}{
\Pbb = \begin{pmatrix}0 & 0 & 0 & 0 \\
0& 0 & 0 & 0 \\
0& 0 & \id & 0 \\
0 & 0 & 0& \id \end{pmatrix}.
}
\item $\bfr$ is a constant, bilinear map.
\item The non-zero components of $\Rc_{\ep,\vec{\yv}}$ consist of the differences $\frac{1}{\ep}\tau^{0J}_{\ep,\vec{\yv}}-\taut^{\ep,\vec{\yv}}_{0J}$
and $\del{0}\Phi^{\ep,\vec{\yv}}_J -  \del{0}\Phit^{\ep,\vec{\yv}}_J$, and $\lf$ is the projection map on to the component containing
$\del{0}\Phi^{\ep,\vec{\yv}}_J -  \del{0}\Phit^{\ep,\vec{\yv}}_J$.
\item $V_{\ep,\vec{\yv}}$ is defined by the terms that are either not given by (a) $\Rc_{\ep,\vec{\yv}}$, (b) $X_{\ep,\vec{\yv}}$ or $Z_{\ep,\vec{\yv}}$ multiplied by
an $x^0$-dependent, spatially constant matrix, or (c) a term with a explicit power of $\ep$ in front of it. All of the remaining terms on the righthand side of \eqref{Zev.1}
can, via a simple application of Taylor's theorem, be expressed in the form $\Bc(V_{\ep,\vec{\yv}},Z_{\ep,\vec{\yv}})$
for a suitable constant bilinear map $\Bc$.
\end{enumerate}
Moreover, it not difficult to verify, using the
estimates from Proposition \ref{uniprop} and Lemma \ref{Nlem}, the expansions \eqref{cevC.6}, \eqref{tauexpA.1}-\eqref{tauexpA.3} and
\eqref{tauexpC.1}-\eqref{tauexpC.2}, and the calculus inequalities from Appendix \ref{calc},
the validity of the following estimates for all $(x^0,\ep,\vec{\yv})\in [0,T)\times (0,\ep_0)\times \Rbb^{3N}$:
\lgath{coef}{
\Af^0_{\ep,\vec{\yv}}(x^0) \geq \kappa \id \quad \text{ for some $\kappa > 0$,} \label{coef.1}\\
\norm{\Af^k_{\ep,\vec{\yv}}(x^0)}_{L^\infty} + \frac{1}{\ep}\norm{\Pbb^\perp D\Af^0_{\ep,\vec{\yv}}(x^0)}_{H^{s-1}}+\norm{D\Af^k_{\ep,\vec{\yv}}(x^0)}_{H^{s-1}} \lesssim 1,
\label{coef.2}\\
\norm{V_{\ep,\vec{\yv}}(x^0)}_{H^{s-1}}+\norm{\Rf_{\ep,\vec{\yv}}(x^0)}_{H^{s-1}}+
|W_\ep(x^0)|+|\rfr(x^0)| \lesssim 1, \label{coef.3} \\
\norm{Z_{\ep,\vec{\yv}}(x^0)}_{L^6}+\norm{\del{0}Z_{\ep,\vec{\yv}}(x^0)}_{H^{s-1}} + \norm{DZ_{\ep,\vec{\yv}}(x^0)}_{H^{s-1}} \lesssim 1, \label{coef.4} \\
\norm{X_{\ep,\vec{\yv}}(x^0)}_{H^s} \lesssim 1 \label{coef.5}
\intertext{and}
\norm{D\Rc_{\ep,\vec{\yv}}(x^0)}_{H^{s-2}} \lesssim \norm{DX_{\ep,\vec{\yv}}(x^0)}_{H^{s-2}} + |Y_\ep(x^0)|+
\norm{D Z_{\ep,\vec{\yv}}(x^0)}_{H^{s-2}}
     +\ep , \label{coef.6}
}
where
\eqn{Pbbperpdef}{
\Pbb^\perp = \begin{pmatrix}\id & 0 & 0 & 0 \\
0& \id & 0 & 0 \\
0& 0 & 0 & 0 \\
0 & 0 & 0& 0 \end{pmatrix}.
}

Differentiating \eqref{Zev.1} and \eqref{Zev.3} spatially, we find that
\lgath{DZev}{
\Af^0_{\ep,\vec{\yv}}\del{0}\del{I}Z_{\ep,\vec{\yv}} +  \frac{1}{\ep}\Cf^K\del{K}\del{I}Z_{\ep,\vec{\yv}} + \Af^K_{\ep,\vec{\yv}}\del{K}\del{I}Z_{\ep,\vec{\yv}}
 = \Ff_{\ep,\vec{\yv}} \label{DZev.1}
\intertext{and}
\del{0} \del{I}X_{\ep,\vec{\yv}} = \lf \del{I}\Rc_{\ep,\vec{\yv}},\label{DZev.2}
}
where
\alin{Ffrdef}{
\Ff_{\ep,\vec{\yv}} =&  \del{I}\Af^K_{\ep,\vec{\yv}} \del{K}Z_{\ep,\vec{\yv}} + \Lc_\ep \del{I}Z_{\ep,\vec{\yv}}
 + \del{I}R_{\ep,\vec{\yv}}  +  \Lf_\ep \del{I}X_{\ep,\vec{\yv}}  +  \Bf(V_{\ep,\vec{\yv}},\del{I}Z_{\ep,\vec{\yv}})\notag\\
&+ \Bf(\del{I}X_{\ep,\vec{\yv}},Z_{\ep,\vec{\yv}})
-\del{I}\Af^0_{\ep,\vec{\yv}}\Pbb\del{0}Z_{\ep,\vec{\yv}}+ \ep \biggl(- \frac{1}{\ep} \Pbb^\perp\del{I}\Af^0_{\ep,\vec{\yv}}\del{0}Z_{\ep,\vec{\yv}} + \del{I}\Rf_{\ep,\vec{\yv}}\biggr).
}

We can bound the individual terms in $\Ff_{\ep,\vec{\yv}}$ using
the calculus inequalities from Appendix \ref{calc} and the estimates \eqref{ZevdefC} and \eqref{coef.2}-\eqref{coef.6}. For example,
we can estimate the term $\Bf(V_{\ep,\vec{\yv}},Z_{\ep,\vec{\yv}})$ as follows:
\alin{DXZest}{
\norm{\Bf(V_{\ep,\vec{\yv}},Z_{\ep,\vec{\yv}})}_{H^{s-2}}
&\lesssim \norm{V_{\ep,\vec{\yv}}}_{H^{s-2}}\norm{Z_{\ep,\vec{\yv}}}_{K^{s-1}}  \\
&\lesssim \norm{Z_{\ep,\vec{\yv}}}_{W^{1,6}}+\norm{
DZ_{\ep,\vec{\yv}}}_{H^{s-2}} \notag \\
&\lesssim \norm{DZ_{\ep,\vec{\yv}}}_{H^{s-2}},
}
where in deriving this result we have used the multiplication inequality from Theorem \ref{calcpropD}, two applications
of Soblev's inequality, and the bound \eqref{coef.3} on $V_{\ep,\vec{\yv}}$. Estimating the remaining terms of $\Ff_{\ep,\vec{\yv}}$ in
a similar fashion, we find that
\leqn{Ffbound}{
\norm{\Ff_{\ep,\vec{\yv}}(x^0)}_{H^{s-2}} \lesssim \norm{DX_{\ep,\vec{\yv}}(x^0)}_{H^{s-2}}+|Y_\ep(x^0)|+ \norm{DZ_{\ep,\vec{\yv}}(x^0)}_{H^{s-2}} + \ep
}
for all $(x^0,\ep,\vec{\yv})\in [0,T)\times (0,\ep_0)\times \Rbb^{3N}$.

Due to the estimates \eqref{ZevdefC}, \eqref{coef.1}-\eqref{coef.6},  and \eqref{Ffbound} and the fact that the matrices $\Cf^K$ are constant, the system
consisting of \eqref{Zev.2}, \eqref{DZev.1} and \eqref{DZev.2} is a singular symmetric hyperbolic system of the type analyzed
in \cite{BrowningKreiss:1982,KlainermanMajda:1981,KlainermanMajda:1982,Kreiss:1980}. The energy estimates from these works, for example, see Theorem 1' in \cite{KlainermanMajda:1982},
then imply that
\eqn{ZboundA}{
 \norm{DX_{\ep,\vec{\yv}}(x^0)}_{H^{s-2}}+ |Y_\ep(x^0)| + \norm{DZ_{\ep,\vec{\yv}}(x^0)}_{H^{s-2}}  \lesssim  \norm{DX_{\ep,\vec{\yv}}(0)}_{H^{s-2}}+ |Y_\ep(0)| + \norm{DZ_{\ep,\vec{\yv}}(0)}_{H^{s-2}}
}
holds for all $(x^0,\ep,\vec{\yv})\in [0,T)\times (0,\ep_0)\times \Rbb^{3N}$. As discussed earlier, the key to these estimates is the constancy of the
matrices $\Cf^K$, since the only place that the singular term $\frac{1}{\ep}\Cf^K$ appears in the energy estimates is in the form $\frac{1}{\ep}\del{K}\Cf^K$,
which vanishes.

 Noting the bound \eqref{XYZidata} on the initial data, we conclude,
via an application of Sobolev's inequality, that
\eqn{ZboundC}{
\norm{X_{\ep,\vec{\yv}}(x^0)}_{K^{s-1}\cap L^6} + |Y_\ep(x^0)| + \norm{Z_{\ep,\vec{\yv}}(x^0)}_{K^{s-1}\cap L^6} \lesssim \ep \quad \forall \; (x^0,\ep,\vec{\yv})\in [0,T)\times (0,\ep_0)\times \Rbb^{3N},
}
and the proof is complete.
\end{proof}
We conclude with a few remarks:
\begin{enumerate}[(i)]
\item  The inhomogeneous component of the proper energy density $\mu_{\ep,\vec{\yv}}$, see \eqref{mudef}, \eqref{Nfldef.1}-\eqref{Nfldef.2}, and \eqref{totalB.8},
satisfies
\gath{mainrem1}{
\mu_{\ep,\vec{\yv}} \in \bigcap_{\ell=0}^1 C^\ell([0,T),K^{s-\ell}(\Rbb^3)\cap L^6(\Rbb^3))
\intertext{and}
\norm{\mu_{\ep,\vec{\yv}}-\mut_{\ep,\vec{\yv}}}_{L^\infty([0,T),K^{s-1}\cap L^6)} \lesssim \ep \qquad \forall (\ep,\yv)\in (0,\ep_0)\times \Rbb^{3N}.
}
This follows directly from Theorem \ref{mainthm} and the calculus inequalities from Appendix \ref{calc}.
\item From  \eqref{gbdef}, \eqref{utog}, \eqref{ghdef}, \eqref{udefA}, \eqref{nbdef}, \eqref{zetabdef}, \eqref{zbdefA}, \eqref{betadef}, \eqref{Nfldef.2},
\eqref{Nfldef.3}, \eqref{Nfldef.4}, \eqref{zetahdef} and \eqref{totalB.8}, it is clear that the collection of fields $\{u^{ij}_{j,\ep,\vec{\yv}},\delta\zeta_{\ep,\vec{\yv}}, z_J^{\ep,\vec{\yv}},\beta^\ep, \psi^{\ep,\vec{\yv}}\}$
determine a solution $\{\grave{g}_{\ep,\vec{\yv}}^{ij},\vb_i^{\ep,\vec{\yv}},\rhob_{\ep,\vec{\yv}}\}$ via
the formulas \eqref{maininf5.1}-\eqref{maininf5.3} to the Einstein-Euler equations \eqref{EE.1}-\eqref{EE.2} on the spacetime region $M=[0,T)\times \Rbb^3$.
\end{enumerate}

%% file: calc.tex
\sect{calc}{Calculus Inequalities}

In this section, we collect, for the convenience of the reader, a number of calculus inequalities that we employ throughout this article.

\subsect{sobcalc}{Sobolev inequalities} The proof of the following inequalities are well known and may be found, for example, in
the books \cite{AdamsFournier:2003}, \cite{Friedman:1976} and \cite{TaylorIII:1996}.

\begin{thm}{\emph{[H\"{o}lder's inequality]}} \label{Holder}
\begin{itemize}
\item[(i)] If $0< p,q,r \leq \infty$ satisfy $1/p+1/q = 1/r$, then
\eqn{Holder1}{
\norm{uv}_{L^r} \leq \norm{u}_{L^p}\norm{v}_{L^q}
}
for all $u\in L^p(\Rbb^n)$ and $v\in L^q(\Rbb^n)$.
\item[(ii)] If $1\leq p,q,r \leq \infty$, $0\leq \theta \leq 1$ and
\eqn{Holder2}{
\frac{1}{r} = \frac{\theta}{p} + \frac{1-\theta}{q},
}
then
\eqn{Holder3}{
\norm{u}_{L^r} \leq \norm{u}^\theta_{L^p}\norm{u}^{1-\theta}_{L^q} \lesssim  \norm{u}_{L^p} +  \norm{u}_{L^q}
}
for all $u\in L^p(\Rbb^n)\cap L^q(\Rbb^n)$.
\end{itemize}
\end{thm}


\begin{thm}{\emph{[Gagliardo-Nirenberg-Sobolev inequalities]}} \label{Sobolev}
\begin{itemize}
\item[(i)] If $1\leq p < \infty$, then
\eqn{Sobolev0}{
\norm{u}_{L^{p*}} \lesssim \norm{Du}_{L^p} \qquad p^* = \frac{np}{n-p}
}
for all $u\in \{ v \in L^{p*}(\Rbb^n)\cap W^{1,p}_{\emph{loc}}(\Rbb^n) \: | \: \norm{Dv}_{L^p} < \infty\}$.
\item[(ii)] If $s\in \Zbb_{\geq 1}$,
$1\leq p < \infty$ and $sp<n$, then
\eqn{Sobolev1}{
\norm{u}_{L^q} \lesssim \norm{u}_{W^{s,p}} \qquad p\leq q \leq \frac{np}{n-s p}
}
for all $u\in W^{s,p}(\Rbb^n)$.
\item[(iii)] 
If $s\in \Zbb_{\geq 1}$,
$1\leq p < \infty$ and $sp > n$, then
\eqn{Sobolev3}{
\norm{u}_{L^\infty} \lesssim \norm{u}_{W^{s,p}} 
}
for all $u\in W^{s,p}(\Rbb^n)$.
\end{itemize}
\end{thm}

\newpage

\begin{thm}{\emph{[Product and commutator estimates]}} \label{calcpropB} $\;$

\begin{enumerate}[(i)]
\item
Suppose $1\leq p_1,p_2,q_1,q_2\leq \infty$, $s=|\alpha|$, and
\leqn{calcpropB.1}{
\frac{1}{p_1}+\frac{1}{p_2} = \frac{1}{q_1} + \frac{1}{q_2} = \frac{1}{r}.
}
Then
\alin{calcpropB.2}{
\norm{D^\alpha(uv)}_{L^r(\Rbb^n)} \lesssim \norm{D^s u}_{L^{p_1}(\Rbb^n)}\norm{v}_{L^{q_1}(\Rbb^n)} + \norm{u}_{L^{p_2}(\Rbb^n)}\norm{D^s v}_{L^{q_2}(\Rbb^n)} \label{clacpropB.2.1}
\intertext{and}
\norm{D^\alpha(uv)-uD^\alpha v}_{L^r(\Rbb^n)} \lesssim \norm{Du}_{L^{p_1}(\Rbb^n)}\norm{v}_{W^{s-1,q_1}(\Rbb^n)} + \norm{Du}_{
W^{s-1,p_2}(\Rbb^n)}\norm{v}_{L^{q_2}(\Rbb^n)}
}
for all $u,v \in C^\infty_0(\Rbb^n)$.
\item[(ii)]  If $s_1,s_2\geq s_3\geq 0$, $1\leq p \leq \infty$, and $s_1+s_2-s_3 > n/p$, then
\eqn{calcpropB.3}{
\norm{uv}_{W^{s_3,p}(\Rbb^n)} \lesssim \norm{u}_{W^{s_1,p}(\Rbb^n)}\norm{v}_{W^{s_2,p}(\Rbb^n)}
}
\end{enumerate}
\end{thm}



\begin{thm}{\emph{[Moser's estimates]}}  \label{calcpropC}
Suppose $s\in \Zbb_{\geq 1}$, $1\leq p \leq \infty$, $|\alpha|\leq s$, $f\in C^s(\Rbb)$, $f(0) = 0$,
 $g\in C^{s+1}(\Rbb)$  and $V$ is open and bounded in $\Rbb$. Then
\eqn{calcpropC.1}{
\norm{D^\alpha f(u)}_{L^{p}(\Rbb^n)} \leq C\bigl(\norm{f}_{C^s(\overline{V})}\bigr)(1+\norm{u}^{s-1}_{L^\infty(\Rbb^n)})\norm{u}_{W^{s,p}(\Rbb^n)}
}
and
\eqn{calcpropC.2}{
\norm{D^\alpha ( g(u)- g(v))}_{L^{p}(\Rbb^n)} \leq C\bigl(\norm{g}_{C^{s+1}(\overline{V})}\bigr)(1+\norm{u}^{s-1}_{L^\infty(\Rbb^n)}+ \norm{v}^{s-1}_{L^\infty(\Rbb^n)} )\norm{u-v}_{W^{s,p}(\Rbb^n)}
}
for all $u,v \in C^0(\Rbb^n)\cap L^\infty(\Rbb^n)\cap W^{s,p}(\Rbb^n)$ with
$u(\xv), v(\xv) \in V$ for all $\xv\in \Rbb^n$.
\end{thm}

\begin{rem} \label{Moserrem}
Theorems \ref{calcpropB} and \ref{calcpropC} also hold for the obvious vector/matrix valued generalizations, that
is when $u$ is matrix valued and $v$ is vector valued in Theorem \ref{calcpropB}, and $u,v$ are vector valued in
Theorem \ref{calcpropC}.
\end{rem}

\subsect{Zhidcalc}{Zhidkov inequalities}
The spaces
\eqn{ZhidkovA}{
K^s(\Rbb^n) = \{\, u\in L^\infty(\Rbb^n) \: | \: Du \in H^{s-1}(\Rbb^n) \, \}
}
with norm
\eqn{ZhidkovB}{
\norm{u}_{K^s(\Rbb^n)} = \norm{u}_{L^\infty(\Rbb^n)} + \norm{Du}_{H^{s-1}(\Rbb^n)},
}
are known as the \emph{Zhidkov spaces}. On these spaces, there are analogous calculus inequalities that can be proved using a straightforward
adaptation of the proofs of the Sobolev space estimates.

\begin{thm}{\emph{[Product  estimates]}} \label{calcpropD}
If $s_1,s_2\geq s_3\geq 0$ and $s_1+s_2-s_3 > n/2$, then
\eqn{calcpropD.1}{
\norm{uv}_{K^{s_3}} \lesssim \norm{u}_{K^{s_1}}\norm{v}_{K^{s_2}}
}
for all $u\in K^{s_1}(\Rbb^n)$ and $v \in K^{s_2}(\Rbb^n)$, and
\eqn{calcpropD.2}{
\norm{uv}_{H^{s_3}} + \norm{uv}_{K^{s_3}} \lesssim \norm{u}_{H^{s_1}}\norm{v}_{K^{s_2}}
}
for all $u\in H^{s_1}(\Rbb^n)$ and $v \in K^{s_2}(\Rbb^n)$.
\end{thm}

\begin{thm}{\emph{[Moser's estimates]}}  \label{calcpropE}
Suppose $s\in \Zbb_{\geq 1}$, $|\alpha|\leq s$, 
$F\in C^s(\Rbb^2)$,
$g\in C^{s+1}(\Rbb)$,
$G\in C^{s+1}(\Rbb^2)$, $f(\xi,0)=0$ for all $\xi \in \Rbb$,
$f(0)=0$
and $V$ is open and bounded in $\Rbb$.  Then
\alin{calcpropE.1}{
\norm{f(u)}_{K^{s}} &\leq C\bigl(\norm{f}_{C^s(\overline{V})}\bigr)(1+\norm{u}^{s-1}_{L^\infty})\norm{u}_{K^s}, \\
\norm{F(u,v)}_{H^{s}} &\leq C\bigl(\norm{F}_{C^s(\overline{V}^2)}\bigr)(1+\norm{u}^{s-1}_{L^\infty}+ \norm{v}^{s-1}_{L^\infty})(\norm{u}_{K^s} +  \norm{v}_{H^s}\bigl)
}
for all  $u \in C^0(\Rbb^n)\cap K^{s}(\Rbb^n)$, $v \in C^0(\Rbb^n)\cap L^\infty(\Rbb^n)\cap H^{s}(\Rbb^n)$ with
$u(\xv), v(\xv) \in V$ for all $\xv\in \Rbb^n$, and
\alin{calcpropE.2}{
&\norm{g(u_1)-g(u_2)}_{K^{s}}
 \leq C\bigl(\norm{g}_{C^{s+1}(\overline{V})}\bigr)\Biggl(1+\sum_{i=1}^2\norm{u_i}^{s-1}_{L^\infty} \Biggr)\norm{u_1-u_2}_{K^{s}},\\
&\norm{G(u_1,v_1)-G(u_2,v_2)}_{H^{s}}
 \leq C\bigl(\norm{G}_{C^{s+1}(\overline{V}^2)}\bigr)  \\
 &\qquad \times \Biggl(1+\sum_{i=1}^2\Biggl[\norm{u_i}^{s-1}_{L^\infty} + \norm{v_i}^{s-1}_{L^\infty}\Biggr] \Biggr)\bigl(\norm{u_1-u_2}_{K^{s}} + \norm{v_1-v_2}_{H^{s}}\bigr)
}
for all $u_i \in C^0(\Rbb^n)\cap K^{s}(\Rbb^n)$, $v_i \in C^0(\Rbb^n)\cap L^\infty(\Rbb^n)\cap H^{s}(\Rbb^n)$ with
$u_i(\xv), v_i(\xv) \in V$ for all $\xv\in \Rbb^n$ and $i=1,2$.
\end{thm}

\begin{cor} \label{calccorA}
Suppose $s\in \Zbb_{\geq 1}$, $|\alpha|\leq s$, $f,F\in C^s(\Rbb^2)$, $g,G\in C^{s+1}(\Rbb^2)$, and  $V$ is open and bounded in $\Rbb$.
\begin{enumerate}[(i)]
\item If $|f(u,v)|\lesssim |v|^3$ for all $(u,v)\in \overline{V}^2$, then
\eqn{calccorA.1}{
\norm{f(u,v)}_{H^{s}} \leq C\bigl(\norm{f}_{C^s(\overline{V}^2)}\bigr)(1+\norm{u}^{s-1}_{L^\infty}+ \norm{v}^{s-1}_{L^\infty}+\norm{v}^2_{L^6})(\norm{u}_{K^s} +  \norm{v}_{K^s\cap L^6}\bigl)
}
for all $u \in C^0(\Rbb^n)\cap K^{s}(\Rbb^n)$, $v\in C^0(\Rbb^n)\cap K^{s}(\Rbb^n)\cap L^6(\Rbb^n)$
with $u(\xv), v(\xv) \in V$ for all $\xv\in \Rbb^n$.
\item If $|g(u,v)|\lesssim |v|^3$ for all $(u,v)\in \overline{V}^2$, then
\alin{calcorA.2}{
&\norm{g(u_1,v_1)-g(u_2,v_2)}_{H^{s}}
 \leq C\bigl(\norm{g}_{C^{s+1}(\overline{V}^2)}\bigr)  \\
 &\qquad \times \biggl(1+\sum_{i=1}^2\bigl[\norm{u_i}^{s-1}_{L^\infty} + \norm{v_i}^{s-1}_{L^\infty}+\norm{v_i}_{L^6}^3\bigr] \biggr)\bigl(\norm{u_1-u_2}_{K^{s}} + \norm{v_1-v_2}_{K^{s}\cap L^6}\bigr)
}
for all $u_i \in C^0(\Rbb^n)\cap K^{s}(\Rbb^n)$, $v_i\in C^0(\Rbb^n)\cap K^{s}(\Rbb^n)\cap L^6(\Rbb^n)$
with $u_i(\xv), v_i(\xv) \in V$ for all $\xv\in \Rbb^n$ and $i=1,2$.
\item If $|F(u,v)|\lesssim |u|^2|v|$ for all $(u,v)\in \overline{V}^2$, then
\eqn{calccorA.3}{
\norm{F(u,v)}_{K^{s}\cap L^{\frac{6}{5}}} \leq C\bigl(\norm{F}_{C^s(\overline{V}^2)}\bigr)(1+\norm{u}^{s-1}_{L^\infty}+ \norm{v}^{s-1}_{L^\infty}+\norm{u}_{L^6}^2+\norm{v}_{L^2}^2)(\norm{u}_{K^s\cap L^6} +  \norm{v}_{H^s}\bigl)
}
for all $u \in C^0(\Rbb^n)\cap K^{s}(\Rbb^n)\cap L^6(\Rbb^n)$, $v\in C^0(\Rbb^n)\cap H^{s}(\Rbb^n)$
with $u(\xv), v(\xv) \in V$ for all $\xv\in \Rbb^n$.
\item If $|G(u,v)|\lesssim |u|^2|v|$ for all $(u,v) \in \overline{V}^2$, then
\alin{calccorA.4}{
&\norm{G(u_1,v_1)-G(u_2,v_2)}_{K^{s}\cap L^{\frac{6}{5}}}
 \leq C\bigl(\norm{G}_{C^{s+1}(\overline{V}^2)}\bigr)  \\
 &\qquad \times \biggl(1+\sum_{i=1}^2\bigl[\norm{u_i}^{s-1}_{L^\infty} + \norm{v_i}^{s-1}_{L^\infty}
 +\norm{u_i}_{L^6}^3+\norm{v_i}_{L^2}^3\bigr] \biggr)\bigl(\norm{u_1-u_2}_{K^{s}\cap L^6} + \norm{v_1-v_2}_{H^{s}}\bigr)
}
for all $u_i \in C^0(\Rbb^n)\cap K^{s}(\Rbb^n)\cap L^6(\Rbb^n)$, $v_i\in C^0(\Rbb^n)\cap H^{s}(\Rbb^n)$
with $u_i(\xv), v_i(\xv) \in V$ for all $\xv\in \Rbb^n$ and $i=1,2$.
\item If $|\Fc(u,v)|\lesssim |u||v|$ for all $(u,v)\in \overline{V}^2$, then
\eqn{calccorA.5}{
\norm{\Fc(u,v)}_{K^{s}\cap L^{\frac{6}{5}}} \leq C\bigl(\norm{\Fc}_{C^s(\overline{V}^2)}\bigr)(1+\norm{u}^{s-1}_{L^\infty}+ \norm{v}^{s-1}_{L^\infty}+\norm{v}_{L^\frac{6}{5}})(\norm{u}_{K^s} +  \norm{v}_{H^s\cap L^{\frac{6}{5}}}\bigl)
}
for all $u \in C^0(\Rbb^n)\cap K^{s}(\Rbb^n)$, $v\in C^0(\Rbb^n)\cap H^{s}(\Rbb^n)\cap
L^{\frac{6}{5}}(\Rbb^n)$
with $u(\xv), v(\xv) \in V$ for all $\xv\in \Rbb^n$.
\item If $|\Gc(u,v)|\lesssim |u||v|$ for all $(u,v) \in \overline{V}^2$, then
\alin{calccorA.6}{
&\norm{\Gc(u_1,v_1)-\Gc(u_2,v_2)}_{K^{s}\cap L^{\frac{6}{5}}}
 \leq C\bigl(\norm{\Gc}_{C^{s+1}(\overline{V}^2)}\bigr)  \\
 &\qquad \times \biggl(1+\sum_{i=1}^2\Bigl[\norm{u_i}^{s-1}_{L^\infty} + \norm{v_i}^{s-1}_{L^\infty}
 +\norm{v_i}_{L^{\frac{6}{5}}}^2\Bigr] \biggr)\bigl(\norm{u_1-u_2}_{K^{s}} + \norm{v_1-v_2}_{H^{s}\cap L^{\frac{6}{5}}}\bigr)
}
for all $u_i \in C^0(\Rbb^n)\cap K^{s}(\Rbb^n)$, $v_i\in C^0(\Rbb^n)\cap H^{s}(\Rbb^n)
\cap L^{\frac{6}{5}}(\Rbb^n)$
with $u_i(\xv), v_i(\xv) \in V$ for all $\xv\in \Rbb^n$ and $i=1,2$.
\end{enumerate}
\end{cor}
\begin{proof}
We only prove statements (i) and (ii) as (iii)-(vi) follow from similar arguments.
\smallskip

\noindent\textit{(i)}: We estimate
\alin{calccorA.7}{
&\norm{f(u,v)}_{H^s} \lesssim \norm{f(u,v)}_{K^s} + \norm{f(u,v)}_{L^2} \\
&\qquad \leq C\bigl(\norm{f}_{C^s(\overline{V}^2)}\bigr)\Bigl[(1+\norm{u}^{s-1}_{L^\infty}+ \norm{v}^{s-1}_{L^\infty})(\norm{u}_{K^s} +  \norm{v}_{K^s}\bigl) + \norm{v^3}_{L^2} \Bigr] && \text{by Theorem \ref{calcpropE}}\\
&\qquad \leq C\bigl(\norm{f}_{C^s(\overline{V}^2)}\bigr)\Bigl[(1+\norm{u}^{s-1}_{L^\infty}+ \norm{v}^{s-1}_{L^\infty})(\norm{u}_{K^s} +  \norm{v}_{K^s}\bigl) + \norm{v}_{L^6}^3 \Bigr] && \text{by Theorem \ref{Holder}}\\
&\qquad \leq C\bigl(\norm{f}_{C^s(\overline{V}^2)}\bigr)(1+\norm{u}^{s-1}_{L^\infty}+ \norm{v}^{s-1}_{L^\infty}+\norm{v}_{L^6}^2)(\norm{u}_{K^s} +  \norm{v}_{K^s\cap L^6}).
}

\noindent\textit{(ii)}: Since $|g(u,v)|\lesssim |v|^3$, it follows from a Taylor expansion that we can write
$g(u,v) = \gt(u,v)v^3$ for some $\gt \in C^{1}(\Rbb^2)$. Writing the difference
$g(u_1,v_2)-g(u_2,v_2)$ as
\eqn{calccorA.8}{
g(u_1,v_1)-g(u_2,v_2) = \bigl[\gt(u_1,v_1)-\gt(u_2,v_2)\bigr]v_1^3 + \gt(u_2,v_2)(v_1^2+v_2^2+v_1v_2)(v_1-v_2),
}
we see that, using Holder's inequality and
\eqn{calccorA.9}{
|\gt(u_1,v_2)-\gt(u_2,v_2)|\leq C\bigl(\norm{\gt}_{C^1(\overline{V}^2)}\bigr)
(|u_1-u_2|+|v_1-v_2|) \quad \forall u_1,u_2,v_1,v_2 \in \overline{V},
}
that
\lalign{calccorA.10}{
\norm{g(u_1,v_1)-g(u_2,v_2)}_{L^2} \leq  C\bigl(\norm{\gt}_{C^1(\overline{V}^2)}&\bigr)\bigl(1+ \norm{v_1}_{L^6}^3
+\norm{v_2}_{L^6}^3\bigr)\bigl(\norm{u_1-u_2}_{L^\infty} \notag \\
&+\norm{v_1-v_2}_{L^\infty}
+ \norm{v_1-v_2}_{L^6}\bigr). \label{calccorA.10.1}
}
So then
\alin{calccorA.11}{
&\norm{g(u_1,v_1)-g(u_2,v_2)}_{H^s} \lesssim \norm{g(u_1,v_1)-g(u_2,v_2)}_{K^s} + \norm{g(u_1,v_1)-g(u_2,v_2)}_{L^2} \\
& \qquad \lesssim  C\bigl(\norm{g}_{C^{s+1}(\overline{V}^2)}\bigr)\biggl(1+\sum_{i=1}^2\bigl[\norm{u_i}^{s-1}_{L^\infty} + \norm{v_i}^{s-1}_{L^\infty}+\norm{v_i}_{L^6}^3\bigr] \biggr)\bigl(\norm{u_1-u_2}_{K^{s}} + \norm{v_1-v_2}_{K^{s}\cap L^6}\bigr)
}
by Theorem \ref{calcpropC} and estimate \eqref{calccorA.10.1}.
\end{proof}

\begin{rem} \label{MoserZidrem}
Theorem \ref{calcpropE} and Corollary \ref{calccorA} also hold for the obvious vector valued generalizations.
\end{rem}

%% file: elliptic.tex
\sect{elliptic}{Elliptic estimates}

In this appendix, we state some well known elliptic estimates and use these to establish related elliptic
estimates that will be used throughout this article.


Letting
\leqn{lap}{
\Delta = \delta^{IJ}\del{I}\del{J} \qquad (I,J=1,2,3)
}
denote the flat Laplacian on $\Rbb^3$, we recall that the Riesz potential of a function $f(\xv)$ that decays sufficiently rapidly at infinity
is defined by
\leqn{Rieszpotdef}{
(-\Delta)^{-\frac{\alpha}{2}}(f)(\xv) = 2^{-\alpha}\pi^{-\frac{3}{2}}
\frac{\Gamma\left(\frac{3-\alpha}{2}\right)}{\Gamma\left(\frac{\alpha}{2}\right)}\int_{\Rbb^3} \frac{f(\yv)}{|\xv-\yv|^{3-\alpha}} d^3 \yv,
}
which we note coincides with the negative of the Newtonian potential
\leqn{Newtpot}{
\Delta^{-1}(f)(\xv) = -\frac{1}{4\pi} \int_{\Rbb^3} \frac{f(\yv)}{|\xv-\yv|} d^3 \yv
}
for $\alpha=2$. The following mapping property of
the Riesz potential \cite[Theorem 6.1.3]{Grafakos:2009} will be needed below and elsewhere.
\begin{thm} \label{Rieszpotthm}
Suppose $0<\alpha < 3$ and $1<p<3/\alpha$. Then
\eqn{Rieszptthm}{
\norm{(-\Delta)^{-\frac{\alpha}{2}}(f)}_{L^{\frac{3p}{3-\alpha p}}}\lesssim \norm{f}_{L^p}
}
for all $f\in L^p(\Rbb^3)$.
\end{thm}
We also require the following estimates for the Riesz transform $\Rf_I$ \cite[Corollary 4.2.8]{Grafakos:2008}, which is defined by
\leqn{Riesztransdef}{
\Rf_I = -\del{I}(-\Delta)^{-\frac{1}{2}}.
}
\begin{thm} \label{Riesztransthm}
Suppose $1<p<\infty$ and $s\in \Zbb_{\geq 0}$. Then
\gath{Riesztransthm.1}{
\norm{\Rf_I(f)}_{W^{s,p}} \lesssim \norm{f}_{W^{s,p}}
}
for all $f\in W^{s,p}(\Rbb^3)$.
\end{thm}
When defined, the Riesz potential and transform satisfy the identities
\lalign{Rieszid}{
\Rf_{I} (-\Delta)^{-\frac{1}{2}}(f) &= (-\Delta)^{-\frac{1}{2}} \Rf_{I}(f) = \del{I} \Delta^{-1}(f) = \Delta^{-1}\del{I}(f),
\label{Rieszid.1}\\
\Rf_{I}\Rf_{J}(f) &= -\del{I}\del{J}\Delta^{-1}(f) = -\Delta^{-1}\del{I}\del{J}(f) = -\del{I} \Delta^{-1} \del{J}(f)
\label{Rieszid.2}
\intertext{and}
\delta^{IJ}\Rf_{I}\Rf_{J}(f) &= -\Delta \Delta^{-1}(f) = -\Delta^{-1}\Delta (f)=-f. \label{Rieszid.3}
}

Next, we prove a variation of the $L^\infty$ estimate for the Newtonian potential  from \cite[Appendix 1]{Rendall:1994}  that will be needed below.
\begin{prop} \label{ellipCprop}
If $1\leq p < 3/2$ and $3/2 < q \leq \infty$, then
\eqn{ellipCprop1}{
\norm{\Delta^{-1}f}_{L^\infty} \lesssim \norm{f}_{L^p} + \norm{f}_{L^q}
}
for all $f\in L^p(\Rbb^3)\cap L^q(\Rbb^3)$.
\end{prop}
\begin{proof}
First, we split $-4\pi \Delta^{-1}(f)$ as
\eqn{ellipCprop3}{
-4\pi \Delta^{-1}(f)(\xv) = \int_{|\xv-\yv|<R} \frac{f(\yv)}{|\xv-\yv|} d^3 \yv + \int_{|\xv-\yv|>R} \frac{f(\yv)}{|\xv-\yv|} d^3 \yv.
}
Estimating the first piece, we find, using H\"{o}lder's inequality, that
\lalign{ellipCprop4}{
\left|\int_{|\xv-\yv|<R} \frac{f(\yv)}{|\xv-\yv|} d^3 \yv \right|  &\leq \left(\int_{|\xv-\yv|<R} \frac{1}{|\xv-\yv|^{q'}} d^3 \yv \right)^{1/q'} \norm{f}_{L^q(B_R(\xv))} \notag \\
&\leq \left(\int_{|\yv|<R} \frac{1}{|\yv|^{q'}} d^3 \yv \right)^{1/q'} \norm{f}_{L^q}\notag \\
& \lesssim  R^{(3-q')/q'} \norm{f}_{L^q}, \label{ellipCprop4.1}
}
where
\leqn{ellipCprop5}{
q' < 3  \AND  \frac{1}{q}+\frac{1}{q'} = 1.
}
A similar estimate for the second piece, again using H\"{o}lder's inequality, shows that
\leqn{ellipCprop6}{
\left|\int_{|\xv-\yv|>R} \frac{f(\yv)}{|\xv-\yv|} d^3 \yv \right| \lesssim  R^{(3-p')/p'} \norm{f}_{L^p}
}
where
\leqn{ellipCprop7}{
p' > 3  \AND  \frac{1}{p}+\frac{1}{p'} = 1.
}
Setting
\eqn{ellipCprop8}{
R = \left(\frac{\norm{f}_{L^p}}{\norm{f}_{L^q}}\right)^{\frac{1}{3\left(\frac{1}{p}-\frac{1}{q}\right)}}
}
in \eqref{ellipCprop4.1} and \eqref{ellipCprop6} gives
\eqn{ellipCprop9}{
\norm{\Delta^{-1}f}_{L^\infty} \lesssim \norm{f}^{1-\sigma}_{L^q} \norm{f}^\sigma_{L^p} \qquad \sigma = \frac{1}{3\left(\frac{1}{p}-\frac{1}{q}\right)}\left[2-\frac{3}{q}\right],
}
where $(q,p)$ satisfy $1\leq p < 3/2$ and $3/2 < q \leq \infty$
by \eqref{ellipCprop5} and \eqref{ellipCprop7}. Finally, applying Young's inequality, i.e. $ab\leq a^r/r+b^s/s$ with $a,b\geq 0$, $r,s>0$ and $1/r+1/s=1$,  yields the desired estimate
\eqn{ellipCprop10}{
\norm{\Delta^{-1}f}_{L^\infty} \lesssim \norm{f}_{L^q} +  \norm{f}_{L^p}.
}
\end{proof}

We are now in the position to establish the key elliptic estimate that will be crucial for the construction of initial data carried out in Section \ref{idata}.
\begin{thm} \label{ellipDthm}
Suppose $1<p<3/2$, and
\eqn{ellipDthm1}{
 a,c\in L^\infty(\Rbb^3),\quad   b\in L^\infty(\Rbb^3)\cap L^3(\Rbb^3), \quad Dw \in L^{\frac{3p}{3-p}}(\Rbb^3), \quad f,D^2 v \in L^p(\Rbb^3)\cap L^{\frac{3p}{3-p}}(\Rbb^3).
}
Then
\eqn{ellipDthm1a}{
u = \Delta^{-1}\bigl(a^{IJ}\del{I}\del{J}v + b^I\del{I}w + cf \bigr)
}
is the unique solution in $L^{\frac{3p}{3-2p}}(\Rbb^3)\cap W^{2,p}_{\emph{loc}}(\Rbb^3)$
of
\eqn{elliptA}{
\Delta u = a^{IJ}\del{I}\del{J} v  + b^I \del{I}w +cf
}
and satisfies the estimate
\alin{ellipDthm2}{
&\norm{u}_{L^\infty} + \norm{u}_{L^{\frac{3p}{3-2p}}} + \norm{Du}_{W^{1,\frac{3p}{3-p}}} + \norm{D^2 u}_{L^p} \lesssim
\norm{a}_{L^\infty}\Bigl(\norm{D^2 v}_{L^p} \\
 &\qquad + \norm{D^2 v}_{L^{\frac{3p}{3-p}}}\Bigl) +  \Bigl(\norm{b}_{L^\infty}+\norm{b}_{L^3}\Bigr)\norm{Dw}_{L^{\frac{3p}{3-p}}} + \norm{c}_{L^\infty} \bigl(\norm{f}_{L^p}
 + \norm{f}_{L^{\frac{3p}{3-p}}}\bigr).
}
In particular for $p=6/5$,
\alin{ellipDthm3}{
&\norm{u}_{L^\infty} + \norm{u}_{L^{6}} + \norm{Du}_{H^{1}} + \norm{D^2 u}_{L^{6/5}} \lesssim
\norm{a}_{L^\infty}\Bigl(\norm{D^2 v}_{L^{6/5}} \\
 &\qquad + \norm{D^2 v}_{L^{2}}\Bigl) +  \Bigl(\norm{b}_{L^\infty}+\norm{b}_{L^3}\Bigr)\norm{Dw}_{L^{2}} + \norm{c}_{L^\infty} \bigl(\norm{f}_{L^{6/5}}
 + \norm{f}_{L^{2}}\bigr).
}
\end{thm}
\begin{proof}
Setting
\eqn{ellipDthm4}{
F = a^{IJ}\del{I}\del{J} v  + b^I \del{I}w +cf,
}
it follows easily from H\"{o}lder's inequality that
\alin{ellipDthm5}{
\norm{F}_{L^p} &\lesssim \norm{a}_{L^\infty}\norm{D^2 v}_{L^p} + \norm{b}_{L^3}\norm{Dw}_{L^{\frac{3p}{3-p}}} + \norm{c}_{L^\infty}\norm{f}_{L^p} 
\intertext{and}
\norm{F}_{L^{\frac{3p}{3-p}}} &\lesssim \norm{a}_{L^\infty}\norm{D^2 v}_{L^{\frac{3p}{3-p}}} + \norm{b}_{L^\infty}\norm{Dw}_{L^{\frac{3p}{3-p}}} +
\norm{c}_{L^\infty}\norm{f}_{L^{\frac{3p}{3-p}}}.  
}
Since $1<p<3/2$, it follows from Theorems \ref{Rieszpotthm} and \ref{Riesztransthm}, Proposition \ref{ellipCprop},
the identities
\eqref{Rieszid.1}-\eqref{Rieszid.3}, and the above estimates that
\eqn{ellipDthm6}{
u = \Delta^{-1}(F)
}
is well defined and satisfies
\eqn{ellipDthm7}{
\norm{u}_{L^\infty}+\norm{u}_{ L^{\frac{3p}{3-2p}}}+ \norm{Du}_{W^{1,\frac{3p}{3-p}}}+ \norm{D^2 u}_{L^p} \lesssim \norm{F}_{L^p}+\norm{F}_{L^{\frac{3p}{3-p}}}
}
and
\eqn{ellipDthm8}{
\Delta u = F.
}
\end{proof}

%% file: lr.bbl
\providecommand{\bysame}{\leavevmode\hbox to3em{\hrulefill}\thinspace}
\providecommand{\MR}{\relax\ifhmode\unskip\space\fi MR }
\providecommand{\MRhref}[2]{%
  \href{http://www.ams.org/mathscinet-getitem?mr=#1}{#2}
}
\providecommand{\href}[2]{#2}
\begin{thebibliography}{10}

\bibitem{AdamsFournier:2003}
R.A. Adams and J.~Fournier, \emph{Sobolev spaces}, $2^{\text{nd}}$ ed.,
  Academic Press, 2003.

\bibitem{Blanchet_et_al:2005}
L.~Blanchet, G.~Faye, and S.~Nissanke, \emph{On the structure of the
  post-{N}ewtonian expansion in general relativity}, Phys. Rev. D \textbf{72}
  (2005), 44024.

\bibitem{Blanchet:2014}
Luc Blanchet, \emph{Gravitational radiation from post-{N}ewtonian sources and
  inspiralling compact binaries}, Living Reviews in Relativity \textbf{17}
  (2014), no.~2.

\bibitem{BrowningKreiss:1982}
G.~Browning and H.O. Kreiss, \emph{Problems with different time scales for
  nonlinear partial differential equations}, SIAM J. Appl. Math. \textbf{42}
  (1982), 704--718.

\bibitem{BuchertRasanen:2012}
T.~Buchert and S.~R\"{a}s\"{a}nen, \emph{Backreaction in late-time cosmology},
  Annual Review of Nuclear and Particle Science \textbf{62} (2012), 57--79.

\bibitem{Chandrasekhar:1965}
S.~Chandrasekhar, \emph{The post-{N}ewtonian equations of hydrodynamics in
  general relativity}, Ap. J. \textbf{142} (1965), 1488--1512.

\bibitem{ChoquetBruhat:2009}
Y.~Choquet-Bruhat, \emph{General relativity and the {E}instein equations},
  Oxford University Press, 2009.

\bibitem{Clarkson_etal:2011}
C.~Clarkson, G.~Ellis, J.~Larena, and O.~Umeh, \emph{Does the growth of
  structure affect our dynamical models of the universe? the averaging,
  backreaction and fitting problems in cosmology}, Rept. Prog. Phys.
  \textbf{74} (2011), 112901.

\bibitem{Dautcourt:1964}
G.~Dautcourt, \emph{Die {N}ewtonsche {G}ravitationstheorie als strenger
  {G}renzfall der {A}llgemeinen {R}elativit\"{a}tstheorie}, Acta Phys. Polonica
  \textbf{25} (1964), 637--646.

\bibitem{Ehlers:1986}
J.~Ehlers, \emph{On limit relations between, and approximative explanations of,
  physical theories}, Logic, methodology and philosophy of science VII
  (B.~Marcus, G.J.W. Dorn, and P.~Weingartner, eds.), North Holland, 1986,
  pp.~387--403.

\bibitem{Einstein_et_al:1938}
A.~Einstein, L.~Infeld, and B.~Hoffmann, \emph{The gravitational equations and
  the problem of motion}, Ann. Math. \textbf{39} (1938), 65--100.

\bibitem{Ellis:2011}
G.~F.~R. Ellis, \emph{Inhomogeneity efffects in cosmology}, Class. Qauntum
  Grav. \textbf{28} (2011), 164001.

\bibitem{Friedman:1976}
A.~Friedman, \emph{Partial differential equations}, Krieger Publishing Company,
  1976.

\bibitem{FutamaseItoh:2007}
Toshifumi Futamase and Yousuke Itoh, \emph{The post-{N}ewtonian approximation
  for relativistic compact binaries}, Living Reviews in Relativity \textbf{10}
  (2007), no.~2.

\bibitem{Grafakos:2008}
L.~Grafakos, \emph{Classical {F}ourier analysis}, $2^{\text{nd}}$ ed.,
  Springer, 2008.

\bibitem{Grafakos:2009}
\bysame, \emph{Modern {F}ourier analysis}, $2^{\text{nd}}$ ed., Springer, 2009.

\bibitem{Green_Wald:2011}
S.R. Green and R.M. Wald, \emph{A new framework for analyzing the effects of
  small scale inhomogeneities in cosmology}, Phys. Rev. D \textbf{83} (2011),
  084020.

\bibitem{Green_Wald:2012}
\bysame, \emph{Newtonian and relativistic cosmologies}, Phys. Rev. D
  \textbf{85} (2012), 063512.

\bibitem{HwangNoh:2013}
J.~Hwang and H.~Noh, \emph{Newtonian limit of fully nonlinear cosmological
  perturbations in {E}instein's gravity}, JCAP \textbf{04} (2013), 035.

\bibitem{Hwangetal:2008}
J.~Hwang, H.~Noh, and D.~Puetzfeld, \emph{Cosmological non-linear hydrodynamics
  with post-{N}ewtonian corrections}, JCAP \textbf{03} (2008), 010.

\bibitem{KlainermanMajda:1981}
S.~Klainerman and A.~Majda, \emph{Singular perturbations of quasilinear
  hyperbolic systems with large parameters and the incompressible limit of
  compressible fluids}, Comm. Pure Appl. Math. \textbf{34} (1981), 481--524.

\bibitem{KlainermanMajda:1982}
\bysame, \emph{Compressible and incompressible fluids}, Comm. Pure Appl. Math.
  \textbf{35} (1982), 629--651.

\bibitem{KopeikinPetrov:2013}
S.M. Kopeikin and A.N. Petrov, \emph{Post-{N}ewtonian celestial dynamics in
  cosmology: {F}ield equations}, Phys. Rev. D \textbf{87} (2013), 044029.

\bibitem{KopeikinPetrov:2014}
\bysame, \emph{Dynamic field theory and equations of motion in cosmology},
  Annals of Physics (to appear) (2014).

\bibitem{Kreiss:1980}
H.O. Kreiss, \emph{Problems with different time scales for partial differential
  equations}, Comm. Pure Appl. Math. \textbf{33} (1980), 399--439.

\bibitem{Kunzle:1972}
H.P. K\"{u}nzle, \emph{Galilei and {L}orentz structures on space-time:
  comparison of the corresponding geometry and physics}, Ann. Inst. Henri
  Poincar\'{e} \textbf{17} (1972), 337--362.

\bibitem{Kunzle:1976}
\bysame, \emph{Covariant {N}ewtonian limit of {L}orentz space-times}, Gen. Rel.
  Grav. \textbf{7} (1976), 445--457.

\bibitem{KunzleDuval:1986}
H.P. K\"{u}nzle and C.~Duval, \emph{Relativistic and non-relativistic classical
  field theory on five-dimensional spacetime}, Class. Quantum Grav. \textbf{3}
  (1986), 957--974.

\bibitem{Lottermoser:1992}
M.~Lottermoser, \emph{A convergent post-{N}ewtonian approximation for the
  constraint equations in general relativity}, Annales de l'institut Henri
  Poincar\'{e} (A) Physique th\'{e}orique \textbf{57} (1992), 279--317.

\bibitem{Majda:1984}
A.~Majda, \emph{Compressible fluid flow and systems of conservation laws in
  several space variables}, Springer, 1984.

\bibitem{MatarreseTerranova:1996}
S.~Matarrese and D.~Terranova, \emph{Post-{N}ewtonian cosmological dynamics in
  {L}agrangian coordinates}, Mon. Not. Roy. Astron. Soc. \textbf{283}
  (400-418), 1996.

\bibitem{Oliynyk:CMP_2007}
T.~A. Oliynyk, \emph{The {N}ewtonian limit for perfect fluids}, Commun. Math.
  Phys. \textbf{276} (2007), 131--188.

\bibitem{Oliynyk:CMP_2009}
\bysame, \emph{Post-{N}ewtonian expansions for perfect fluids}, Commun. Math.
  Phys. \textbf{288} (2009), 847--886.

\bibitem{Oliynyk:CMP_2010}
\bysame, \emph{Cosmological post-{N}ewtonian expansions to arbitrary order},
  Commun. Math. Phys. \textbf{295} (2010), 431--463.

\bibitem{Oliynyk:JHDE_2010}
\bysame, \emph{A rigorous formulation of the cosmological {N}ewtonian limit
  without averaging}, JHDE \textbf{7} (2010), 405--431.

\bibitem{Oliynyk:PRD_2014}
T.A. Oliynyk, \emph{Cosmological {N}ewtonian limit}, Phys. Rev. D \textbf{89}
  (2014), 124002.

\bibitem{Rasanen:2010}
S.~R\"{a}s\"{a}nen, \emph{Applicability of the linearly perturbed {FRW} metric
  and {N}ewtonian cosmology}, Phys. Rev. D \textbf{81} (2010), 103512.

\bibitem{Rendall:1992}
A.D. Rendall, \emph{The initial value problem for a class of general
  relativistic fluid bodies}, J. Math. Phys. \textbf{33} (1992), 1047--1053.

\bibitem{Rendall:1994}
\bysame, \emph{The {N}ewtonian limit for asymptotically flat solutions of the
  {V}lasov-{E}instein system}, Comm. Math. Phys. \textbf{163} (1994), 89--112.

\bibitem{TaylorIII:1996}
M.E. Taylor, \emph{Partial differential equations iii: nonlinear equations},
  Springer, 1996.

\end{thebibliography}
